\documentclass[
twocolumn,
showpacs,
prb,
 superscriptaddress,
 nofootinbib,eqsecnum,floatfix
 ]{revtex4}%

\usepackage{color}

\newcommand{\bpsi}{\mbox{\boldmath $\psi$}}
\newcommand{\nnabla}{\boldsymbol{\nabla}}

\newcommand{\req}[1]{Eq.~(\ref{#1})}
\newcommand{\reqs}[1]{Eqs.~(\ref{#1})}
\newcommand{\rref}[1]{(\ref{#1})}

\newcommand{\n}{\mathbf{n}}
\newcommand{\nn}{\mathbf{n}}
\newcommand{\mm}{\mathbf{m}}
\renewcommand{\r}{\mathbf{r}}

\renewcommand{\k}{\mathbf{k}}

\newcommand{\beq}{\begin{equation}}
\newcommand{\eeq}{\end{equation}}
\newcommand{\be}{\begin{equation}}
\newcommand{\ee}{\end{equation}}
\newcommand{\beqa}{\begin{eqnarray}}
\newcommand{\eeqa}{\end{eqnarray}}
\newcommand{\bea}{\begin{eqnarray}}
\newcommand{\eea}{\end{eqnarray}}
\usepackage{amssymb}
\usepackage{array}
\usepackage{amsmath}
\usepackage{graphicx}
\usepackage{dcolumn}
\usepackage{bm}
\usepackage[mathscr]{eucal}

\usepackage{amsfonts}%

\begin{document}
\title{Spontaneous symmetry breakings  in graphene 
subjected to in-plane magnetic field}
\author{I.L. Aleiner}
\affiliation{ Physics Department, Columbia University, New York, NY
  10027, USA}

\author{D. E. Kharzeev}
\affiliation{Department of Physics,
Brookhaven National Laboratory, Upton, NY 11973-5000, USA}

%\author{S. A. Reyes}
%\affiliation{Department of Condensed Matter Physics and Materials
%  Science, Brookhaven National Laboratory, Upton, NY 11973-5000, USA}
%\affiliation
%{Department of Physics and Astronomy, 
%SUNY at Stony Brook, Stony Brook, NY 11794-3800, USA}

\author{A. M. Tsvelik}
\affiliation{Department of Condensed Matter Physics and Materials
  Science, Brookhaven National Laboratory, Upton, NY 11973-5000, USA}
\affiliation
{Department of Physics and Astronomy, 
SUNY at Stony Brook, Stony Brook, NY 11794-3800, USA}
\date{\today}

\begin{abstract}
Application of the magnetic field parallel to the plane
of the graphene sheet leads to the formation of 
electron- and hole-like Fermi surfaces. Such situation is shown
to be unstable with respect to the formation of an excitonic condensate even for an
arbitrary weak magnetic field and interaction strength. 
At temperatures lower than the mean-field temperature
the order parameter amplitude is formed.  
The order parameter itself is a $U(2)$ matrix
allowing for the combined rotations in the spin and valley spaces. 
These rotations smoothly interpolate between site and bond centered 
spin density waves and spin flux states. The trigonal warping,
short range interactions, and the three particle Umklapp processes
freeze some degrees of freedom at temperatures much smaller than
the mean-field transition temperature and make either
Berezinskii-Kosterlitz-Thouless 
(driven either by vortices or half-vortices)
or Ising type transitions possible.
Strong logarithmic renormalization for the coupling constants
of these terms by the Coulomb interaction are calculated
within one-loop renormalization group. It is found that 
 in the presence of the Coulomb interaction some
short range interaction terms become  much greater than one
might  expect from the naive dimensionality counting.

\end{abstract}

\pacs{73.63.-b,81.05.Uw, 72.15.Rn }
\maketitle

\section{Introduction}

%\vspace*{5cm}

The fabrication of the graphene (graphite monolayers) \cite{novo,review} and subsequent
observation of the Integer Quantum Hall \cite{novohall,zhang} in this
layer produced a splash of theoretical and experimental activity.

Though the transport properties of graphene
are controlled by the impurities, it is still worthwhile to understand
the phase diagram of the completely clean graphene.

The main purpose of this paper is to point out that
 an application of magnetic field in  graphene plane facilitates
a spontaneous symmetry breaking. Though the dimensionless Coulomb
 coupling $e^2/\hbar v$ in graphene is large at large energies,
 it undergoes strong downward renormalization at small energies so
 that the analysis for weak magnetic fields can be safely 
carried out without resorting to any uncontrollable approximations
 \cite{khv}.

To achieve this goal  and understand the effects of the naively
dimensionally irrelevant terms (such as trigonal warping, short
range part of the interaction, Umklapp terms) we considered their
logarithmic renormalization by the long-range Coulomb interaction and
found some un-expected  results. Earlier the effect of the
Coulomb interaction was considered for the isotropic terms only \cite{Abrikosov,vozmediano,son}.

 The remainder of the paper is organized as follows. In Section II we discuss
 symmetries of a two-dimensional graphene sheet  and write down its 
low energy Hamiltonian. In Section III we describe  physical reasons
 for the instability and suggest  the order parameter. 
In Section IV we write down  the Landau-Ginzburg free energy functional,
 discuss  thermal fluctuations and describe the phase diagram.
 Section V is devoted to the analysis of the microscopics: we study
 the energy dependence of the effective coupling and 
renormalization of the leading anisotropies in graphene. Section VI
 contains the summary and conclusions.
Some auxiliary material is relegated into
two Appendices.  

\section{ Symmetries of the system and
the model low energy Hamiltonian. } 
\label{sec:2}\textsc{}

The purpose of this section is to write down the low energy
energy field theory 
to describe the electron-electron interaction in graphene. Our consideration will be based on the
discrete symmetries of the lattice only and we will not appeal to
any microscopic model \cite{Pikus}.

\begin{figure}

\includegraphics[width=0.45\textwidth]{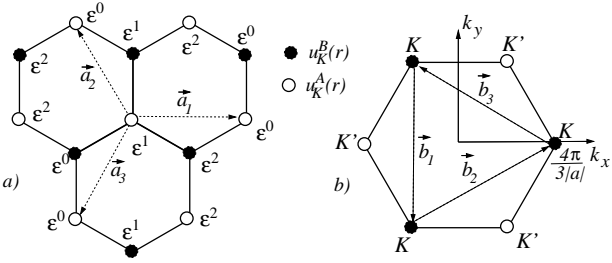}

\caption{a)
The hexagonal lattice of graphene with the shortest
translation vectors $\vec{a}_{1,2,3}$. Circles shows the
positions of the maxima of the absolute values of Bloch functions
$u^{A,B}_K(\r)$. Their relative phases are shown, $\varepsilon=\exp(i2\pi/3)$.
b) The first Brillouin zone and the shortest translation vectors of the
reciprocal lattice $\vec{b}_{1,2,3}$. Two non-equivalent Dirac cones
are formed in the vicinity of points $K$ and $K'$.
}
\label{fig1}
\end{figure}

The effective low energy field theory of graphene is
constructed by the  
factorization of the original fermionic fields
$\Psi_\sigma(\r;\tau), \sigma=\uparrow,\downarrow$ in terms of
the oscillatory Bloch functions corresponding
to the $K,K'$ points of the Brillouin zone, see Fig.~\ref{fig1},
\be
\Psi_\sigma(\r;\tau)=\vec{\psi}_\sigma(\r,\tau)*\vec{u}(\r).
\label{eq:1}
\ee
where  $\psi_\sigma(\r,\tau)$ is the four component
fermionic field which can vary only over distances much larger
than the lattice constant,  and $\vec{u}(\r)$ is
the four-dimensional vector of the Bloch functions
whose structure is described below.

Consider the Bloch functions
$(u^A(\r)_{K}, u^B(\r)_{K})$ forming a basis for two-dimensional
irreducible
representation of the wave-vector symmetry group ${\cal C}_{3v}$.
(In the tight-binding picture those Bloch functions are peaked
on the corresponding sub-lattices, see Fig.~\ref{fig1}).
The overall point symmetry group is ${\cal C}_{6v}$ and
thus the wave-functions $u^A_{K'}
=\left[u^A_{K}\right]^*, u^B_{K'}
=\left[u^B_{K}\right]^*$ also have to be included as points $K',K$
are connected to each other by $C_2$ rotation and by time reversal symmetry.

They  are conveniently joined in a vector \cite{Efetov}
\be
\vec{u}^T=\begin{pmatrix}
\begin{pmatrix} u^A_{K}; 
u^B_{K}
\end{pmatrix}_{AB}
\begin{pmatrix}
u^B_{K'}; -u^A_{K'}
\end{pmatrix}_{AB}
\end{pmatrix}_{KK'}
\label{eq:2}
\ee 
forming the basis of the four-dimensional irreducible representation 
of the planar symmetry group of graphene $\sum_{j_1,j_2}
{\cal C}_{6}^v t_{j_1\vec{a}_1+j_2\vec{a}_2}$,
with the normalization condition
\be
 \int_{uc}d\r\vec{u}(\r)\cdot \vec{u}*(\r)
=4|\vec{a}_1\times\vec{a}_2|.
\label{normalization}
\ee
Hereinafter, $\int_{uc}$ denotes  integration within
the unit cell.

Thus, the fermionic  field describing all relevant degrees of freedom,
$\bpsi^T=(\psi_\uparrow,\psi_\downarrow )$, has  eight
components.
This eight-dimensional space is  represented, see \req{eq:2},
as a direct product of the valley, $(KK')$, the sub-lattice, $(AB)$,
and the spin, $(s)$, two-dimensional spaces. We will
use standard $2\times 2$ Pauli matrices $\hat{\tau}_{x,y,z}$, 
with the corresponding
subscripts to parametrize all  $8\times 8$ matrices 
describing the Hamiltonian and the symmetry properties

The partition function describing the low energy
properties of the interacting electrons in a clean graphene is given by
($\hbar=1$)
\be
{\cal Z}\!\! =\!\!\int\!\! {\cal D}{\bpsi}^\dagger{\cal D}{\bpsi}
\exp\left[-\int\limits_0^{1/T}\! d\tau\! \int\!\! d\r
\left[{\bpsi}^\dagger\frac{\partial\bpsi}{\partial\tau}+H({\bpsi}^\dagger,\bpsi)\right]
\right],
\label{eq:3}
\ee

The Hamiltonian of the system has to satisfy all the discrete
symmetries of the clean graphene, and to remain invariant with respect
to transformations of the fields generated
by the rotation $C_3$, two reflections $\sigma_{v}^{x,y}$, and translations ($t$);
\begin{subequations}
\label{discrete}
\bea
C_3:& \bpsi(\r)\to 
-
\exp\left[\theta^C
\left(\r\times\nnabla + 
\frac{i}{2}
\hat{\Sigma}_z
\right)
\right]
\bpsi(\r)
\label{discretea}
\\
\sigma_v^x:&
 \bpsi(x,y)\to 
\hat{\Sigma}_x\hat{\Lambda}_z
\bpsi(x,-y)
\label{discreteb}
\\
\sigma_v^y:&
 \bpsi(x,y)\to 
\hat{\Sigma}_z\hat{\Lambda}_x
\bpsi(-x,y)\label{discretec}
\\
t:&
 \bpsi(\r)\to 
\exp\left[
i\theta^t
\hat{\Lambda}_z
\right] \bpsi(\r).
\label{discreted}
\eea
\end{subequations}
where $\theta^{C,t}=\pm 2\pi/3$.
and we introduced the matrices
\be
\begin{split}
&\hat{\Sigma}_\alpha=\hat{\tau}_{\alpha}^{AB}\otimes \openone^{KK'}\otimes
\openone^{s};\ 
\hat{\Lambda}_\alpha=
 \openone^{AB}\otimes \hat{\tau}_{\alpha}^{KK'}\otimes \openone^{s}\\
&\hat{S}_\alpha=
 \openone^{AB}\otimes  \openone^{KK'}\otimes\hat{\tau}_{\alpha} ^{s}
\end{split}
\label{matrices}
\ee
$\alpha=x,y,z$.

Continuous $U(1)$ rotations in the spin space are given
by 
\be
U(1): \quad \bpsi\to 
\exp\left(i\theta_s\hat{S}_z/2\right)\bpsi
\label{U1}
\ee
where we choose $z$-direction of the spin to be along the magnetic field.

Time reversal symmetry for the parametrization \rref{eq:2} acquires
a natural form
\be
{\cal T}: \quad \psi(\tau) \to  
\hat{\tau}_{y}^{AB} \otimes
\hat{\tau}_{y}^{KK'} \otimes 
\hat{\tau}_{y}^{s}\psi^*(-\tau),\ {\cal B}\to -{\cal B}
\label{time}
\ee
where ${\cal B}$ is the magnetic field
acting in our case only on electron spin.

Having listed the important symmetries of the problem, we
present the Hamiltonian in the form
\be
H=H_D+H_C+H_w+H_{sr}+H_{u}.
\label{Hamiltonian}
\ee

The first term describes the   Dirac-type kinetic energy
and the Zeeman energy
\be
H_D=-iv(r_c)\psi^\dagger\nnabla\cdot\hat{\vec{\Sigma}}\psi
+ {\cal B}\psi^\dagger\hat{S}_z\psi.
\label{Dirac}
\ee
where 
$\nnabla=(\partial_x,\partial_y)$, $\hat{\vec{\Sigma}}$ is defined in
\req{matrices}, the Bohr magneton and the $g$-factor are included into
the definition of $B$,
and $r_c$ is the minimal linear scale present  in the problem.
As it was pointed out in Ref. \onlinecite{Abrikosov} ( see also
Refs.~\onlinecite{vozmediano,son}),  
the velocity
$v(r_c)$ becomes scale dependent due to the 
Coulomb interaction 
\be
H_C=\frac{e^2}{2}\int {dr_1}\frac{
\left(\psi^\dagger(\r)\psi(\r)\right)
\left(\psi^\dagger(\r_1)\psi(\r_1)\right)
}{|\r-\r_1|},
\label{Coulomb}
\ee
whose strength $e^2$ cannot be renormalized
as it is the only nonlocal term in the system.

Though  the terms described by  \reqs{Dirac} and \rref{Coulomb} are
the most important ones on the dimensional grounds,
they are not sufficient to define the problem completely since 
their symmetries are much higher than of (\ref{discrete}):
\begin{subequations}
\begin{align}
&C_\infty:\ \bpsi(\r)\to 
\exp\left[\theta^C
\left(\r\times\nnabla + 
\frac{i}{2}
\hat{\Sigma}_z
\right)
\right]
\bpsi(\r)
\label{continousa}
\\
&\sigma_v: \
 \bpsi(x,y)\to 
\hat{\Sigma}_x
\bpsi(x,-y)
\label{continousb}
\\
&U(1)\otimes SU(2)\otimes SU(2):\
 \bpsi(\r)\to 
\hat{U}\left(\alpha^{\pm},\theta^s;\nn_{1,2}\right) \bpsi(\r),
\nonumber
\\ 
&
\hat{U}=
\exp\left[\frac{i\alpha^-\nn_2\cdot\vec{\hat\Lambda}\hat{S}_z }{2}
\right]
 \exp\left[
\frac{i\theta^s\hat{S}_z}{2}
\right]
\exp\left[
\frac{i\alpha^+\nn_1\cdot\vec{\hat\Lambda}}{2}
\right] 
\label{continousc}
\end{align}
than it is allowed
by \reqs{discrete}--\rref{time}.
Here $\theta^{\cdot}$ are the continuous real variables
and $\nn_{1,2}$ are three-dimensional unit vectors.
\label{continuous}
\end{subequations}

The terms lowering the symmetry of the Hamiltonian can appear
both in the kinetic energy and in the interaction Hamiltonian.
For instance, the trigonal warping of the one-electron spectrum is given by
\be
H_w=\lambda_w(r_c) r_c v(r_c)
\psi^\dagger\left[\partial_+^2
\hat{\Sigma}_+\hat{\Lambda}_z+h.c.\right]\psi
\label{warping}
\ee
where $\partial_+\equiv\partial_x+i\partial_y$,
$\hat{\Sigma}_+\equiv (\hat{\Sigma}_x+i\hat{\Sigma}_y)/2 $.
Dimensionless coupling $\lambda_w(r_c)$ is of the order of unity
at $r_c$ of the order of the lattice constant and scales
down at larger distances.

In writing down the short-range interaction, one 
can neglect the effect of the Zeeman term on
the scale of the order of the lattice constant. Thus
the $SU(2)$ invariance in the spin space must be preserved and
\footnote{All the other short range spin rotational invariant
  interaction terms can be reduced to those of \reqs{sr}
by using the identity $2\delta_{\sigma_1\sigma_2}\delta_{\sigma_3\sigma_4}
=\delta_{\sigma_1\sigma_4}\delta_{\sigma_2\sigma_3}
+ \hat{\vec{\tau}}_{\sigma_1\sigma_4} \hat{\vec{\tau}}_{\sigma_2\sigma_3}
$.
}
\begin{subequations}
\label{sr}
\bea
&&\frac{2}{r_c v(r_c)} H_{sr}=
\sum_{\alpha,\beta=x,y,z}
F_{\alpha\beta}(r_c)
\left(\psi^\dagger \Sigma_{\alpha}\Lambda_{\beta}\psi\right)^2
\\
&&+
\sum\limits_{\alpha=x,y,x}\left[
J^{\Sigma}_{\alpha}(r_c)
\left(\psi^\dagger \Sigma_{\alpha}
\psi\right)^2
+J^{\Lambda}_{\alpha}(r_c)
\left(\psi^\dagger \Lambda_{\alpha}\psi\right)^2
\right].
\nonumber
\eea

The symmetries \rref{discrete} immediately yield
the relation
\be
\begin{split}
&F_{zz}=2F_-^z+F_+^z;\ \ F_{xz}=F_{yz}=-F_-^z+F_+^z;\\  
&F_{zx}=F_{zy}=2F_-^\perp+F_+^\perp; \\ 
&F_{xx}=F_{yy}=F_{xy}=F_{yx}=
-F_-^\perp+F_+^\perp;\\
&J^{\Sigma,\Lambda}_z
=2J_-^{\Sigma,\Lambda}+J_+^{\Sigma,\Lambda}; \ \
 J^{\Sigma,\Lambda}_{x}=
J^{\Sigma,\Lambda}_{y}=
-J_-^{\Sigma,\Lambda}+J_+^{\Sigma,\Lambda}.
\end{split}
\raisetag{1.6cm}
\ee
\end{subequations}
A reason for introducing $F_\pm,J_\pm$ couplings will become
clear later in Sec.~\ref{sec:log}.
As for the numerical values of the couplings,
a very crude estimate at the scale $r_c$ of the order of
the lattice constants can be obtained by calculating
the matrix elements of the bare interaction potential:
\be
\begin{split}
&F_{\alpha\beta}=
\left(\frac{e^2}{4 v(r_c)}\right)
\int d r_1\int_{uc}
\frac{ dr_2 \rho_{\alpha\beta}(\r_1)\rho_{\alpha\beta}(\r_2)
}
{r_c|\vec{a}_1\times\vec{a}_2| |r_1-r_2|}
\\
&\rho_{\alpha\beta}(\r) \equiv \left(\vec{u}(\r)^\dagger
\Sigma_\alpha\Lambda_\beta\vec{u}(\r)
\right),
\end{split}
\label{Festimate}
\ee
and $J^{\Sigma,\Lambda}=0$.
(Obtaining  finite values of
$J^{\Sigma,\Lambda}$ requires the virtual processes  at least of the second order.)
As $\rho_{\alpha\beta}(\r)$ contains only oscillatory components,
see \reqs{eq:2} and \rref{matrices}, the integral
in \req{Festimate} is determined only by the distances  
of the order of the lattice constant. Thus, the parameters are extremely
sensitive to the details at short distances and should
be treated as entries for the low-energy theory.

Though the warping and the short range interaction
\rref{warping}-\rref{sr} lift most of the spurious symmetries
\rref{continuous}, the extra continuous  $U(1)$ symmetry is still
present corresponding to \req{discreted} with continuous $\theta^t$.
It is related  to conservation of quasimomentum which
is violated  only by the Umklapp processes.
The lowest order term satisfying the symmetries \rref{discrete}
has the form
\be
\begin{split}
H_u&=\frac{r_c^3 v(r_c)}{6}
\sum_{\alpha\beta\gamma}
{\cal F}_{\alpha\beta\gamma}(r_c)
\\
&\times\left[
\left(\bpsi^\dagger \Sigma_\alpha\hat{\Lambda}_+\bpsi\right)
\left(\bpsi^\dagger \Sigma_\beta\hat{\Lambda}_+\bpsi\right)
\left(\bpsi^\dagger \Sigma_\gamma\hat{\Lambda}_+\bpsi\right)
+h.c.\right]
\end{split}
\raisetag{1.4cm}
\label{umklapp}
\ee
where 
$\hat{\Lambda}_+\equiv (\hat{\Lambda}_x+i\hat{\Lambda}_y)/2 $,
${\cal F}_{\alpha\beta\gamma}$ is symmetric with respect to permutations of the
indices and
\be
\begin{split}
{\cal F}_{xyz}=0;\ \ {\cal F}_{zzz}=2{\cal F}_- + {\cal F}_+;\\
 {\cal F}_{zxx}={\cal F}_{zyy}=-{\cal F}_- + {\cal F}_+.
\end{split}
\label{Fumklapp}
\ee

Finally, we notice that all the listed terms
\rref{Dirac}--\rref{umklapp} remain invariant under the electron-hole
transformation
\be
e-h:\quad \bpsi\to \hat{\Sigma}_z\hat{S}_x\bpsi^*.
\label{e-h}
\ee
This electron-hole correspondence will be important for the discussion
of the instability arising in the in-plane magnetic field which we
will discuss now\footnote{The leading irrelevant term lifting
the e-h symmetry $\propto \nnabla\bpsi^\dagger\nnabla\bpsi$ 
does not break any other interesting symmetries and will not be important
for our purposes.}.

\section{Physical reasons for the  instability and the order parameter}
\label{sec:3}

\begin{figure}[h]
\includegraphics[width=0.4\textwidth]{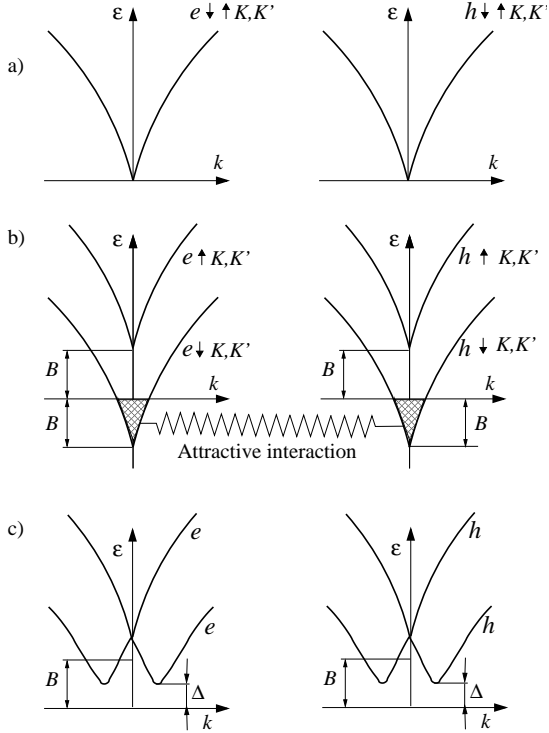}
\caption{
The mechanism of instability in parallel magnetic field, ${\cal B}$.
a) The spectra of the one-particle excitation at ${\cal B}=0$;
b) The shift of the spectra by finite ${\cal B}$ and the formation
of the electron and hole Fermi seas;
c) The  attractive Coulomb interaction between electrons and holes
leads to the instability towards a formation of the excitonic condensate
\cite{KeldyshKopaev} creating a gap in the one particle spectrum.
}
\label{fig2}
\end{figure}

Having established the form of the Hamiltonian consistent
with the symmetries of the lattice, we turn to the qualitative discussion
of the instability and determine  the target space of the order parameter.
The symmetry arguments will allow us to do this  without any actual
calculation,

Assume that no symmetries are broken at ${\cal B}=0$.
Then, the low lying excitations are  fermionic electron- and hole-
like excitations with spin $1/2$ and the dispersion 
$\epsilon(k)=v(|k|)|k|$
as shown in Fig.~\ref{fig2} (a). 
The magnetic field parallel to the plane acts  only on the spin
and hence shifts the spectrum of the excitations making the creation
of the Fermi seas for the electrons and holes energetically
favorable, Fig.~\ref{fig2} b). As the electrons and holes
have the opposite charge, the Coulomb interaction makes them to  attract each other. On the other hand, 
existence of the finite Fermi-surface
 leads to the Cooper-like instability first discovered in 
Ref.~\onlinecite{KeldyshKopaev}. This instability occurs  even for an arbitrary weak interaction
potential.

As the result, gap $\Delta_0(T)$ is formed in the one-particle
spectrum, Fig.~\ref{fig2} c). 
The resulting state is incompressible excitonic insulator with gapped 
charge excitations. The neutral excitations, however, are still quite interesting.

As the electrons and holes can be paired
with different phases and different valley indices can be
involved,
the resulting order parameter has a non-trivial matrix structure which
 will be discussed now.
As the exciton condensate is created by  pairing of an 
electron and  a hole with  opposite momenta, the order parameter $\Delta$
has to be of the form $\langle\psi_e\psi_h\rangle$. On the other 
hand the electron-hole transformation is defined in \req{e-h}.
Thus, we obtain 
\be
\hat{\Delta}=\langle\bpsi\otimes\bpsi^\dagger\rangle
=\Delta_0(T)\hat{\Sigma}_z\hat{S}_x
\label{delta0}
\ee
Here ${\Delta}$ is $8\times 8$ matrix acting in the space
discussed after \req{normalization} and
the $\hat{\Sigma}_z,\hat{S}_x$ are defined in \req{matrices}.

If  only symmetric terms \rref{Dirac}
and \rref{Coulomb} were present, the energy of the system
would be invariant with respect to the replacement
$\Delta\to \hat{U}\Delta \hat{U}^\dagger$.
where $\hat{U}$ is given by \req{continousc}.
Using \reqs{delta0} and \rref{continousc}, we obtain the most general 
form of the matrix order parameter
\begin{subequations}
\label{Q}
\be
\hat{\Delta}=\Delta_0(T)\hat{\tau}_z^{AB}\otimes \hat{Q},
\label{Qa}
\ee
where $\hat{Q}$ is the $4\times 4$ Hermitian matrix 
acting in spin and valley spaces subjected
to the following constraints
\be
\begin{split}
&\hat{Q}=\hat{Q}^\dagger;\
\hat{Q}^2=\hat{\openone}^{KK'}\otimes \hat{\openone}^{AB}:
\\
&\left(\openone^{KK'}\otimes \hat{\tau}_{z}^s\right)
\hat{Q}
\left(\openone^{KK'}\otimes \hat{\tau}_{z}^s\right)
=-\hat{Q}.
\end{split}
\label{qconstraints}
\ee
The corresponding mean field single particle spectrum consists of four branches (see Fig. 2c):
\bea
E^2_{\pm} = [\epsilon(k) \pm {\cal B}]^2 + |\Delta|^2
\eea

In terms of the angles in \reqs{continousc}, the $Q$-matrix
can be re-written as ($\alpha_-,\nn_2\to\alpha,\nn$)
\be
\hat{Q}=
\hat{\openone}^{KK'}\!\!\!\otimes
\left({\mathbf e}_1 \vec{\hat{\tau}}^{s}\right)\,\cos\alpha
+ \left(\nn \vec{\hat{\tau}}^{KK'}\right)\otimes
\left({\mathbf e}_2\vec{\hat{\tau}}^{s}\right)
 \sin\alpha
\label{Qb}
\ee
where ${\mathbf e}_{1,2}$ are two mutually orthogonal unit vectors in the plane
perpendicular to the spin quantization axis ${\mathbf e}_1=
(\cos\theta_s,\sin\theta_s,0),\ {\mathbf e}_2=
(-\sin\theta_s,\cos\theta_s,0)$.

Another way to parametrize $Q$ from \req{qconstraints} is to write
\be
\hat{Q}=\begin{pmatrix}
0 & \hat{V}\\
\hat{V}^\dagger & 0
\end{pmatrix}_s,
\quad \hat{V}^\dagger\hat{V}=\openone^{KK'},\label{V}
\ee
where $\hat{V}$ is a unitary $2\times 2$ matrix in the valley space.
Therefore, the order parameter is described
by $U(2)=SO(3)\times U(1)$ matrices. 

\end{subequations}

Before writing down the effective action or the free energy functional,
it is better to explain  the physical meaning of different
angles in \req{Qb}. To do so we will introduce
the spin density and the ``spin flux'',
 which in terms of the original (not
smooth) fermionic fields have the form
\be
\begin{split}
&\vec{s}(\r)=\frac{1}{2}\langle\Psi_\sigma^\dagger(\r)
\vec{\tau}_{\sigma\sigma'}^s
\Psi_{\sigma'}(\r)\rangle
\\
&\vec{\Phi}(\r)=\frac{i}{9}\!\!
\sum_{j_{1},j_2=1}^3\sin\frac{2\pi j_{12}}{3}
\langle\Psi_\sigma^\dagger(\r+\vec{a}_{j_2})
\vec{\tau}_{\sigma\sigma'}^s
\Psi_{\sigma'}(\r+\vec{a}_{j_1})\rangle,
\end{split}
\raisetag{1.8cm}
\label{physics}
\ee
where the translation vectors $\vec{a}_{1,2,3}$ are shown on
Fig.~\ref{fig1},
and $j_{12}=j_1-j_2$.

\begin{figure}[h]
\includegraphics[width=0.18\textwidth]{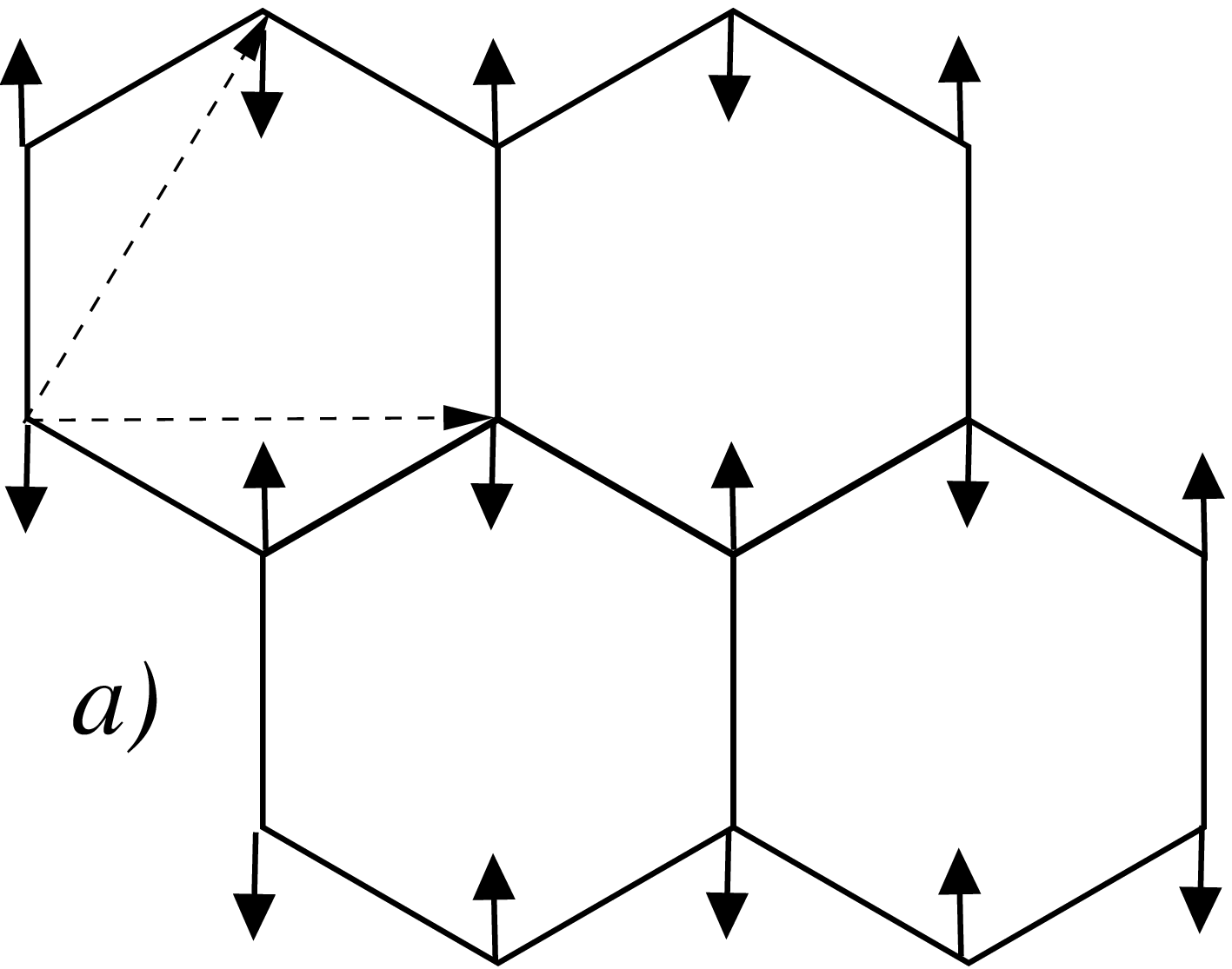}
\hspace*{0.3cm}\includegraphics[width=0.2\textwidth]{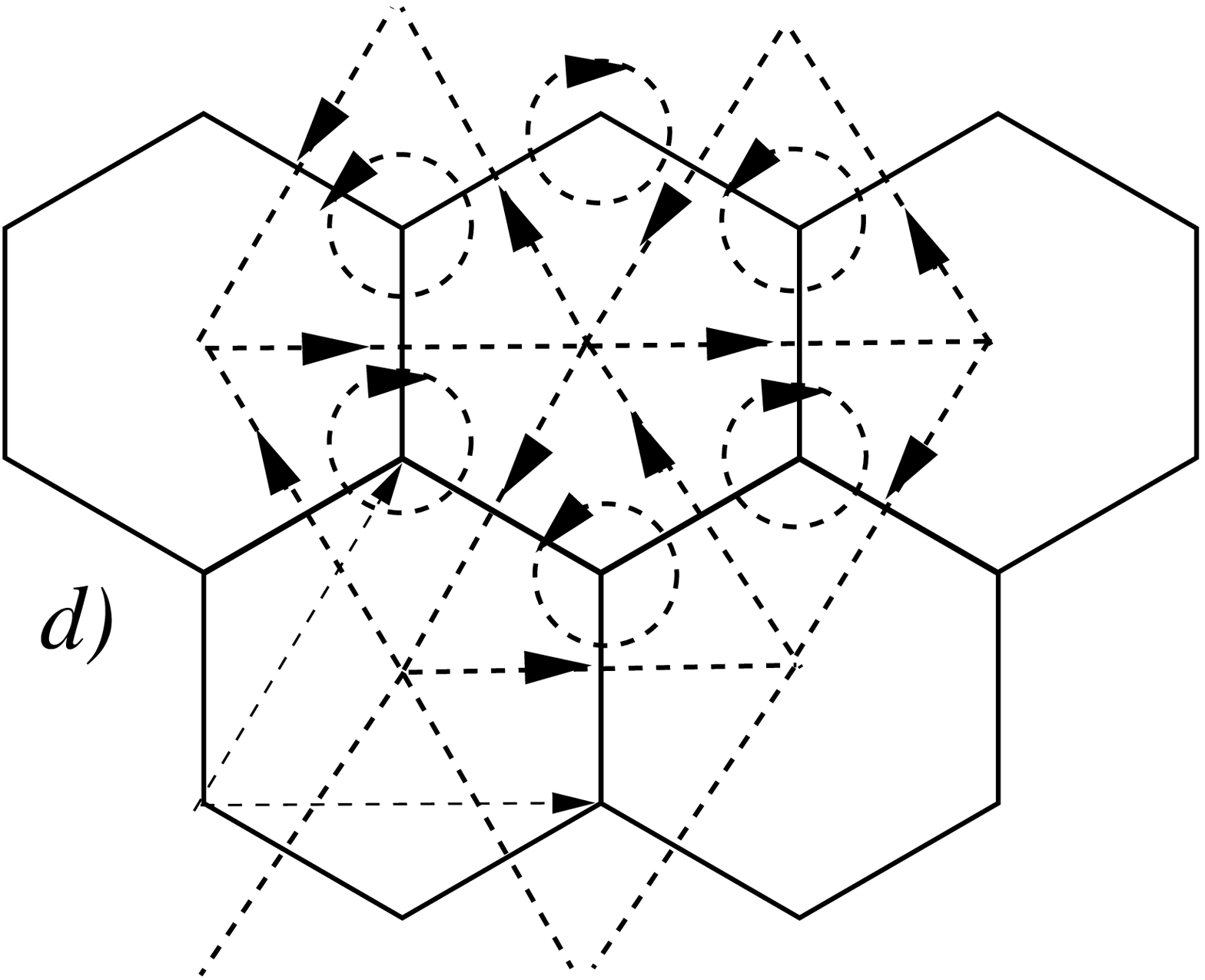}
\\[0.3cm]
\includegraphics[width=0.3\textwidth]{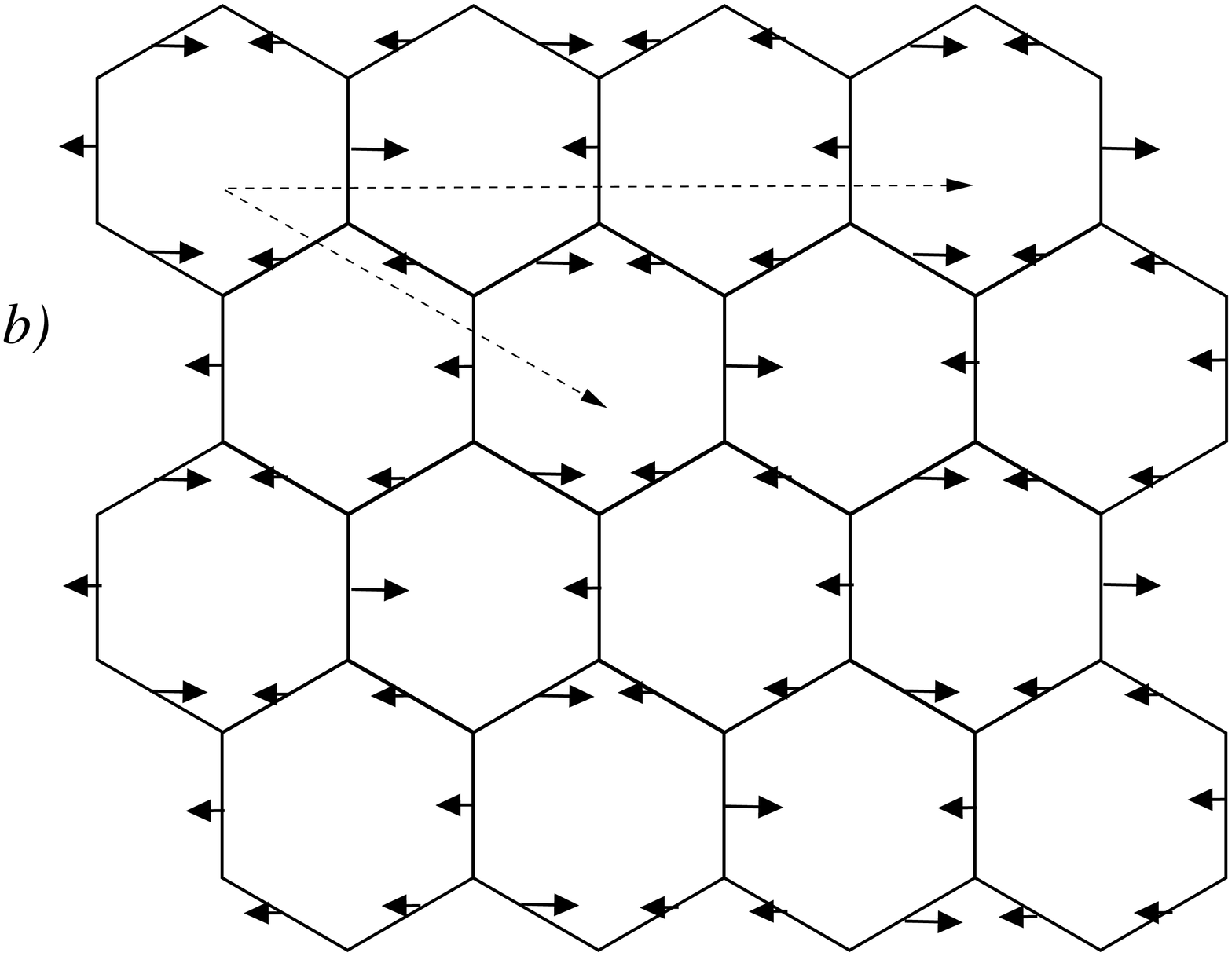}\\[0.3cm]
\includegraphics[width=0.3\textwidth]{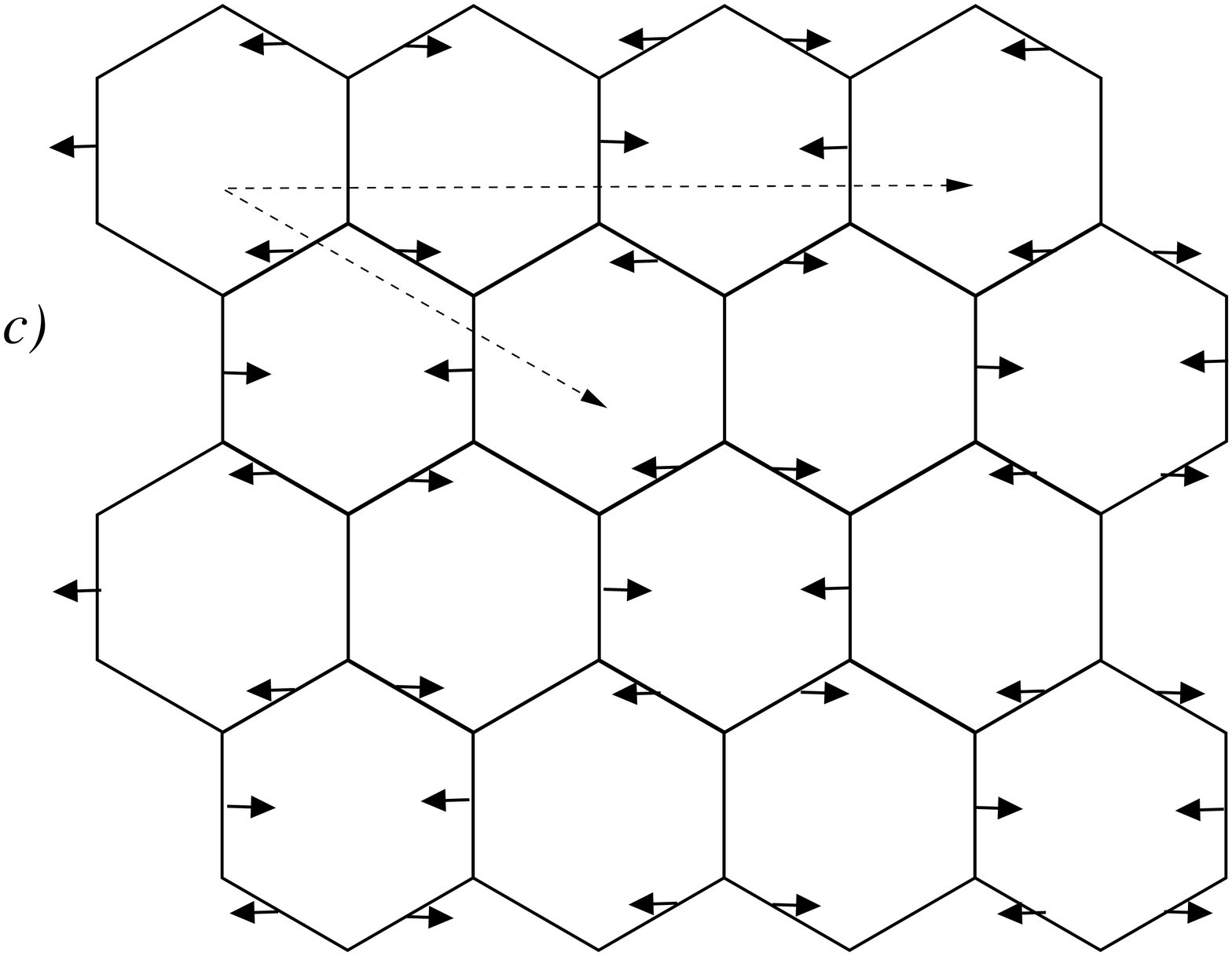}
\caption{a)-c) Sketches of the spin density
from \req{spindensity}, for $\alpha=\pi/2$,
and a) $n_z=1$ [$N_4=1$ in notation of \req{NN}], b) $n_x=1$ ($N_2=1$) and c) $n_y=1$ ($N_3=1$).
Dashed arrowed lines show the primitive translation
vectors of the resulting superstructure.
Here, we choose ${\mathbf e}_{2}
=(1,0,0)$, ${\mathbf e}_{1}
=(0,0,1)$, notice that the coordinate frame for the spin
is rotated with respect to spatial frame (Axis $x$ for the spin
is the direction perpendicular to the plane and $y$ is in plane
axis perpendicular to the magnetic field.);
d) Lines of the constant spin flux $\vec{\Phi}(\r)\cdot{\mathbf e}_1$
(dashed lines) from \req{fluxdensity}, $\alpha=0$ ($N_1=1$). Those lines
can be also thought of as the spin current lines.
}
\label{fig3}
\end{figure}

Using \reqs{eq:1}, \rref{delta0}, and \rref{Q}, we find
\begin{subequations}
\be
\begin{split}
\vec{s}(\r)& \propto {\mathbf e}_2 \sin\alpha 
\Big[n_z(|u_A(\r)|^2-|u_B(\r)|^2)
\\
&+2 n_x {\rm Re}\, u_A(\r)u_B(\r)
+ 2 n_y {\rm Im}\, u_A(\r)u_B(\r)
\Big].
\end{split}
\label{spindensity}
\ee
The corresponding spin density configurations consistent
with the phase factors of the Bloch function of Fig.~\ref{fig1}
are shown on Fig.~\ref{fig3} a)-c).
Configuration of $\alpha=\pi/2,n_z=1$ corresponds to the
site centered spin density wave. It does not change the periodicity
of the original lattice, so  the Bragg peaks in the neutron
scattering will remain  at the same positions
$\vec{q}=j_1\vec{b}_1+j_2\vec{b}_2$, and the ordering will affect only their internal structure. On the other hand, $\alpha=\pi/2,n_z=0$,
see Fig.~\ref{fig3} b,c), corresponds to the link centered density
waves. It is easy to see that in that case the unit cell is tripled so that
 additional Bragg peaks  at the positions
$\vec{q}=\pm K + j_1\vec{b}_1+j_2\vec{b}_2$ will emerge.
Rotation in $n_x$--$n_y$ plane corresponds to a smooth transition
between  Fig.~\ref{fig3} b,c) configurations. Such smooth
rotation can be also understood as a continuous sliding
of the superimposed spin-density wave with respect to the crystal lattice.

The spin flux, Fig.~\ref{fig3} d) is maximal at $\alpha=0,\pi$
\be
\vec{\Phi}(\r)\propto {\mathbf e}_1 \cos\alpha 
\, (|u_A(\r)|^2-|u_B(\r)|^2).
\label{fluxdensity}
\ee
Unfortunately, it is not coupled to any obvious physical field which makes its direct
experimental observation  unlikely.
\label{densities}
\end{subequations}

\section{Free energy, thermal fluctuations, and phase diagram}
\label{sec:4}

Order parameter \rref{Q} allows for  smooth rotations between
all the states of Fig.\ref{fig3}, and thus is subjected to  strong thermal
fluctuations. Such fluctuations are governed by 
the  Landau free energy functional which we are about to describe.

First we introduce a new definition of $\Lambda,S$-matrices \req{matrices}, as
\be
\hat{\Lambda}_\alpha=
\hat{\tau}_{\alpha}^{KK'}\otimes \openone^{s},
\quad
\hat{S}_\alpha=
   \openone^{KK'}\otimes\hat{\tau}_{\alpha} ^{s},
\label{4.3}
\ee
where $\alpha=x,y,z$,
and $\hat{\Lambda}_\pm\equiv (\hat{\Lambda}_x\pm i\hat{\Lambda}_y)/2
$.
This modification is convenient because there are no gapless rotations possible in the
sub-lattice space after the value of the mean-field order parameter is
established.

The only possible form for the free energy 
compatible with the symmetries of the system
\rref{discrete} is given by
\be
\begin{split}
&{\cal Z}\propto \sum_v 
\dots \mu^{N_v}
 \int {\cal D}{\hat Q}
\exp\left[-\frac{1}{T}\int d^2\r\,  {\mathbb F}\left\{\hat{Q}\right\}\right];
\\
&
{\mathbb F}={\mathbb F}_\circ+{\mathbb F}_\perp+{\mathbb
  F}_\rhd,
\end{split}
\label{4.1}
\ee
where symbol $\sum_v \dots \mu^{N_v}$ stands for a summation over
the topological defects, see subsection \ref{sec:symmetric}, and $\mu$
denotes a fugacity of such defects. A description of the
thermodynamics of in terms of matrix $Q$ 
subject to hard constraints \rref{qconstraints} is valid only at the
distances larger than the spatial scale
\be
\xi_{MF} \simeq \frac{v(R_B)}{\Delta_0(T)},
\label{xiMF}
\ee
where the scale dependence of the velocity and the length $R_B$ is defined
in microscopic theory section, see Sec.~\ref{sec:log}.

\begin{subequations}
\label{4.2}

The dominant term in the free energy,
\be
\begin{split}
{\mathbb F}_\circ
& = \frac{1}{8}
\Bigg\{
\rho_K(r_0){\mathrm Tr}\left(\partial_\nu \hat{Q}\right)^2 
\\
&\qquad + 
\frac{1}{4}\left[\rho_s(r_0)-\rho_K(r_0)\right]
\left[-i {\mathrm Tr}\hat{S}_z \hat{Q} \partial_\nu \hat{Q}\right]^2
\Bigg\}
\end{split}
\label{4.2a}
\ee
[new notation is defined in \req{4.3}]
has the symmetry which is higher than the symmetry of the original
problem, see \req{continuous}. Hereinafter, we will imply the summation
over the repeating index $\nu=x,y$.

To facilitate the further analysis, we rewrite \req{4.2a} using
 parametrisation \rref{V}

\be
\begin{split}
{\mathbb F}_\circ
& = \frac{1}{4}
\Bigg\{
\rho_K(r_0){\mathrm Tr}\left(\partial_\nu  \hat{V}^\dagger\partial_\nu \hat{V}\right) 
\\
&\qquad + 
\frac{1}{2}\left[\rho_s(r_0)-\rho_K(r_0)\right]
\left[-i {\mathrm Tr}
\hat{V}^\dagger\partial_\nu \hat{V}
\right]^2
\Bigg\}
\end{split}
\tag{\ref{4.2a}$^\prime$}
\label{4.2aprime}
\ee
where $\hat{V}$ is a unitary $2\times 2$ matrix.
Further investigation of free energy \rref{4.2a}
is postponed until subsection \ref{sec:symmetric}.

The remaining contributions, though may be small, are crucial because
they remove artificial symmetries of the system.
The following terms suppress $SU(2)$ rotations in the valley space: 
\be
{\mathbb F}_\perp
= \frac{1}{8 r_0^2}
\sum_{\alpha=x,y,z}\eta_\alpha (r_0)
{\mathrm Tr}
\hat{Q}\hat{\Lambda}_\alpha\hat{Q}\hat{\Lambda}_\alpha,
\quad \eta_x=\eta_y=\eta_\perp.
\label{4.2b}
\ee
\be
{\mathbb F}_\Box
= \frac{\kappa(r_0)}{16 r_0^2}
\sum_{\alpha,\beta =x,y}
{\mathrm Tr}
\hat{Q}\hat{\Lambda}_\beta\hat{Q}\hat{\Lambda}_\beta
\hat{Q}\hat{\Lambda}_\alpha\hat{Q}\hat{\Lambda}_\alpha,
\label{4.2c}
\ee
Finally, the term 
\be
{\mathbb F}_\rhd
= \frac{\zeta (r_0)}{4 r_0^2}
{\mathrm Tr}
\left[
\left(\hat{Q}\hat{\Lambda}_+\right)^6
+\left(\hat{Q}\hat{\Lambda}_-\right)^6
\right]
\label{4.2d}
\ee
generated by the Umklapp processes, reduces the $U(1)$ rotations in the valley space,
to discrete rotations \rref{discreted}.
\end{subequations}

A discussion of the role played by anisotropic terms \rref{4.2b} --
\rref{4.2d} will be continued in subsection \ref{sec:anisotropies}.

\subsection{Isotropic part of the action and topological defects}
\label{sec:symmetric}

 The sum over discrete topological defects can be replaced by a path integral over continuous variables - 
%Perturbations to the effective action generated by topological defects are expe%cted to be nonlocal
%in $Q$. This non-locality is removed by introduction of the so-called 
dual fields. 
Introduction of dual fields become more transparent when one  uses  the following  
parametrisation of the order parameter (\ref{V}):
\begin{eqnarray}
&& V = e^{i\theta_s}g, ~~ g =\left(
\begin{array}{cc}
N_1 +iN_4 & iN_2 + N_3\\
iN_2 - N_3 & N_1 - i N_4
\end{array}
\right) \label{NN}
%=\nonumber\\
%&& e^{i\theta_s}e^{i\vec\sigma{\bf n}/2}\left(
%\begin{array}{cc}
%e^{i\alpha} & 0\\
%0 & e^{-i\alpha}
%\end{array}
%\right)e^{-i\vec\sigma{\bf n}/2}\label{NN}
\end{eqnarray}
where $g$ is an SU(2) matrix and $\sum_{i=1}^4 N_i^2 =1$. 
%Phase fields $\theta_s$ and $\alpha$ are related to the group invariants:
%\be
%e^{2i\theta_s} = \mbox{det} V, ~~ \cos(\theta_s)\cos(\alpha) = \frac{1}{2}\mbox%{Tr}(V + V^+)
%\ee
%which guarantees that the choice of the dual fields is also group invariant. 
 Then 
%In parameterization  \rref{NN}, 
free energy density
\rref{4.2aprime} acquires the form
\be
{\mathbb F}_\circ
 = 
\frac{\rho_K}{2}\sum_{i=1}^4\left( \partial_\nu N_i\right)^2 
+ \frac{\rho_s}{2}
\left( \partial_\nu \theta_s\right)^2,
\tag{\ref{4.2a}$^{\prime\prime}$}
\label{4.2aprimeprime}
\ee
So it appears that  the $U(1)$ and  the $SU(2)$ sectors of the theory 
are decoupled. 
This decoupling, however, breaks down when one takes  into account the topological defects. As we shall demonstrate, the non-Abelian sector changes the selection rules for the vortices winding numbers.

As usual for compact theories, the free energy density
expression obtained in continuous limit has to be supplemented by  topological defects to take into account a behavior  
of the order parameter in the vicinity of some singular points
$\r_j$,
where it vanishes: $\hat{Q}^2(\r_j) = 0;
\ \hat{V}(\r_j)\hat{V}^\dagger(\r_j) = 0$. 
The absolute value of the  order parameter is established at the
distances of the order of $\xi_{MF}$, see \req{xiMF}, and at
large distances, the defect is characterized by a 
contour integral around it.

The topological defects are characterized by the winding number 
\be
q =  \frac{i}{4\pi}\oint dx_\nu Tr(V^+\partial_\nu V) \label{q}
\ee
The most standard approach would be to keep the $g$ matrix in \req{NN} single-valued and non-singular and create $2\pi$- vortex configuration in $U(1)$
field $\theta_s$: 
\be
\begin{split}
&e^{i\theta_s}=\frac{(x-x_i)\pm i (y-y_i)}
{\sqrt{(x-x_i)^2+ (y-y_i)^2}};
\\
&\frac{1}{2\pi}\oint dx_\nu \partial_\nu\theta_s = \frac{i}{4\pi}\oint dx_\nu Tr(V^+\partial_\nu V) = \pm 1,
\end{split}
\label{1v}
\ee 
If only such excitations were present, $U(1)$ and  $SU(2)$ sectors 
would remain decoupled and the standard
Berezinskii-Kosterlitz-Thouless transition\cite{Berezinskii,KT} would occur in the $U(1)$
sector.

There are, however, other configurations which are energetically
more profitable than those of \rref{1v}. Indeed, consider the configuration of the form
\be
\hat{V}_{1/2} =
e^{i\nn\cdot\hat{\boldsymbol{\sigma}}/2}
\begin{pmatrix}
\frac{(x-x_i)\pm i (y-y_i)}
{\sqrt{(x-x_i)^2+ (y-y_i)^2}};
& 0\\
0; & 1
\end{pmatrix}
e^{-i\nn\cdot\hat{\boldsymbol{\sigma}}/2}
,
\label{Vhalfv}
\ee
where $\n(\r)$ is a smooth
three-dimensional unit vector $\sum_{i=1}^3n_i^2 =1$, and
$\hat{\sigma}_i$
are the Pauli matrices. 

Re-writing \req{Vhalfv} in the form of \req{NN},
we obtain instead of \req{1v}
\be
q= \frac{i}{4\pi}\oint dx_\nu Tr(V^+\partial_\nu V) \pm \frac{1}{2}.
\label{halfv}
\ee 
i.e. configuration \rref{Vhalfv} corresponds to a $\pi$ or {\em half-vortex}
in the $U(1)$ sector glued with a {\em half-vortex} in the $SU(2)$ -sector. 
 Since the stiffness in the SU(2) sector vanishes at large distances, the change of sign of $g$ in large defects does not require energy. So in the absence of anisotropy the main effect of the SU(2) sector is the change in selection rules of the vortices reflected in their  topological invariant  (\ref{q}). 

%It is clear from \req{Vhalfv} that the half-vortex is characterized
%not-only by its vorticity but also by its direction in  three-dimensional space. 

To anticipate a role of the vortex and half-vortex configurations in
the thermodynamic properties of the system we evaluate their energies.
Substituting configurations \rref{1v} and \rref{Vhalfv} into \reqs{4.2aprime}
or \rref{4.2aprimeprime}, we find
\be
\begin{split}
&E_{1/2}=\left(\frac{1}{2}\right)^2\pi\left(\rho_s+\rho_K\right)
\ln\left(\frac{L}{\xi_{MF}}\right);\quad 
\\
&
E_{1}=\pi\rho_s
\ln\left(\frac{L}{\xi_{MF}}\right);
\end{split}
\label{energies}
\ee
where $L$ is the system size. As $\rho_K \leq \rho_s$, one can see that
the half-vortices are in fact the main configurations responsible
for the disordering of the $U(1)$ sector, which we will in due course  incorporate
into the corresponding RG equations.

To treat the contributions of the topological configurations  \rref{1v} and
\rref{halfv} systematically, we rewrite the partition function
\rref{4.1} (still neglecting the anisotropic parts) in the form
\be
\begin{split}
&{\cal Z}\propto
 \int {\cal D}{\hat{V}}{\cal D}{\hat{h}}
\exp\left[-\frac{1}{T}\int d^2\r\,  
{\mathbb F}\left\{\hat{V}, \hat{V}^\dagger,\hat{h}\right\}\right];
\\
&
\frac{{\mathbb F}}{T} =\frac{1}{4}
 {\mathrm Tr}
\left\{
\frac{\rho_K }{T} \partial_x
  \hat{V}^\dagger\partial_x 
\hat{V}+ \frac{T}{\rho_K}\left(\partial_x \hat{h}\right)^2
+2\hat{V}^\dagger\partial_y
  \hat{V}\partial_x \hat{h}
\right\}
\\
&
+\frac{\left[\rho_s-\rho_K\right]}{8T}
\left[-i {\mathrm Tr}
\hat{V}^\dagger\partial_{x}\hat{V}
\right]^2 
+
\left[\frac{T}{8\rho_s}-\frac{T}{8\rho_K}\right]
\left[{\mathrm Tr}
\partial_x \hat{h}
\right]^2
\\
& 
- \frac{2\mu_1}{r_0^2}\cos 2\pi {h}_s
- \frac{2\mu_{1/2}}{r_0^2}\frac{\sin\pi |\vec{h}|}{\pi\,|\vec{h}|}\cos\pi h_s ;
\\
&
\hat{V}\hat{V}^\dagger=\openone;
\quad
\hat{h}=\hat{h}^\dagger=h_s\openone^s + \vec{h}\cdot\hat{\vec{\tau}}^{\, s}
\end{split}
\raisetag{0.5cm}
\label{dual}
\ee
where $2\times 2$ matrix $\hat{h}$ is the dual field, $\hat{\tau}^s_{x,y,z}$
are the Pauli matrices in the spin space,
and $\mu_1$, $\mu_{1/2}$, are fugacities for the vortices and
half-vortices respectively. Though it may appear that term 
\req{dual}  
breaks the rotational symmetry, all the physical correlation functions
calculated with using \req{dual} are rotationally symmetric.
Derivation of \req{dual} is relegated to Appendix~\ref{appendixA}

Summing up the leading logarithmic series within the first loop
renormalization group scheme, we find the following equations for the corresponding
fugacities:
\begin{subequations}
\be
\begin{split}
\frac{d\mu_{1/2}}{d\ln r_0}
=\left[2-
\left(\frac{1}{2^2}\right)
\frac{\pi}{T}
\left(\rho_K+\rho_s\right)
\right]\mu_{1/2}.
\end{split}\label{fugas}
\ee
They are valid for $\mu_{1/2},\mu \ll 1$ and also $\rho_K \gg
T$.

The fugacity for the conventional vortices evolves as
\be
\begin{split}
\frac{d\mu_{1}}{d\ln r_0}
=\left(2-
\frac{\pi\rho_s}{T}
\right)\mu_1
\end{split}\label{fugas1}
\ee
for $\mu_1\ll 1$.
Equations \rref{fug}
are analogous for the simple estimate of the energy of the
defects \rref{energies}, however, they allow for a renormalization
of the stiffness $\rho_s$ caused by the bound pairs of (half)vortices
and (half)antivortices,
and of the stiffness $\rho_K$ caused by  the bound pairs 
 the half-vortices and anti-half-vortices as well as by  thermal
fluctuations of the order parameter.
\label{fug}
\end{subequations}
\begin{subequations}
The latter renormalization, for  $\mu_{1/2},\mu \ll 1$  and $\rho_K
\gg T$, can be represented in the form
\begin{align}
&\frac{d\rho_K}{d\ln r_0}
=
-  \frac{T}{\pi} - \mu_{1/2}^2\beta_K^{(1/2)}(\rho_K,\rho_s);
\label{stiffnessK}\\
&\frac{d\rho_s}{d\ln r_0}
=
- \mu_{1/2}^2\beta_s^{(1/2)}(\rho_K,\rho_s)- \mu^2_1\beta_s^{(1)}(\rho_s).
\label{stiffnessS}
\end{align}
\label{stiffness}
Functions $\beta$ in \reqs{stiffness} are
difficult to obtain explicitly for arbitrary stiffnesses. However, such forms 
will not be necessary for the further consideration.  
\end{subequations}

As the initial fugacities of the half-vortices, $\mu_{1/2}$ are not much smaller than the fugacities for the 
vortices $\mu_1$, the latter ones being less relevant can be
neglected in the description of the phase transition.
In the limit of $\mu_{1/2} \ll 1$ one can also neglect the half-vortices
in the renormalization of $\rho_K$. Then, from
\req{stiffnessK} we obtain  the following equations
\be
\rho_K(r_0)=\frac{T}{\pi}\ln\frac{\xi_K}{r_0};
\quad
\xi_K=\xi_{MF}\exp\left(\frac{\pi\rho(\xi_{MF})}{T}\right),
\label{rhoK}
\ee
for $\xi_{MF}\lesssim r_0\lesssim \xi_K$. At ${r_0} \sim \xi_K$ the perturbative renormalization group
is no longer valid, and the $SU(2)$ sector enters into the strongly
disordered regime.  Hence  length $\xi_K$ has a meaning
of the correlation length of the order parameter.

\begin{subequations}
At $r_0 \gg \xi_K$  the non-Abelian stiffness vanishes: $\rho_K \to 0$. We
can then use \req{fugas} to obtain the value of the Kosterlitz
jump at the phase transition:
\be
\frac{\pi\rho_s(T_{KT}-0)}{T}
=2*2^2;
\label{Kjump}
\ee
This value is modified in comparison with the pure $U(1)$ model
\be
\frac{\pi\rho_s(T_{KT}-0)}{T}
=2;
\label{KjumpU1}
\ee
due to  the presence of half-vortices and  the $SU(2)$ sector being disordered.
\label{Kjump-both}
\end{subequations}

Assuming that $\mu_{1/2}(\xi_{MF}) \ll 1$, we can also use \req{Kjump}
 to  estimate $T_{KT}$:
\be
\frac{\pi\rho_s(T_{KT})}{2T_{KT}}
=2^2 + {\cal O}\left[\mu_{1/2}(\xi_{MF})\right];
\label{TKT}
\ee 
In Sec.~\ref{sec:log}  we will use \req{TKT} to relate $T_{KT}$ 
to mean-field transition temperature $T_{MF}$.

The main conclusions of these  subsections are: (i) 
spin $U(1)$ and valley  $SU(2)$ rotations are coupled to each other in  the presence
of {\em half-vortices}, and (ii)  the  $U(1)$ sector acquires the
algebraic order at $T<T_{KT}$ whereas the isotropic version of the
valley rotations always remain disordered.

The purpose of the next two subsections is to analyze how 
 the disorder in the valley space is affected when the artificial symmetries are lifted by the leading anisotropies.

\subsection{Effect of weak anisotropies. }
\label{sec:anisotropies}

Let us assume that the BKT transition in the $U(1)$ sector has already
occurred so that the half-vortices are not relevant, and $SU(2)$ sector
is decoupled. 
(This assumption will be lifted in Sec.~\ref{sec:anisotropies-strong}.)
Then the thermal fluctuations lead to the
renormalization of the anisotropies in the same fashion as the first
term in the right-hand-side of \req{stiffnessK}. In the first loop
approximation,
we find 
\be
\begin{split}
&\frac{d\eta_\alpha}{d\ln r_0}
=\left[2-
\frac{2 T}{\pi\rho_K }
\right]\eta_\alpha;
\\
&\frac{d\kappa}{d\ln r_0}
=\left[2-
\frac{8 T}{\pi\rho_K }
\right]\kappa;
\\
&\frac{d\zeta}{d\ln r_0}
=\left[2-
\frac{12 T}{\pi\rho_K }
\right]\zeta;
\\
\end{split}
\label{RGan}
\ee
Solving \req{RGan} with the help of \req{rhoK},
we obtain
\be
\begin{split}
\eta_\alpha(r_0)&=\eta_\alpha(\xi_{MF})
\left(\frac{r_0}{\xi_{MF}}\right)^2
\left(\frac{\ln \frac{\xi_K}{r_0}}
{\ln \frac{\xi_K}{\xi_{MF}}}
\right)^2\\
\kappa(r_0)&=\kappa(\xi_{MF})
\left(\frac{r_0}{\xi_{MF}}\right)^2
\left(\frac{\ln \frac{\xi_K}{r_0}}
{\ln \frac{\xi_K}{\xi_{MF}}}
\right)^8\\
\zeta(r_0)&=\zeta (\xi_{MF})
\left(\frac{r_0}{\xi_{MF}}\right)^2
\left(\frac{\ln \frac{\xi_K}{r_0}}
{\ln \frac{\xi_K}{\xi_{MF}}}
\right)^{12}.
\end{split}
\label{RGansol}
\ee
These equations are valid for $r_0 \lesssim \xi_K$.

Thus, the effect of the anisotropies is determined by
their value at distances
of the order of $\xi_K$. If they are still small, i.e.
\be
\begin{split}
&\eta_\alpha(\xi_{MF}) \lesssim
T \left(\frac{\xi_{MF}}{\xi_K}\right)^2
\left(\ln \frac{\xi_K}{\xi_{MF}}\right)^2;
\\
&\kappa(\xi_{MF})
\lesssim
T \left(\frac{\xi_{MF}}{\xi_K}\right)^2
\left(\ln \frac{\xi_K}{\xi_{MF}}\right)^8;\\
&\zeta(\xi_{MF})
\lesssim
T \left(\frac{\xi_{MF}}{\xi_K}\right)^2
\left(\ln \frac{\xi_K}{\xi_{MF}}\right)^{12},
\end{split}
\label{smallanis}
\ee
then the $SU(2)$ sector remains disordered and the conclusions of the
previous subsection remain qualitatively valid. If, however, one or
several  
conditions \rref{smallanis} are violated,
the anisotropies are important in determining  the phase diagram of  the
system.

The phase diagram obviously depends on the relative values of the coupling
constants $\eta,\kappa,\zeta$ and their signs.
The latter are determined by the microscopic coupling constants
\rref{warping} -- \rref{umklapp}, see also Sec.~\ref{sec:log}.
Since such couplings can not be established on a merely symmetry grounds, we will analyze the
phase diagrams in the most possible general case and
then construct the ``physical'' phase diagram in Sec.~\ref{sec:log4}.

Let us assume 
\be
\begin{split}
&\bar{\eta}(\xi_{MF})
\equiv\left[\eta_\alpha^2(\xi_{MF})\right]^{1/2}
\gtrsim
T \left(\frac{\xi_{MF}}{\xi_K}\right)^2
\ln^2\frac{\xi_K}{\xi_{MF}};
\\
&\kappa,\zeta \ll \bar{\eta};
\end{split}
\label{notweak}
\ee

Then, at the distance
\be
R_* \simeq  \xi_{MF}
\left(\frac{T}{\bar{\eta}(\xi_{MF})}\ln^2 \frac{\xi_K}{\xi_{MF}}\right)^{1/2}
\label{R*}
\ee
we have 
\be
\bar{\eta}(R_*)\simeq \rho_K(R_*),
\label{etaR*}
\ee
so that the anisotropy becomes more important than the stiffness,
and the perturbative treatment \rref{RGan} is no longer valid.

Instead, one has to identify the remaining soft-modes still compatible
with the anisotropy potential \rref{4.2b} and consider the fluctuations
for those modes only. To achieve this goal, let us re-write
\req{4.2b}
using the parametrization \rref{NN}:
\be
\label{anisNN}
{\mathbb F}_\perp=
\frac{1}{R_*^2}
\left[\left(\eta_\perp+\eta_z\right) N_1^2 + \left(\eta_z-\eta_\perp\right) N_4^2\right],
\ee
and the isotropic part is given by \req{4.2aprimeprime}, and
$\sum_{i=1}^4 N_i^2 =1$.

\begin{figure}[h]
\includegraphics[width=0.3\textwidth]{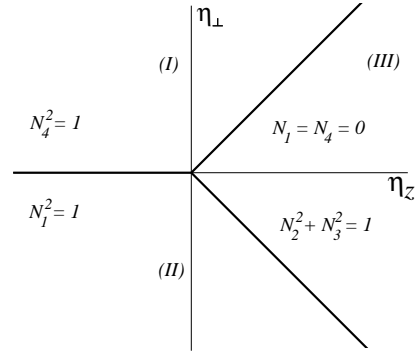}
\caption{Phase diagram in the absence of the thermal fluctuation.
For the pictorial representation of the states in terms of physical spins see Fig.~\ref{fig3}.}
\label{fig4}
\end{figure}

If there were no thermal fluctuations at all, the direction of the order parameter
would be obtained by minimizing the expression \rref{anisNN} with the
results summarized on Fig.~\ref{fig4}.

The regions (I) and (II) on Fig.~\ref{fig4}, correspond to the Ising
type anisotropy and the region (III) corresponds to the $XY$ model.
Hence  deep inside the regions $(I),(II)$ the system is ordered,
whereas the region $(III)$ is characterized by an algebraic order.
At the lines separating those regions the system possesses  extra
degeneracies which will be analyzed separately.

\subsubsection{Vicinity of the line $\eta_\perp=0,\ \eta_z <0$.}
\label{etaperpsmall}

Near this line one can neglect the fluctuations of $N_{2,3}$
and consider $N_{1,4}$ only. 
Then, the relevant part of the free energy, see \reqs{4.2aprimeprime}
and \rref{anisNN}, acquires the form
\be
{\mathbb F}_{14}
 = 
\frac{\rho_K}{2}\left[\left( \partial_\nu  N_1\right)^2 + \left(  \partial_\nu N_4\right)^2 \right]
+  \frac{\eta_\perp}{R_*^2}
\left( N_1^2 - N_4^2\right),
\label{N1N2}
\ee
with $N_{1,4}$ constrained by $N_1^2+N_4^2=1$. 
Representing $N_1=\cos\phi$, $N_4=\sin\phi$ and 
introducing dual field $\theta$ 
%related to $\phi$ via equation 
%\be
%\partial_{\mu}\theta = \epsilon_{\mu\nu}\partial_{\nu}\phi
%\ee
to describe the vortices with a
core size $\lesssim R_*$, we find
\be
\begin{split}
&{\cal Z}\propto \int{\cal D}\phi{\cal D}\theta
\exp\left(-{\mathbb F}_{14}/T\right);
\\
&\frac{{\mathbb F}_{14}}{T}
 = 
\frac{\rho_K}{2T}\left(\partial_{x}  \phi\right)^2 +
\frac{T}{2\rho_K}\left(\partial_x  \theta\right)^2+ 
i\partial_x\theta\partial_y\phi
\\
&\qquad +
\frac{\eta_\perp}{R_*^2 T}\cos 2\phi+
%\frac{\mu_{14}}{R_*^2}\cos \left(2\pi T\theta/\rho_K\right),
\frac{\mu_{14}}{R_*^2}\cos \left(2\pi\theta\right),
\end{split}
\label{N1N2phi}
\ee 
The derivation of \req{N1N2phi} is completely analogous to the derivation
of \req{dual}, see also Appendix.~\ref{appendixA}.

The first loop scaling equations for the anisotropy $\eta_\perp$ and the vortex
fugacities $\mu_{14}$ are
\be
\begin{split}
&\frac{d\mu_{14}}{d\ln r_0}
=\left(2-
\frac{\pi\rho_K}{T}
\right)\mu_{14};
\\
&
\frac{d\eta_\perp}{d\ln r_0}
=\left(2-
\frac{T}{\pi\rho_K}
\right)\eta_\perp.
\end{split}
\label{scaling34}
\ee
They have the obvious solutions
\be
\begin{split}
&\mu_{14}\left(r_0\right)=\mu_{14}\left(R_*\right)
\left(\frac{R_*}{r_0}
\right)^{\frac{\pi\rho_K-2T}{T}};
\\
&
\eta_\perp\left(r_0\right)=\eta_\perp\left(R_*\right)
\left(\frac{R_*}{r_0}
\right)^{\frac{T-2\pi\rho_K}{\pi\rho_K}}.
\end{split}
\label{scaling34sol}
\ee
If $\eta_\perp\left(R_*\right)=0$, the system
undergoes  Berezinskii-Kosterlitz-Thouless transition
at $T_{BKT} = \pi\rho_K/2$; from the high temperature vortex-dominated disordered state to a phase with power 
law correlations. 

 At non-zero $\eta_{\perp}$ the ordered phase has a 
finite correlation length since  below   $\pi\rho_K/2$
the anisotropy $\eta_\perp$ is a relevant perturbation. Therefore, 
 phase $\phi$ is locked and the system is ordered
with a finite correlation length $\xi_{14}$ determined by  the condition $\eta_\perp\left(\xi_{14}\right) \simeq T$ so 
that 
\be
\xi_{14}\simeq R_*\left(\frac{T}{\eta_\perp(R_*)}\right)^{\frac{\pi\rho_K}{2\pi\rho_K-T}}.
\label{xi34}
\ee
From \req{xi34} one may conclude that the  transition between phases (I) and (II) of Fig.~\ref{fig4} is a continuous one 
with smoothly varying correlation
length exponent
along the transition line. However, we will see shortly that this conclusion
is an artifact of neglecting the higher order anisotropy \rref{4.2c}.

At  temperatures above $\pi\rho_K/2$, both the anisotropy (order) and the vortices (disorder) are relevant. The corresponding operators compete with each other since $\theta$ and $\phi$ cannot be locked simultaneously. 
To estimate the location of the transition line and the length scale
$\xi_{I}$ at which criticality becomes important, we require
\[
\frac{\eta_\perp\left(\xi_I\right)}{T}\simeq
\mu_{14}\left(\xi_I\right)\simeq 1,
\]
which yields, assuming as usual that $\mu_{14}(R_*) \ll 1$,
\be
\begin{split}
&\xi_I\simeq R_*\left(\frac{1}{\mu_{14}(R_*)}\right)^\frac{T}{2T-\pi\rho_K};
\\
&\eta_c(\rho_K)=T\left[\mu_{14}(R_*)\right]^{\frac{T}{\pi\rho}
\frac{2\pi\rho_K-T}{2T-\pi\rho_K}
}
\end{split}
\label{criticalline1}
\ee

At distances larger than $\xi_I$, the anisotropy is the largest
scale in the problem, so that  $N_1$ ($N_4$) becomes massive 
and the vortices provide the possibility for $N_4^2$ ($N_1^2$)
to change on the scale of the order of  $\xi_I$.
The resulting transition, therefore, involves only one soft
mode and thus  belongs to the Ising model
universality class, see Appendix~\ref{AppendixB},
for more accurate calculation in the vicinity of $ \pi\rho_K=T$.
The corresponding correlation length $\xi$ is, then, given by
\be
\xi \simeq \xi_I\frac{\eta_\perp^c}{|\eta_\perp-\eta_\perp^c|},
\quad |\eta_\perp-\eta_\perp^c| \lesssim \eta_\perp^c.
\label{xiIsing}
\ee

\begin{figure}[h]
\includegraphics[width=0.45\textwidth]{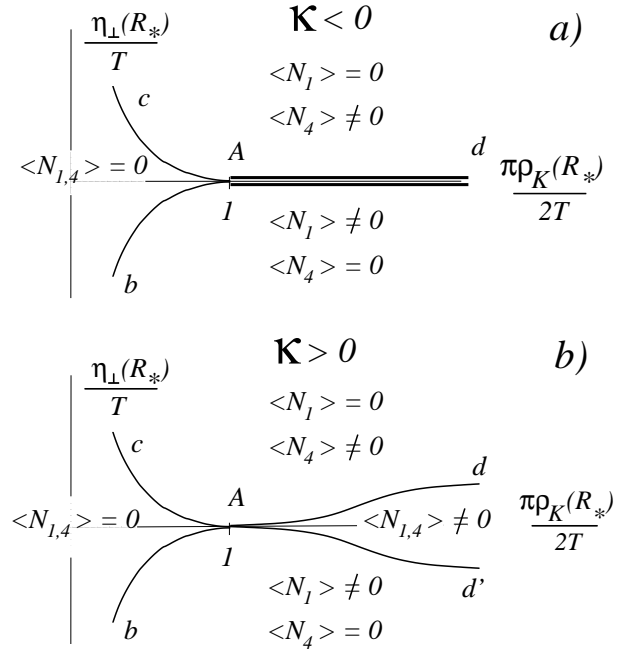}
\caption{Phase diagrams in the vicinity of the line $\eta_\perp=0,\
  \eta_z <0$. Lines $A-b$ and $A-c$ are always of the Ising type.
a) Line $A-d$ is (weak) first order phase transition line at $\kappa <
0$. b)  Lines $A-d$ and $A-d'$ are second order phase transitions
(Ising type)
at $\kappa > 0$.
At $\kappa=0$, the $A-d$ line is a continuous phase transition
with the varying indices given by \req{xi34}. }
\label{fig5}
\end{figure}

So far we ignored the higher order anisotropy term. In fact, those
terms are important only in the vicinity of (I)-(II) transition
line, where they change the order  of the
phase transition.

To see this, we rewrite \req{4.2c} using
parameterizations \rref{V} and \rref{NN}:
\be
\frac{{\mathbb F}_\Box}{T}
= \frac{\kappa(R_*)}{T R_*^2}
{\rm Re}\left(N_x+iN_y\right)^4=
\frac{\kappa(R_*)}{T R_*^2}\cos 4\phi.
\label{quartic}
\ee
Equation \rref{quartic} should be added to \req{N1N2phi}.

The coupling in \req{quartic} is renormalised by thermal
fluctuations. The first loop RG equation and its solution is given by
\be
\begin{split}
&\frac{d\kappa}{d\ln r_0}
=\left(2-
\frac{4T}{\pi\rho_K}
\right)\kappa;
\\
&\kappa\left(r_0\right)=\kappa\left(R_*\right)
\left(\frac{R_*}{r_0}
\right)^{\frac{4T-2\pi\rho_K}{\pi\rho_K}}.
\end{split}
\label{scaling34an}
\ee
At  $\pi\rho_K>2T$
the quartic anisotropy $\kappa$ is a relevant perturbation,
so that the phase $\phi$ becomes locked and the system is ordered
in locally stable state even for $\eta_\perp=0$,
so that the correlation length $\xi_{14}$ is always limited by $\tilde{\xi}_{14}$, that is
found from the condition $\kappa\left(\tilde\xi_{14}\right) \simeq T$.
It yields
\be
\xi_{14} \lesssim \tilde{\xi}_{14} 
\simeq R_*\left(\frac{T}{|\kappa (R_*)|}\right)^{\frac{\pi\rho_K}{2\pi\rho_K-4T}}.
\ee
The length $\tilde{\xi}_{14}$ diverges when the approaching the
multi-critical point  $\pi\rho_K=2T,\ \eta_\perp=0$.
At length scale large than $\tilde{\xi}_{14}$ the non-vanishing
order parameter is found by minimisation of
the expression
\be
\frac{\tilde{\mathbb F}}{T}=\frac{\eta_\perp(\tilde{\xi}_{14})}{ \tilde{\xi}_{14}^2T}
\cos 2\phi+ {\rm sgn \kappa} \cos 4\phi
\ee
with respect to $\phi$ which produces two second
order (Ising type) phase transitions for $\kappa >0$, and the first
order phase transition for $\kappa <0$ [in the former case
 $\tilde{\xi}_{14}$ serves as pre-factor for the diverging
correlation length similarly to \req{xiIsing}].

At $\pi\rho_K<2T$ the quartic anisotropy becomes irrelevant and we
obtain
from \reqs{scaling34an} and \rref{criticalline1}
\be
\begin{split}
&\kappa\left(\xi_I\right)=\kappa\left(R_*\right)
\left(\frac{1}{\mu_{14}}
\right)^{\frac{2}{\pi\rho_K}} < T,
\end{split}
\label{scaling34an1}
\ee
i.e. it can not affect the position and the universality class of the
Ising phase transitions.

The resulting structure of the phase diagram 
in the vicinity of the line $\eta_\perp=0,\ \eta_z <0$
is summarised on Fig.~\ref{fig5}.

\subsubsection{Vicinity of the lines $\eta_z > 0, \
  \eta_\perp=\pm \eta_z$.}
\label{Heisenberg}

In the vicinity of the line  $\eta_z > 0$, $\eta_\perp=-\eta_z$
anisotropy \rref{anisNN} generates mass for the  field $N_4$ 
so that the free energy for the remaining soft modes, see \reqs{4.2aprimeprime}
and \rref{4.2d}, is
\be
\begin{split}
&{\mathbb F}
 = 
\frac{\rho_K(R_*)}{2}\sum_{i=1}^3\left(\partial_\nu  N_i\right)^2 
\\
&\quad
+
\frac{\Delta\eta(R_*)}{R_*^2} N_1^2  + \frac{\zeta (R_*)}{R_*^2}
{\mathrm Re}\left(N_2+i N_3\right)^6
;
\\
&N_1^2+N_2^2+N_3^2=1; \quad \Delta\eta(R_*)\equiv \eta_z(R_*)
+ \eta_\perp(R_*),
\end{split}
\raisetag{1.7cm}
\label{F123}
\ee
where length scale $R_*$  and the coupling constants 
are defined in \reqs{R*}, \rref{rhoK}, and \rref{RGansol}.
The vicinity of the line  $\eta_z > 0$, $\eta_\perp=\eta_z$
does not require a separate consideration as it 
is also described by the  free energy density \rref{F123}
after the replacement
\be
N_1\to N_4; \quad \eta_\perp(R_*) \to - \eta_\perp(R_*).
\label{replacement}
\ee 

At $r_0 > R_*$, the scaling of the couplings in \req{F123}
is governed by the first loop renormalization group equations
\begin{subequations}\label{stiffnessH}
\begin{align}
&\frac{d\rho_K}{d\ln r_0}
=
-  \frac{T}{2\pi}
\label{stiffnessKH}\\
&\frac{d\Delta\eta}{d\ln r_0}
=\left[2-
\frac{3 T}{2\pi\rho_K }
\right]\Delta\eta;
\label{deltaRG}
\\
&\frac{d\zeta}{d\ln r_0}
=\left[2-
\frac{21 T}{2\pi\rho_K }
\right]\zeta,
\label{zetaRG}
\end{align}
where we neglect the effect of the anisotropies on the
renormalization of the stiffness which, however, will be sufficient for
our purposes.
\end{subequations}

The solution of \req{stiffnessKH} is
\begin{subequations}
\be
\rho_K(r_0)=\frac{T}{2\pi}\ln\frac{\xi_H}{r_0};
\quad
\xi_H=R_*\exp\left(\frac{2\pi\rho(R_{*})}{T}\right),
\label{rhoKH}
\ee
which is consistent with  the well known fact that the classical
$SU(2)/U(1)$ sigma model is always disordered by the thermal fluctuations with 
$\xi_H$ being  the correlation length for these fluctuations. 

Anisotropies may lead to ordering.
To estimate positions of the phase transition 
lines, we solve \reqs{deltaRG} and \rref{zetaRG} with the help
of \req{rhoKH} and find
\be
\begin{split}
\Delta\eta(r_0)&=\Delta\eta(R_*)
\left(\frac{r_0}{R_*}\right)^2
\left(\frac{\ln \frac{\xi_H}{r_0}}
{\ln \frac{\xi_H}{R_*}}
\right)^3\\
\zeta(r_0)&=\zeta(R_*)
\left(\frac{r_0}{R_*}\right)^2
\left(\frac{\ln \frac{\xi_H}{r_0}}
{\ln \frac{\xi_H}{R_*}}
\right)^{21}.
\end{split}
\label{RGansolH}
\ee
\end{subequations}

\begin{figure}[h]
\includegraphics[width=0.45\textwidth]{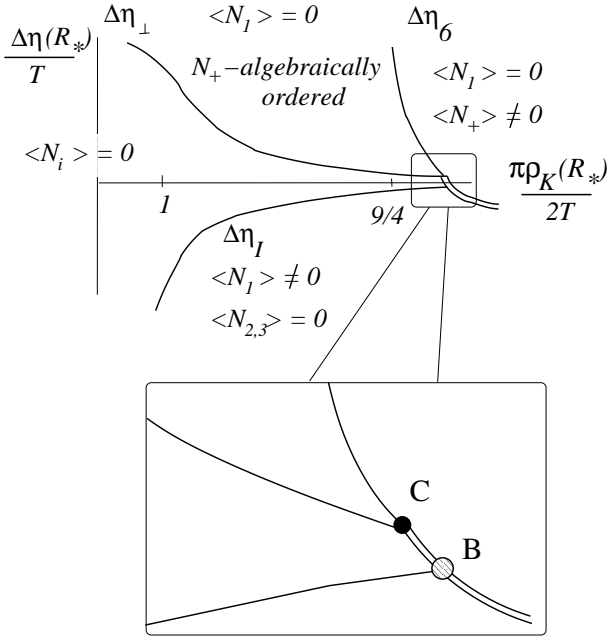}
\caption{Phase diagrams in the vicinity of the line $\Delta\eta
=\eta_z+\eta_\perp=0,\ \eta_z >0$.
Here $N_+=N_2+iN_3$. Lines $\Delta\eta_{\perp,6}$ are the
Berezinskii-Kosterlitz-Thouless phase transitions. Line $\eta_I$ is the Ising type phase
transition.
Double line is the first order phase transition. Point $C$ is
the bi-critical point and point $B$ is the critical end point.
}
\label{fig6}
\end{figure}

First, let us neglect the hexadic anisotropy, $\zeta=0$.
Then, the anisotropy is important if 
$|\Delta\eta(r_0\simeq 2\xi_H)| > T$. Equations \rref{RGansolH}
and \rref{rhoKH} give us approximate positions of the
phase transition lines
\be
\begin{split}
&\frac{\Delta\eta_\perp(R_*)}{T}
= \left(\alpha_\perp \frac{2\pi\rho_K(R_*)}{T}\right)^3
\exp\left(-\frac{4\pi\rho_K(R_*)}{T}\right);\\
&\frac{\Delta\eta_I(R_*)}{T}
= -\left(\alpha_I \frac{2\pi\rho_K(R_*)}{T}\right)^3
\exp\left(-\frac{4\pi\rho_K(R_*)}{T}\right),
\end{split}
\label{positions}
\ee
shown on Fig.~\ref{fig6}. The numerical pre-factors $\alpha_{I,\perp}$
of  order of unity can not be obtained within a perturbative
RG scheme.

Line $\Delta\eta_I$ corresponds to the second order
phase transition from the disordered
phase to the phase characterised by $\langle N_1 \rangle \neq 0$.
This phase transition belongs to the universality class of the
two dimensional Ising model. Additional hexadical anisotropy,
(finite but small $\zeta$) can not affect this transition.

Line $\Delta\eta_\perp$ corresponds to the 
phase transition from the disordered
phase to the phase characterised by  algebraic
correlations for $N_{2,3}$, i.e. to the Berezinskii-Kosterlitz-Thouless
transition.
Contrary to the Ising transition, the hexadic term
may cause a further ordering and an additional
Berezinskii-Kosterlitz-Thouless
type transition between the algebraically ordered to the long-range
ordered phase.

To describe those two transitions, we put $N_1=0$
in \req{F123}, introduce
the field $\phi$, so that $N_2=\cos\phi,\ N_3=\sin\phi$,
and the dual field $\theta$ to describe the vortices.

The singular part of the partition function is given by
\be
\begin{split}
&{\cal Z}\propto \int{\cal D}\phi{\cal D}\theta
\exp\left(-{\mathbb F}_{23}/T\right);
\\
&\frac{{\mathbb F}_{23}}{T}
 = 
\frac{\rho_K}{2T}\left(\partial_{x}  \phi\right)^2 +
\frac{T}{2\rho_K}\left(\partial_x  \theta\right)^2+ 
i\partial_x\theta\partial_y\phi
\\
&\qquad +
\frac{\zeta}{R_\perp^2 T}\cos 6\phi+
%\frac{\mu_{14}}{R_\perp^2}\cos \left(2\pi T\theta/\rho_K\right),
\frac{\mu_{23}}{R_\perp^2}\cos \left(2\pi\theta\right),
\end{split}
\label{N2N3phi}
\ee 
where the coupling constant are described by
\reqs{rhoK} and \rref{RGansolH}
and $R_\perp < R_H$ found from the requirement $\rho_K(R_\perp)\simeq
\Delta\eta(R_\perp)$.
This yields
\be
\frac{R_\perp}{\xi_H}\ln\frac{R_\perp}{\xi_H}
=
\left(\frac{\Delta\eta_\perp(R_*)}{\alpha_\perp^3\Delta\eta(R*)}\right)^{1/2}.
\label{Rperp}
\ee

\begin{subequations}
Once again, the scaling of the vortex fugacity, $\mu_{23}$, and of the
hexadic anisotropy, $\zeta$, can be determined from the first loop
renormalization group equations
\begin{align}
&\frac{d\mu_{23}}{d\ln r_0}
=\left[2-
\frac{\pi\rho_K }{T}
\right]\mu_{23};
\label{mu23eq}\\
&
\frac{d\zeta}{d\ln r_0}
=\left[2-
\frac{9 T}{\pi\rho_K }
\right]\zeta.
\label{zeta23eq}
\end{align}
\label{23eq}
\end{subequations}
In the limit of $\mu_{23}, \zeta \ll 1$, 
\reqs{23eq} enable us to refine the definitions
of $\alpha_{\perp}$ (requiring $\pi\rho_K(R_\perp)=2T$) and to find the
boundary of the ordered phase $\eta_6(R_*)$   (requiring $\pi\rho_K(R_\perp)=9T/2$).
With the help of \reqs{Rperp} and \rref{rhoKH},
we find
\be
\begin{split}
&\alpha_\perp\approx \frac{e^2}{2^{4/3}} \approx 2.9;\\
&\eta_6\approx\left(\frac{4e^9}{9e^4}\right)^2 \eta_{\perp}\approx
4.3*10^3 \eta_{\perp},
\end{split}
\label{alphaperp}
\ee
i.e. even though the boundaries of two phases have the same functional
form, they are very well separated due to the numerical reasons.

In the previous discussion, we implied that $\zeta(R_\perp)\ll 1$.
This condition is clearly violated on the lines $\eta_{6},\
\eta_{\perp},\ \eta_{I}$ if $\pi\rho_K/T \to \infty$, see the last
of \reqs{RGansolH}. Large power of the logarithm
allows one to use the saddle point expression
and we obtain from  condition $\zeta(R_\perp)\gtrsim 1$:
\be
\frac{\pi\rho(R_*)}{2T} > \frac{21}{8}
+\left(\frac{21}{32}\ln \frac{T}{\zeta(R_*)}\right)^{1/2}.
\label{condLargezeta}
\ee
At larger stiffness, the phase transition occurs by   
the locking of the order parameter due to the hexadic term
and the fluctuations become unimportant. Minimising
the potential part of free energy \rref{F123}, we obtain
a first order phase transition at $\eta(R_*) \simeq \zeta(R_*)$.

The only  way of connecting the critical lines on the phase diagram,
allowed by the symmetries of the system is shown in the inset of Fig.~\ref{fig6}.

\subsection{Effect of strong anisotropies.}
\label{sec:anisotropies-strong}

In Sec.~\ref{sec:anisotropies} we assumed that $U(1)$ sector was ordered and the half-vortices 
were not important in ordering the $SU(2)$ sector. It was justified by the
assumption of the weak enough anisotropy so that the parameter $\bar{\eta}$, see \req{notweak},
is constrained by $\bar{\eta}(\xi_{MF}) \ll T$. In this case, the renormalization of
the stiffness $\rho_K$ is so strong that $\rho_K(R_*)\ll \rho_s$ [see \req{R*}]
and separation of the scale was justified. We will see, however, in Sec.~\ref{microscopic-coefficients}
that for some ranges of the in-plane magnetic field the opposite limits of the strong
anisotropies,  $\bar{\eta}(\xi_{MF}) \gtrsim T$ are more relevant.

In this case the logarithmic renormalization of the stiffness $\rho_K$, see \req{stiffnessK},
and of the anisotropy parameters, see \req{RGan} are no longer strong and one has to
investigate the soft modes in \req{anisNN} already on the scale of the order of $\xi_{MF}$.
Similarly to the approach of \ref{sec:anisotropies} only the lines of extra degeneracies
requires  further investigation. The difference is that $U(1)$ sector and the {\em half-vortex} configurations
have to be taken into account.
To simplify further manipulations,
we introduce the parameter
\be
\gamma\equiv\frac{\rho_K(R_*)}{\rho_s(R_*)} < 1.
\label{gamma-def}
\ee
All the results of the previous subsection correspond to the limiting case $\gamma \to 0$.

\subsubsection{The vicinity of line $\eta_\perp=0,\ \eta_z <0$.}
\label{etaperpsmall-2}

Let us generalise \req{N1N2phi} by including the half-vortices.
Substituting parametrisation \rref{NN} with
$N_{2,3}=0,\ N_1=\cos\phi$,\ $N_4=\sin\phi$ into  \req{dual},
representing $2\hat{h}={\rm diag}\,\left[\theta+h_s; \, \theta-h_s\right]$,
we obtain
\be
\begin{split}
&{\cal Z}\propto \int{\cal D}\phi{\cal D}\theta{\cal D}\phi_s{\cal D}\theta_s
\exp\left(-{\mathbb F}_{14}/T\right);
\\
&\frac{{\mathbb F}_{14}}{T}
 = 
[\frac{\rho_K}{2T}\left(\partial_{x}  \phi\right)^2 +
\frac{T}{2\rho_K}\left(\partial_x  \theta\right)^2+ 
i\partial_x\theta\partial_y\phi
\\
&\quad \frac{\mu_{14}}{R_*^2}\cos \left(2\pi\theta\right) + \frac{\eta_\perp}{R_*^2 T}\cos 2\phi ] +
\\
&
\quad + [\frac{\rho_K}{2\gamma T}\left(\partial_{x}  \theta_s\right)^2 +
\frac{T\gamma }{2\rho_K}\left(\partial_x  h_s\right)^2+ 
i\partial_x h_s\partial_y\theta_s
\\
&+
%\frac{\mu_{14}}{R_*^2}\cos \left(2\pi T\theta/\rho_K\right),
\frac{\mu_{1}}{R_*^2}\cos \left(2\pi h_s\right)] +
\frac{\mu_{1/2}}{R_*^2}\cos \left(\pi\ h_s\right)\cos \left(\pi\theta\right)
\end{split}
\label{N1N2phi-strong}
\ee 
[to obtain $\mu_{14}$ term by generating vortices is completely analogous to \req{N1N2phi}].
Here  $\gamma$ is given by (\ref{gamma-def}),   $\mu_{14}$ is a
fugacity for creation of a vortex in $\phi$ field 
(we will call them ``$N_1-N_4$ vortices''), the vortices in $\theta_s$ field [we will call them ``vortices in the $U(1)$ sector''] 
are governed by the fugacity $\mu_1$, and $\mu_{1/2}$ is the fugacity for the half-vortices. 

The first loop RG equations are analogous to \reqs{fug}, 
and \rref{scaling34}:
\be
\begin{split}
&\frac{d\mu_{1/2}}{d\ln r_0}
=\left[2-
\frac{\pi\rho_K}{4T}
\left(\frac{1+\gamma}{\gamma}\right)
\right]\mu_{1/2};
\\
&\frac{d\mu_{14}}{d\ln r_0}
=\left(2-
\frac{\pi\rho_K}{T}
\right)\mu_{14};
\\
&\frac{d\mu_{1}}{d\ln r_0}
=\left(2-
\frac{\pi\rho_K}{\gamma T}
\right)\mu_{1};
\\
&
\frac{d\eta_\perp}{d\ln r_0}
=\left(2-
\frac{T}{\pi\rho_K}
\right)\eta_\perp;
\end{split}
\label{scaling34-strong}
\ee
We will see, that $\gamma=1/4$ is a special point where the $N_1-N_4$ vortices have the same scaling dimension
as the half-vortices for the disordered valley sector.
Another special point is  $\gamma=1/3$  where the half-vortices become more relevant in the disordering
the $N_1-N_4$ sector than the simple vortices.

\begin{figure}[ht]

\includegraphics[width=0.40\textwidth]{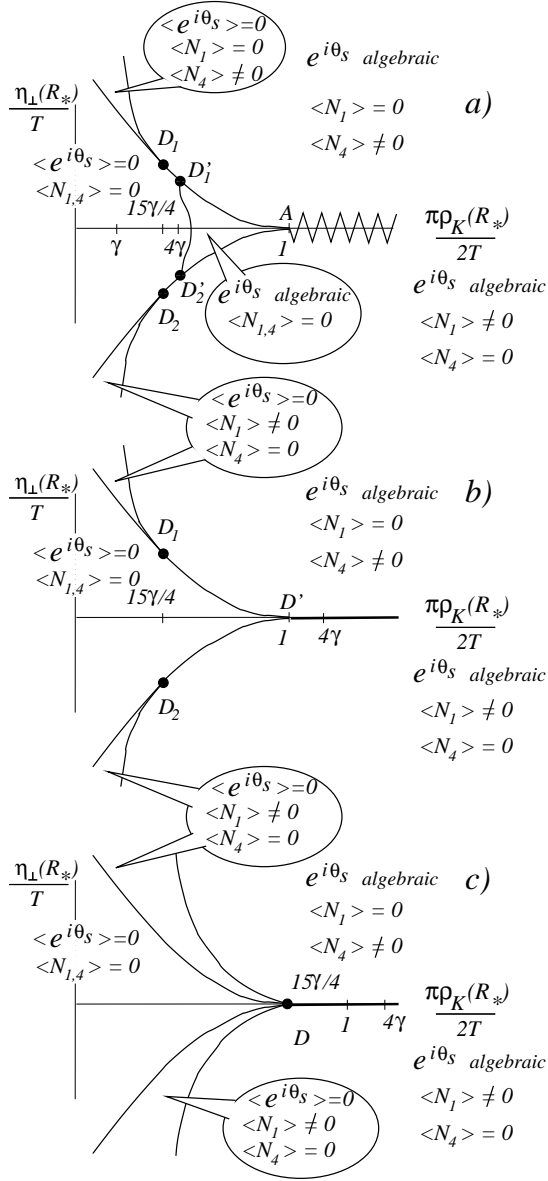}

\caption{The phase diagram in the vicinity of the line $\eta_\perp=0$ for different values of $\gamma$;
a) $\gamma < 1/4$;
Zigzag line denotes either the first order phase transition or pair of the
Ising transition, see Fig.~\ref{fig5}.
The asymptotics of the Berezinskii-Kosterlitz-Thouless lines are given
by \reqs{etaKT} and \rref{nearphoc2} for points $D_{1,2}$ and $D_{1,2}^\prime$ respectively.
 b) $1/4 < \gamma < 4/15$; The asymptotics of the Berezinskii-Kosterlitz-Thouless lines are given
by \reqs{etaKT}  for points $D_{1,2}$.
c) $1/3 <\gamma < 1/4$. The 
difference between the slopes of the Ising and
 the Berezinskii-Kosterlitz-Thouless near multi-critical point $B$
is given by \reqs{etaKT}.  
Finite $\kappa$ leads to the modification of the multi-critical point $B$.
}
\label{fig60}
\end{figure}

The phase diagram for $\gamma < 1/4$ is shown on Fig.~\ref{fig60}. The physics in the vicinity
of the multi-critical point $A$ is still determined by $N_1-N_4$ vortices whereas the half-vortices are irrelevant
in this region. Inside the $N_1-N_4$ disordered region any deformation in $\phi$ causes only
a finite energy, 
\[
\langle \cos \pi\theta\rangle \neq 0.
\]
Therefore, the half-vortices in $U(1)$ sector are allowed and
the condition \rref{Kjump} determines the value of the Kosterlitz jump.
In fact, $\cos \pi\theta $ is proportional to the disorder operator $\mu$ of the Ising model,
see Ref.~\onlinecite{book}. We will use this analogy shortly.

On the other hand, deep in the ordered sector, the half-vortices are confined
and the disordering  the $U(1)$ sector occurs due to the vortices and, thus, the Kosterlitz jump is
determined by \req{KjumpU1}.

Only vicinities of the tricritical points $D_{1,2},\,D'_{1,2}$ on the
Fig.~\ref{fig60} a) deserve a special consideration. In their vicinity
$\mu_1$ term in (\ref{N1N2phi-strong}) is irrelevant and $\mu_{14}$
and $\eta_{\perp}$ terms conspire to produce an Ising critical point.
Therefore at $\mu_{1/2} =0$ we have two decoupled critical theories:
the Ising one described by $\phi,\theta$ fields and the $U(1)$ model
described by $(\theta_s, h_s)$ fields.  At the Ising model critical
point the operator $\cos\theta$ behaves as the disorder parameter
field of the Ising model and has scaling dimension 1/8. Hence the
corresponding RG equation for $\mu_{1/2}$ is 
\be
\frac{d\mu_{1/2}}{d\ln r_0} =\left[2- \frac{1}{8}- \frac{\pi\rho_s
    (r_0)}{4 T} \right]\mu_{1/2},
\label{critpoint-1}
\ee
where we restored $\rho_s$ using \req{gamma-def}.

The scale invariance requirement for the fugacity $\mu_{1/2}$ gives the
stiffness at the multi-critical points $D_{1,2}$
\be
\frac{\pi\rho_s^c(r_0\to\infty)}{2 T_c}=\frac{15}{4} < 4.
\label{crit-point2}
\ee 

If $\rho_s < \rho_s^c$, the half-vortex fugacity $\mu_{1/2}$
grows according to \req{critpoint-1} as
\be
\frac{\mu_{1/2}(r_0)}{\mu_{1/2}(R_{*})}=\left(\frac{r_0}{R_*}
\right)^{\frac{15 (\rho_s^c-\rho_s)}{8 \rho_s^c}}
\label{disordered-scaling}
\ee
The grows \rref{disordered-scaling} has to be stopped at
the correlation length of the Ising transition
\[
\xi_I \simeq \frac{\eta_\perp^c}{|\eta_\perp-\eta_{\perp}^c|}.
\]
At the lengths larger than $\xi^I$ the Ising model becomes ordered, so the half-vortices are confined. However, their fusion produces  usual vortices with the fugacity $\mu_1\simeq \mu_{1/2}^2(\xi_I)$.
The requirement for the resulting parameters of the remaining $U(1)$ theory to be on the line of the 
Berezinskii-Kosterlitz-Thouless
transition leads to the estimate $\mu^{1/2}(\xi_I) \simeq 1$ or, see also Fig.~\ref{fig5} a),
\be
\ln \frac{\eta_c}{\eta_\perp^{KT}-\eta_\perp^c}=\frac{8\rho_s^c}{15(\rho_s^c -\rho_s)}+{\cal O}(1).
\label{etaKT}
\ee

For the spin stiffnesses, in the interval 
\be
\rho_s^c 
< \rho_s < \rho_s^{c,2}, \quad \frac{\pi\rho_s^{c,2}}{2\pi}=4, 
\label{critline3}
\ee
the fugacity for the half-vortices \rref{disordered-scaling}
vanishes at $r_0\to \infty$.
If one tries to deviate from the line $\eta=\eta_c$  
towards the ordered side, those half-vortices fuse into  vortices which are also irrelevant.

Let us now consider the deviation from the tricritical point 
towards the disordered side of the Ising sector. In this region
$\cos\theta$ can be replaced by its average. 
As we have mentioned above, near the Ising model critical line $\cos\theta \sim \mu$ (the disorder operator of the Ising model). 
Hence  we find
\be
\cos\pi\theta\cos\pi\, h_s \to |\eta_\perp-\eta_c^\perp|^{1/8}\cos\pi\, h_s.
\label{half-vortices-disorder}
\ee 
The latter operator is irrelevant in the region \rref{critline3} as well.
Therefore, the line $\eta=\eta_c$,  $\rho_s^c 
< \rho_s < \rho_s^{c,2}$, is the transition line both for the valley and the spin sectors\cite{MF}.

At $\rho > \rho_s^{c,2}$, the half-vortices in the limit of the zero-fugacity $\mu_{1/2}$
become irrelevant and $U(1)$ sector is algebraically ordered. The correction to vertical line
can be evaluated using
\be
\rho_s-\rho_s^{c,2} \simeq \mu_{1/2}(\xi_I) \propto |\eta_\perp-\eta_c^\perp|^{1/8}.
\label{nearphoc2}
\ee 
This estimate is valid near points $D_{1,2}^\prime$.

Case $\gamma >1/4$ needs further investigation as the critical points $D_{1,2}'$ crosses the point $A$,
see Fig.~\ref{fig60} b,c).  For simplicity, we will neglect the quartic anisotropies: $\kappa =0$.

Let us first consider $1/4 < \gamma < 4/15$. If $\eta_\perp(R_*)=0$, the system undergoes the 
Berezinskii-Kosterlitz-Thouless transition and $\pi\rho_K/2T=1$ where $N_{1}-N_4$ vortices become
relevant, $\langle \cos 2\pi\theta \rangle \ne 0$. At the same time $\langle \cos\pi\theta\rangle \ne 0$
and the remaining factor of the half-vortex operator becomes relevant, with the dimensionality $\pi\rho_K/(4\gamma T)$.
As the result, the $U(1)$ sector also becomes disordered.

Initial steps in  the  consideration of $\eta_\perp \neq 0$ are the same as in the derivation
of \reqs{criticalline1} and \rref{xiIsing}, and we obtain two lines of the Ising phase transitions.
The scaling of the half-vortex fugacity on the Ising line  is still governed by \req{critpoint-1}
so that the conclusions of Eqs.~\rref{crit-point2} and \rref{etaKT} remain intact, see Fig.~\ref{fig60} b).

The peculiarity of $4/15 <\gamma < 1/3$ regime is that the half-vortex operator becomes relevant on
the whole Ising line, whereas it is still irrelevant in the ordered region. As the result the points
$D_{1,2}$ on Fig.~\ref{fig60} b) collapse, see Fig.~\ref{fig60} c). The positions of the 
Berezinskii-Kosterlitz-Thouless lines still can be estimated using \reqs{crit-point2} and \rref{disordered-scaling}.

The most delicate case 
which we were not able to solve
is $\gamma > 1/3$. In this situation the {\em half-vortices} are always more relevant than
the vortices and the transitions both in $U(1)$ and in $N_1 -N_4$ sectors occur due to the half-vortices.
The corresponding free energy obtained by keeping only half-vortices in \req{N1N2phi-strong} is
 \be
\begin{split}
&\frac{{\mathbb F}_{14}}{T}
 = 
\frac{\rho_K}{2T}\left(\partial_{x}  \phi\right)^2 +
\frac{T}{2\rho_K}\left(\partial_x  \theta\right)^2+ 
i\partial_x\theta\partial_y\phi
\\
&
\quad + \frac{\rho_K}{2\gamma T}\left(\partial_{x}  \theta_s\right)^2 +
\frac{T\gamma }{2\rho_K}\left(\partial_x  h_s\right)^2+ 
i\partial_x h_s\partial_y\theta_s
\\
&\quad +
\frac{\mu_{1/2}}{R_*^2}\cos \left(\pi\, h_s\right)\cos \left(\pi\theta\right) +
\frac{\eta_\perp}{R_*^2 T}\cos 2\phi.
\end{split}
\label{half-vortices-only}
\ee
We do not know the critical property of this model of this model in the strong coupling lime.
% However, we believe that the
% phase diagram remains qualitatively the same as  in Fig.~\ref{fig60} c), i.e
% with decresing $|\eta_\perp|$ the disordering the $U(1)$ sector occurs first,
% and $\cos\pi\,h_s$ acquires the finite expectation value. The remaining theory would
% corresopond to the second order phase transition with the critical indices different
% from Ising model.

\subsubsection{The vicinity of  lines $\eta_z > 0, \
  \eta_\perp=\pm \eta_z$.}
\label{Heisenberg-strong}

The analysis of those lines relies on the material of Secs.~\ref{etaperpsmall-2} and \ref{Heisenberg}.
Near the line $\eta_z > 0$, $\Delta\eta=|\eta_\perp|-\eta_z \ll \eta_z$, we have to generalise  \req{F123} 
[see also \req{replacement}] to include $U(1)$ sector.
In this subsection we also neglect the hexadic anisotropy $\zeta=0$. The resulting theory
is obtained from \req{dual} by putting extra constraint on the unitary matrix $\hat{V}$
\be
\begin{split}
\begin{matrix}
\hat{V}^\dagger=-\hat{V}, & \eta_\perp<0;\\
\hat{V}^\dagger=\hat{\tau}_y\hat{V}\hat{\tau}_y, & \eta_\perp>0.
\end{matrix}
\end{split}
\label{extra-constraint}
\ee
One case case is mapped to the other by the substitution $\hat{V}\to i\hat{\tau}_y\hat{V}$ and therefore
only one of them has to be studied.

The line $\Delta \eta_I$, see \req{positions} is of the Ising type and all the analysis of 
\reqs{critpoint-1} -- \rref{nearphoc2} is still valid.

\begin{figure}[h]
\includegraphics[width=0.45\textwidth]{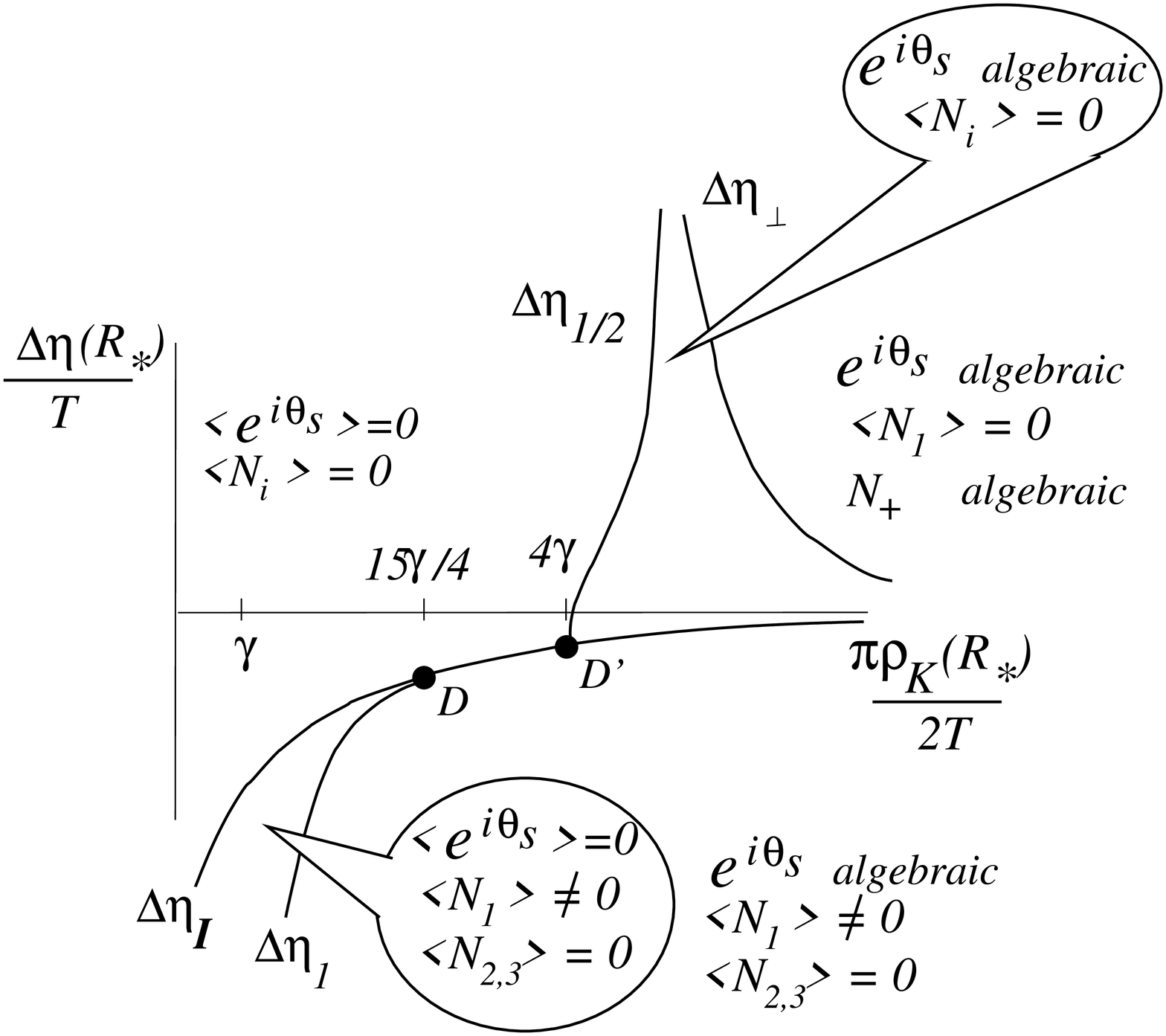}
\caption{
Phase diagrams in the vicinity of the line 
$\Delta\eta=\eta_z+\eta_\perp \ll \eta_z ,\ \eta_z >0$ for finite value of
$\gamma$, see \req{gamma-def}, and in the absence of the hexadic anisotropies,
$\zeta=0$.
Here $N_+=N_2+iN_3$. Lines $\Delta\eta_{\perp,1,1/2}$ are the
Berezinskii-Kosterlitz-Thouless phase transitions. Line $\eta_I$ is the Ising type phase
transition.
The asymptotics of the lines  $\Delta\eta_{1,1/2}$ are given
by \reqs{etaKT} and \rref{nearphoc2} for points $D$ and $D^\prime$ respectively
(those points are equivalent to the multi-critical points of Fig.~\ref{fig60} a) with the
same notation.
The transitions across the lines $\Delta\eta_{1,\perp}$ are   controlled by the unbinding
of the vortices, and $\Delta\eta_{1/2}$ is governed by the half-vortices. 
}
\label{fig61}
\end{figure}

What remains is the vicinity of the Berezinskii-Kosterlitz-Thouless transition line $\Delta\eta_\perp$,
see Fig.~\ref{fig61}. The theory describing phase transition in this case is, compare with \req{N2N3phi},
can be written in terms of dual fields only,
\be
\begin{split}
&{\cal Z}\propto \int{\cal D}h_s{\cal D}\theta
\exp\left(-{\mathbb F}_{23}/T\right);
\\
&\frac{{\mathbb F}_{23}}{T}
 = 
\frac{T}{2\rho_K(R_{**})}\left(\partial_\mu  \theta\right)^2+ 
\frac{T}{2\rho_s}\left(\partial_\mu h_s\right)^2 
\\
&
%\frac{\mu_{14}}{R_\perp^2}\cos \left(2\pi T\theta/\rho_K\right),
+\frac{\mu_{23}(\Delta\eta)}{R_{**}^2}\cos 2\pi\theta
+\frac{\mu_{1/2}(\Delta\eta)}{R_{**}^2}\cos \pi\, h_s \cos\pi\,\theta
,
\end{split}
\label{N2N3phi-strong}
\ee
where $\rho_s \gg \rho_K(R_{**})$ because of the strong logarithmic renormalization at distances where
the valley sector is almost isotropic.
It has a sequence of two phase transitions shown on Fig.~\ref{fig61} by lines
$\Delta \eta_\perp$ and $\Delta\eta_{1/2}$.

%We believe that in this case, the line of the first order transition appears as shown on Fig.~\ref{fig60} b),
%where the coexistences region is shown by dashed line.

% To justify this result, let us assume that $\phi$ in the region
% \be
% 1< \frac{\rho_K}{2T} < 4\gamma
% \ee
% is disordered. Than $\cos \pi\theta $ in the half-vortex term acquires the finite expectation
% value and the scaling of the half-veortex fugacity $\mu_{1/2}$ is determined by
% \be
% \frac{d\mu_{1/2}}{d\ln r_0}
% =\left[2-
%\frac{\pi\rho_K (r_0)}{4\gamma T}
%\right]\mu_{1/2},
%\label{firstorder-1}
%\ee
%i.e. the operator is relevant, and the both fields $h_s,\eta$ become massive. It means that the system
%is stable with respect to switching on small but finite $\eta_\perp$.   

%Let us make now the opposite assumption, let us assume

\subsection{Resulting phase diagram.}

In Secs.~\ref{sec:anisotropies}, \ref{sec:anisotropies-strong},
we  analysed of the vicinities of the degeneracies point and the vicinities
of the degeneracies lines of Fig.~\ref{fig4}. The results of this analysis
enable us to construct the phase diagrams in terms in the plane
defined by anisotropies $(\eta_{z}(R_*), \eta_\perp (R_*)$,
see also \reqs{notweak} and \rref{R*}.  Indeed, there can be no other singularities
than those we have already considered, because for all the other regions
of the phase diagrams the $SU(2)$ sector is massive due to the anisotropies.
Therefore, the character of the singularities along the phase transition
line remain the same.

\begin{figure}[ht]
\includegraphics[width=0.45\textwidth]{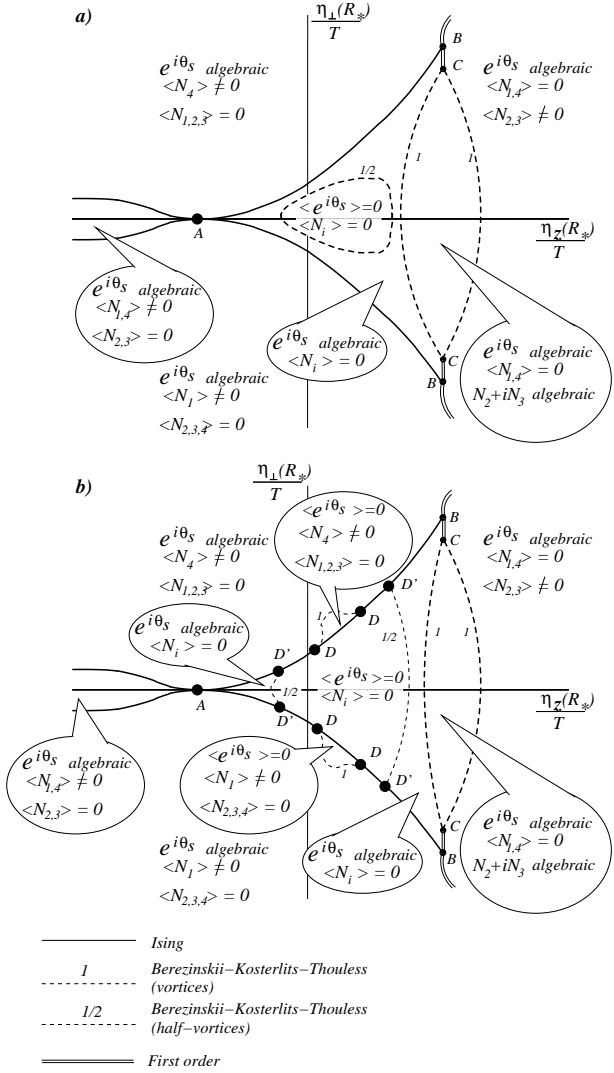}
\caption{The overall phase diagram for (a) decoupled valley and spin sectors, $\gamma \to 0$;
and (b) for $\gamma < 1/4$. 
The positive quartic anisotropy, $\kappa >0$ is assumed.
The more detailed behaviour near the multi-critical points $A,B,C,D$
are shown in more details on Figs.~\ref{fig5}--\ref{fig61}. The notation for those
points here is consistent with that for Figs.~\ref{fig5}--\ref{fig62}.}
\label{fig62}
\end{figure}

Bearing this in mind and expressing $\rho_K(R_*)$ in terms of
anisotropies using \req{etaR*}, we construct the ``global'' phase diagram
shown on Fig.~\ref{fig62}. The relation
of this phase diagram to the physical coordinates {\cal B}, $T$ will
be found in Sec.~\ref{sec:log4} after the microscopic
theory for the parameters of Landau free energy is built.

\section{Logarithmic renormalizations and mean field transition}
\label{sec:log}

The specifics of the problem in hand that it has three
diferent intervel of the logarithmic reormalizations:
(i) energies larger than Zeeman splitting; (ii)
energies larger than temperature but smaller than Zeeman splitting;
(iii) classical renormalizations considered in the previous Section.
As those renormalization are contributed by different degrees of
freedoms they have to be considered separately.
The result will be the microscopic expressions for the coupling
constants in the free energy which will enable us to construct the
physical phase diagram in Sec.~\ref{sec:log4}.

\subsection{Logarithmic renormalizations at energies larger than
  Zeeman splitting.}

\label{sec:log1}

At such energies the Zeeman splitting is not important and
can be considered perturbatively if necessary.
We will also assume that the short range interaction and
umklapp terms are not strong enough to lead to
any reconstruction in the state of the system at high energies,
so they also can be considered within the perturbation
theory,  leading to a simple modification
of the coupling constants which are not well-known anyway.

The only terms which require the special attention are those related to
 the long-range Coulomb interaction. Indeed, 
a simple dimensional analysis of the Hamiltonian \rref{Dirac} --
\rref{Coulomb} points to  logarithmic divergences
in the simple perturbation theory (first identified in
Ref.~\onlinecite{Abrikosov} for the three-dimensional gapless
semiconductors).

As usual, a summation of the leading logarithmic divergences
is performed within the renormalization group scheme.
However, the loop expansion would not be suitable
for the description of the graphene as the dimensionless
interaction strength $e^2/v(r_c)$ is not small at distances
$r_c\simeq a$. 

Instead, the expansion in terms of $N$ -- the number of independent
fermion species entering the Hamiltonian \rref{Dirac} --
\rref{Coulomb} will be used. For the problem at hand, we have
the valley and spin degeneracies so that $N=4$.
The results, which will be obtained, indicate, that $1/N$ corrections
are quite small for $N=4$, so that $1/N$ expansion seems to
be a reasonable approximation.

\begin{figure}[h]
 \includegraphics[width=0.45\textwidth]{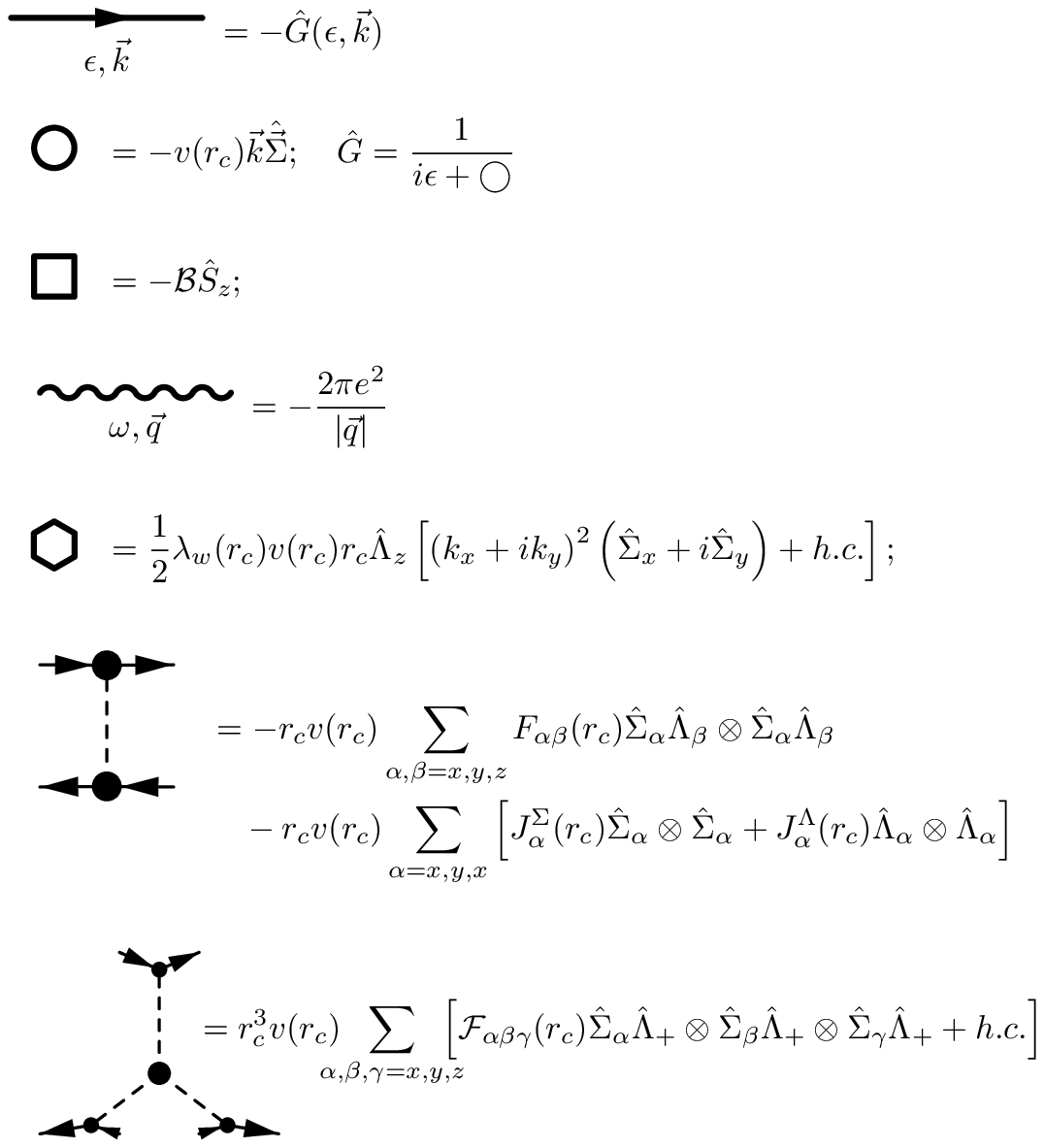}
\caption{Basic elements for the diagrammatic
calculation for Hamiltonian \rref{Hamiltonian}--\rref{Fumklapp},
at distances $r_c\lesssim R_B\equiv v(R_B)/{\cal B}$.}
\label{fig7}
\end{figure}

To make the calculations compact, we will utilise
the standard imaginary time diagrammatic technique\cite{AGD}.
Analytic expressions for the corresponding lines are given on Fig.~\ref{fig7}.

\begin{figure}[ht]
\vspace*{1cm}

\includegraphics[width=0.45\textwidth]{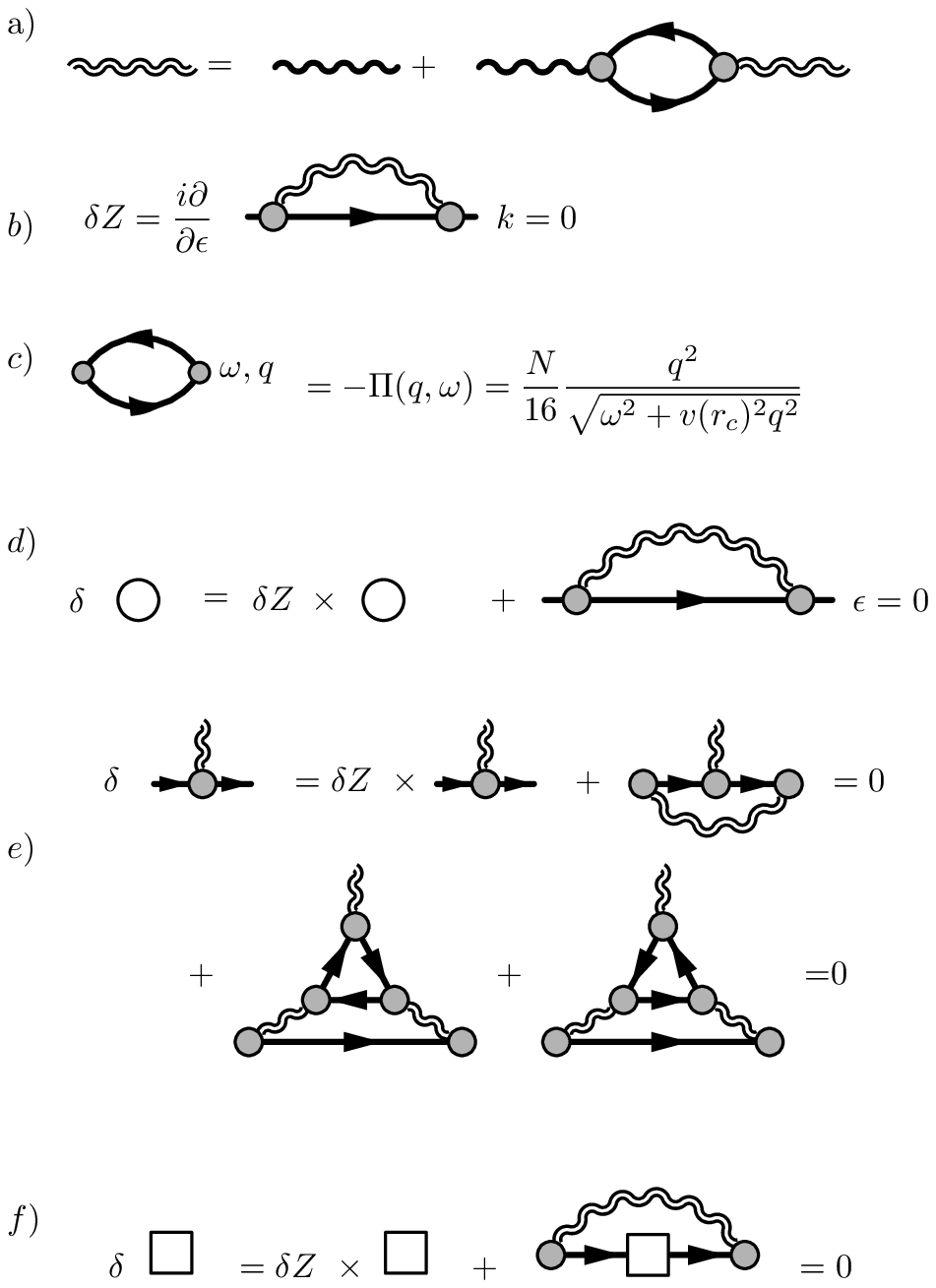}

\vspace*{0.2cm}
\caption{Diagrammatic representation for the
leading in $1/N$ renormalization of the parameters of the
high-symmetric part of the Hamiltonian, see
\req{Dirac}. All the basic elements are defined
on Fig.~\ref{fig7}. The integration
over the momentum $q$ going through the wiggly line
is restricted by $1/r_c^><|q|<1/r_c^<$. 
}
\label{fig8}
\end{figure}

Performing the renormalization group procedure, we
change the smallest spatial scale in the problem from
$r_c^<$ to the exponentially larger scale $r_c^>$,
It amounts to taking into account all the diagrams
where the momenta $q$ going through the Coulomb interaction
propagator, belong to the region $1/r_c^><|q|<1/r_c^<$,
This cut-off procedure, does not violate
the gauge invariance or other symmetries of the problem.
Then, we rescale fermionic fields
as $\psi \to (1+\delta Z/2)\psi,\  \bar{\psi} \to (1+\delta
Z/2)\bar{\psi}$ in order to keep the term
$\bar{\psi}\partial_\tau\psi$ 
intact\footnote{In this scheme the scalar vertex and the Zeeman
  splitting term are not renormalised also,  as a consequence of the
  gauge invariance. This  can be checked by explicit calculation
  of
the diagrams of Fig.~\ref{fig8} e,f).}.

For the high-symmetry part of the Hamiltonian,
see \reqs{Dirac} and \rref{Coulomb}, the only 
coupling which renormalises is the velocity $v(r_c)$. This renormalization is a consequence of non-Lorentz-invariance of the Coulomb interaction.

\begin{figure}[ht]
\includegraphics[width=0.4\textwidth]{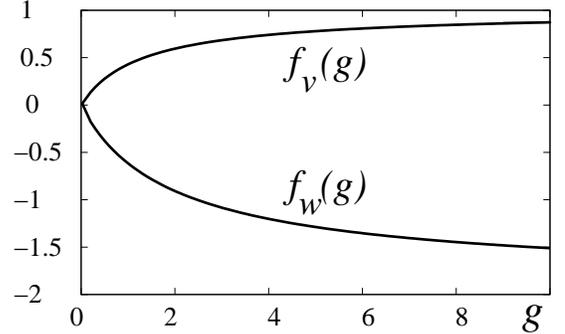}
\caption{Plots of functions $f_v(g)$ and $f_w(g)$ entering into
renormalization group equations.
}
\label{fig80}
\end{figure}

Introducing the dimensionless interaction strength
\be
g(r_c)=\frac{\pi e^2 N}{8v(r_c)},
\label{gdef}
\ee
and calculating the diagrams of Fig.~\ref{fig8} d),
we obtain
\be
\frac{d\ln g}{d\ln r_c}=-\frac{8}{\pi^2 N}f_v(g),
\label{RGg}
\ee
where dimensionless function $f_v$ is given by
\be
f_v(g)=1-\frac{\pi}{2g}+
\left\{
\begin{matrix}
\displaystyle{\frac{\arccos g}{g\sqrt{1-g^2}}}, & g \leq 1;
\\
\\\displaystyle{
\frac{{\rm arccosh}\, g}{g\sqrt{g^2-1}}}
, & g \geq 1.
\end{matrix}
\right.
\label{fvg}
\ee
It is easy to see that the function $f_v(g)$ is monotonously
increasing
and analytic for all $g>0$, see Fig.~\ref{fig80}. 
The asymptotic behaviour of this
function is
\be
f_v(g)\approx\left\{
\begin{matrix}
\displaystyle{
\frac{\pi g}{4}-\frac{2g^2}{3}+
{\cal O}(g^3)}, & g \ll 1;
\\
\\\displaystyle{
1-\frac{\pi}{2 g} + \frac{\ln 2g }{g^{2}}
+ {\cal O}\left(\frac{1}{g^{4}}\right)}
, & g \gg 1.
\end{matrix}
\right.
\tag{\ref{fvg}$^\prime$}
\label{fvgas}
\ee
Equations \rref{RGg} -- \rref{fvg} 
  were first obtained in Ref.~\onlinecite{vozmediano}
for $N=1$ where they are not applicable beyond the first term in the
expansion \rref{fvgas} for $g\ll 1$. The validity of those formulas
 for $N\gg 1$ was pointed out in Ref.~\onlinecite{son}, but the numerical
coefficients here are different from the latter reference.

Solution of \req{RGg} with the help of the asymptotics \rref{fvgas}
yields
\be
\begin{split}
g\left(\frac{r_c}{\cal R}\right) & \approx
\left\{
\begin{matrix}
\displaystyle{
\left[\frac{2}{N\pi}
\ln \left(\frac{r_c}{{\cal R}}\right)+1
\right]^{-1}
}, & r_c \gtrsim
{\cal R};
\\
\displaystyle{\left(\frac{{\cal R}}{r_c}\right)
^{\frac{8}{\pi^2 N}}
\approx\left(\frac{{\cal R}}{r_c}\right)^{0.20},}
& a\ll r_c \lesssim {\cal R};
\end{matrix}
\right..
\end{split}
\label{Solg}
\ee
where ${\cal R}$ is the only relevant spatial scale generated by interaction.
 This scale  can be estimated as
\be
\ln \frac{\cal R}{a} \simeq\left\{  
\begin{matrix} 
\displaystyle{
\frac{\pi^2 N}{8}\ln g(a)}; & g(a) \gtrsim 1\\
-\displaystyle{\frac{\pi N}{2g(a)}}; & g(a) \lesssim 1.
\end{matrix}\right.
\label{calR}
\ee
where $a$ is the scale of the order of the lattice constant at
which the continuous description becomes applicable.

For the graphene sheets, the reported velocity, see e.g. Ref.~\onlinecite{velocity}, is
$v(a)\simeq 10^8\,cm/s$, we estimate $g\simeq 2 \div 4$ (uncertainty is associated with the dielectric
properties of the substrate as well as uncertainty of linear scale at which the
velocity is measured) and we find from \req{calR}
\be
{\cal R}\simeq 10^2\,\div \, 10^3 a \gg a.
\label{calREstimate}
\ee
For all the further consideration we assume that the relation ${\cal R} \gg a$ is fulfilled.

\begin{figure}[ht]
\includegraphics[width=0.45\textwidth]{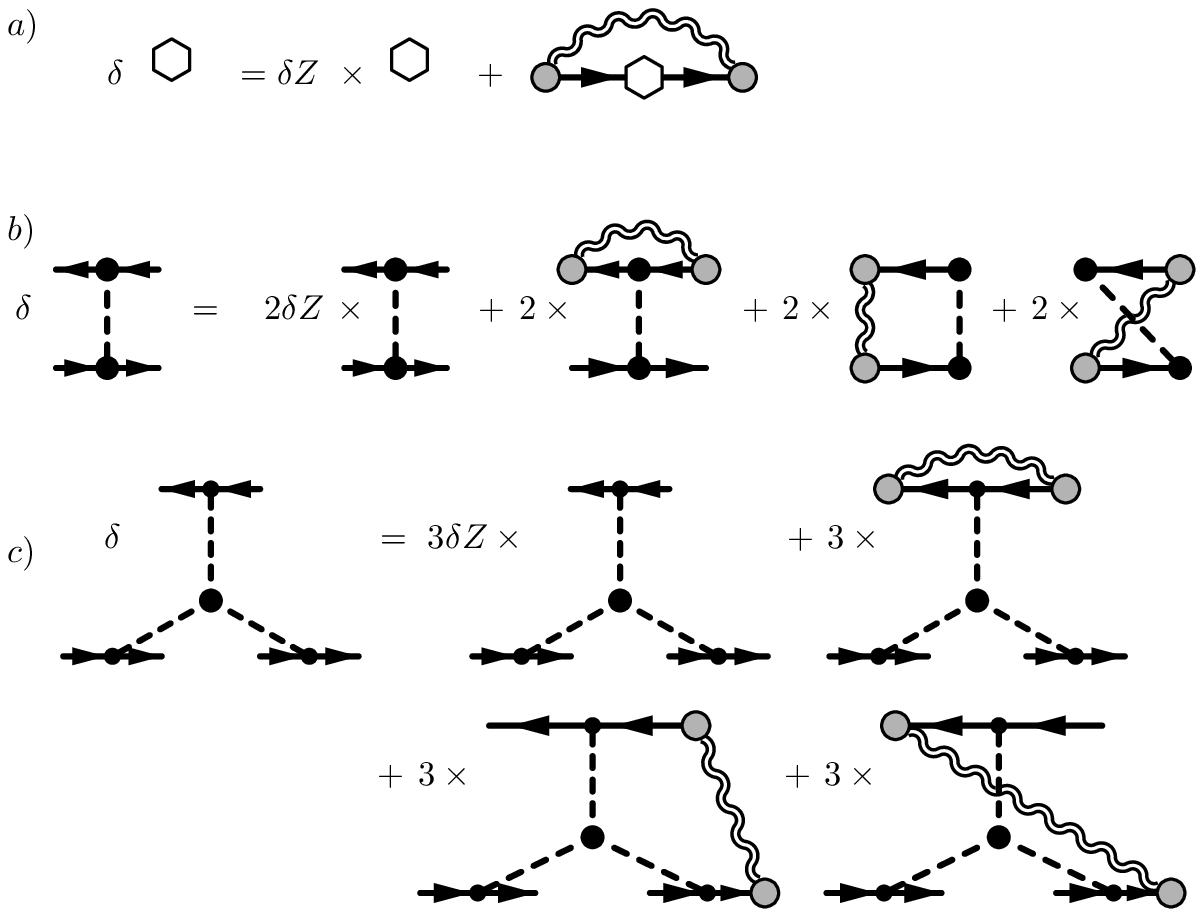}
\caption{Diagrammatic representation for the
leading in $1/N$ renormalization of the parameters of the
low-symmetric part of the Hamiltonian, see
\reqs{warping}, \rref{sr} and \rref{umklapp}. All the basic elements are defined
on Figs.~\ref{fig7}, \ref{fig8}. The integration
over the momentum $q$ going through the wiggly line
is restricted by $1/r_c^><|q|<1/r_c^<$. 
}
\label{fig9}
\end{figure}

The Coulomb interaction strongly affects
the scaling of the low-symmetry terms of the Hamiltonian.
Let us start from the trigonal warping term
\rref{warping}. Calculating diagrams shown in Fig.~\ref{fig9}a,
we find 
\be
\frac{d\ln \lambda_w}{d\ln r_c}=-1+
\frac{4}{\pi^2 N}f_w(g);
\label{warpRG}
\ee
where negative monotonous analytic function
\be
\begin{split}
f_w(g)&=
-\frac{28}{15}+\frac{13\pi}{8g}
+\frac{10}{g^2}-\frac{11\pi}{2g^3}
-\frac{6}{g^4}+\frac{3\pi}{g^5}
\\
&+\left(-\frac{4}{g}+\frac{7}{g^3}-\frac{3}{g^5}\right)\times
\left\{
\begin{matrix}
\displaystyle{\frac{2\arccos g}{\sqrt{1-g^2}}}, & g \leq 1;
\\
\\\displaystyle{
\frac{2\, {\rm arccosh}\, g}{\sqrt{g^2-1}}}
, & g \geq 1;
\end{matrix}
\right.
\end{split}
\label{fwg}
\ee
is also plotted on Fig.~\ref{fig80}.

The asymptotic behaviour of this function is
\be
f_w\approx\left\{
\begin{matrix}
\displaystyle{-
\frac{5\pi g}{16}+\frac{64 g^2}{105}+
{\cal O}(g^3)}, & g \ll 1;
\\
\\\displaystyle{
-\frac{28}{15}+\frac{13\pi}{8 g} + \frac{10-8\ln 2g }{g^{2}}
+ {\cal O}\left(\frac{1}{g^{3}}\right)}
, & g \gg 1.
\end{matrix}
\right.
\tag{\ref{fwg}$^\prime$}
\label{fwgas}
\ee

Solution of \req{warpRG} with the help of \req{fwgas} yields
\be
\begin{split}
\frac{\lambda_w(r_c)}{\lambda_w({\cal R})} & \approx
\left\{
\begin{matrix}
\displaystyle{
\left(\frac{\cal R}{r_c}\right)
\left[\frac{2}{N\pi}
\ln \left(\frac{r_c}{{\cal R}}\right)+1
\right]^{-5/8}
}, 
\\ \qquad\qquad r_c \gtrsim
{\cal R};
\\
\\
\displaystyle{\left(\frac{{\cal R}}{r_c}\right)
^{1+\frac{112}{5 \pi^2 N}}
\approx\left(\frac{{\cal R}}{r_c}\right)^{1.57},}
\\ \qquad\qquad
a\ll r_c \lesssim {\cal R};
\end{matrix}
\right.
\end{split}
\label{Solwarping}
\ee 
where
\be
\lambda_w({\cal R})\simeq \lambda_w(a) \left(\frac{a}{\cal R}\right)^{1.57} 
\simeq 10^{-3}\, \div \, 10^{-5}.
\label{lambdaR}
\ee
Thus, we see that the Coulomb interaction tends to suppress 
drastically the
warping term making the energy surfaces more and more isotropic\cite{AleinerFalko}.
 
On the other hand, the Coulomb interaction leads to
the enhancement on the short-range interaction
terms in the Hamiltonian \rref{sr}.
Calculating diagrams of Fig.~\ref{fig9} b) we
find\footnote{Notice that the ``mean-field analysis'' of
Ref.~\onlinecite{khv} of the excitonic instabilities at zero magnetic
field corresponds to accounting of only the third diagram in the
right-hand-side
of Fig.~\ref{fig9} b) and thus is false even within $1/N$ approximation.}
\be
\begin{split}
&\frac{d \ln F_+^{z,\perp}}{d\ln r_c}=\frac{d \ln J_+^{\Sigma,\Lambda}}
{d\ln r_c}=\frac{d \ln J_-^{\Lambda}}
{d\ln r_c}=
-1;
\\
&\frac{d \ln F_-^{z,\perp}}{d\ln r_c}=\frac{d \ln J_-^{\Sigma}}
{d\ln r_c}=
-1+ \frac{40}{\pi^2 N}f_v(g),
\end{split}
\label{srRG}
\ee
where function $f_v(g)$ is defined in \req{fvg}.

\begin{subequations}
Equations \rref{srRG} can be easily solved
with the help of \req{RGg} and we find
\begin{align}
&\frac{F_+^{z,\perp}(r_c)}{F_+^{z,\perp}(a)}=
\frac{ J_+^{\Sigma,\Lambda}(r_c)}{J_+^{\Sigma,\Lambda}(a)}=
\frac{J_-^{\Lambda}(r_c)}{J_-^{\Lambda}(a}
=\frac{a}{r_c};\label{srRGsolution:a}
\\
&\frac{F_-^{z,\perp}(r_c)}{F_-^{z,\perp}(a)}=
\frac{ J_-^{\Sigma}(r_c)}{J_-^{\Sigma}(a)}=\frac{a}{r_c}
\left(\frac{g(a)}{g(r_c)}\right)^5.\label{srRGsolution:b}
\end{align}
\label{srRGsolution}
\end{subequations}

Couplings in the \req{srRGsolution:a} are irrelevant.
The interactions in \req{srRGsolution:b} are strongly
enhanced by the long-range Coulomb interaction.
This enhancement becomes especially pronounced
at intermediate distances $r_c \lesssim {\cal R}$.
Using \req{Solg} we find
\[
\frac{F_-^{z,\perp}(r_c)}{F_-^{z,\perp}(a)}=
\frac{ J_-^{\Sigma}(r_c)}{J_-^{\Sigma}(a)}\approx
\left(\frac{r_c}{a}\right)^{\frac{40}{\pi^2 N}-1}\approx
\left(\frac{r_c}{a}\right)^{0.01},
\]
i.e. naively dimensionally irrelevant couplings
become weakly relevant. Though, such small value of indices
is clearly beyond the accuracy of the $1/N$ approximation
for $N=4$, this formula indicates, however, that the
effect of the short range interaction lowering
the symmetry of the system is much stronger than it
was thought  before. Whether or not this enhancement
may lead to  instability at zero magnetic field requires
further improvement of the renormalization group
scheme, which is beyond the scope of the present
paper. Here, we simply note that at large
distances. $g\to 1/\ln(r_c)$, see \req{Solg}, so that the coupling
of \req{srRGsolution:b} becomes irrelevant again. It indicates
that at zero magnetic field the excitonic instability can occur
only as a first order phase transition. In all subsequent
consideration, we will assume that such transition does not occur.
This assumption is in accord with all the experimental findings
accumulated so far.

For the further use, let us recast the answer \rref{srRGsolution} for the most important constants,
in the form similar to \rref{Solwarping}. Because of  estimate \rref{calREstimate}, the
second  digits in the exponents are not observable and will be omitted:
\be
\begin{split}
F_-^{z,\perp}(r_c)&\approx
\left\{
\begin{matrix}
F_-^{z,\perp}({\cal R}); & r_c\lesssim {\cal R};\\ \\
F_-^{z,\perp}({\cal R})\frac{{\cal R}}{r_c}\left[
\frac{2}{N\pi}
\ln \left(\frac{r_c}{{\cal R}}\right)+1\right]^5 ; & r_c\gtrsim {\cal R};
\end{matrix}
\right.
\\
F_-^{z,\perp}({\cal R}) & \approx  F_-^{z,\perp}(a) \simeq 1.
\end{split}
\label{FestimateRG}
\ee

Finally, the Umklapp terms are also enhanced by the interaction.
Calculating the contributions shown on Fig.~\ref{fig9} b), we 
obtain the renormalization group equations
\be
\begin{split}
&\frac{d \ln {\cal F}_+}{d\ln r_c}=
-3;
\\
&\frac{d \ln{\cal F}_-}{d\ln r_c}=
-3 + \frac{64}{\pi^2 N}f_v(g).
\end{split}
\label{umklappRG}
\ee
Similarly to the \rref{srRGsolution}, the solution of
\req{umklappRG} is
\be
\frac{{\cal F}_+(r_c)}{{\cal F}_+(a)}=\left(\frac{a}{r_c}\right)^3;
\quad \frac{{\cal F}_+(r_c)}{{\cal
    F}_+(a)}=\left(\frac{a}{r_c}\right)^3
\left(\frac{g(a)}{g(r_c)}\right)^8.
\label{umklappSolution}
\ee

In particular, at the intermediate distances $r_c \lesssim {\cal R}$,
we find
\[
\frac{{\cal F}_+(r_c)}{{\cal
    F}_+(a)}\approx
\left(\frac{a}{r_c}\right)^{3-\frac{64}{\pi^2 N}}\approx
\left(\frac{a}{r_c}\right)^{1.4},
\]
i.e. the three particle Umklapp interaction remain irrelevant though
it is strongly enhanced by the Coulomb interaction.

To conclude this subsection we notice that the short range
interaction terms are vital in the consideration of
the QHE ferromagnets\cite{Levitov,Herbut} and
the effect of warping on the weak localization was considered in
Ref.~\onlinecite{warpingFalko}. The results of this subsection
indicates that the estimates done in those works are hardly reliable.

\subsection{Logarithmic renormalizations at energies smaller than
  Zeeman splitting: separation of the electron-hole and the Cooper channels.}
\label{sec:log2}

In  the previous subsection we considered the
Zeeman splitting as a perturbation. This  is legitimate
to do up to the spatial scale $r_c< R_B$, where the length
$R_B$ is found from the equation
\be
v\left(\frac{R_B}{\cal R}\right)={\cal B}{R_B}.
\label{R_B}
\ee
The scale dependent velocity $v$ is determined from \reqs{gdef}
and \rref{RGg} and we highlighted that this dependence may include only one spatial scale ${\cal R}$.
To solve \req{R_B} and facilitate further discussion, we introduce the natural scale for the Zeeman
splitting,  ${\cal B}_0$, according to
\be
{\cal B}_0=\frac{v(r_c={\cal R})}{\cal R},
\label{B0}
\ee
rough estimate for ${\cal B}_0$ is ${\cal B}_0\simeq 10\div 10^2 K$.
Then the solution of \req{R_B} takes the universal form
\be
R_B={\cal R}f_B\left(\frac{B}{B_0}\right)
\label{Rbsol1}
\ee
where the function $f_B(x)$ is the solution
of the equation
\begin{align}
& v\left[f_B\left(x\right)\right]= x v(1) \nonumber\\
& f_B(x) \approx 1/x^{1.25};\quad x\gtrsim 1; \nonumber\\
& f_B(x) \approx \frac{2}{N\pi x}
\ln \left(\frac{1}{x}\right)+\frac{1}{x}
; \quad x\lesssim 1;
\label{Rbsol2}
\end{align}

At $r_c >R_B$, Zeeman splitting freezes some electronic degrees of
freedom, and, on the other hand, gives rise to the finite density of
state for the other ones, see Fig.~\ref{fig2} b).

To make use of such separation, we will include
the Zeeman splitting in the denominator of the Green function and 
decompose the result as
\be
\begin{split}
&\hat{G}=\frac{1}{i\epsilon-v(R_B)\vec{k}\hat{\vec \Sigma}-
{\cal B}\hat{S}_z
}=\hat{\cal G}+ \delta\hat{\cal G};
\\
&
\hat{\cal G}=\hat{\cal P}(\vec{n})\frac{1}{i\epsilon-\xi\hat{S}_z  };
\quad \delta\hat{\cal G}
=\left(\openone-
\hat{\cal P}(\vec{n})
\right)\frac{1}{i\epsilon+\left(\xi-2{\cal B}\right)\hat{S}_z  };
\\
&\xi\equiv {\cal B}-v(R_B)|\vec{k}|;
\quad \vec{n}=\frac{\vec{k}}{|k|},
\end{split}
\label{calG}
\ee
where
\be
\hat{\cal P}\equiv \frac{1}{2}
\left(1-\vec{n}\cdot\hat{\vec \Sigma}\hat{S}_z\right);
\quad \hat{\cal P}^2=\hat{\cal P};
\label{Projection}
\ee
is the projection operator to the branches of the
electron spectrum which may produce excitations
with the energies smaller than ${\cal B}$.

The $\delta {\cal G}$ component of the Green function \rref{calG} does
not have a resonant denominator and may be neglected. For the
logarithmically divergent contributions only $\xi \lesssim {\cal B}$
are important so that the integration over the momentum can be
replaced by
\be
\int \frac{d^2k}{\left(2\pi\right)^2}
\dots \to  \int \frac{d\vec{n}}{2\pi}\int_{-{\cal B}}^{\cal B}
\frac{d\xi}{2\pi v^2(R_B)}\dots.
\label{integration}
\ee

\begin{figure}[h]
\includegraphics[width=0.48\textwidth]{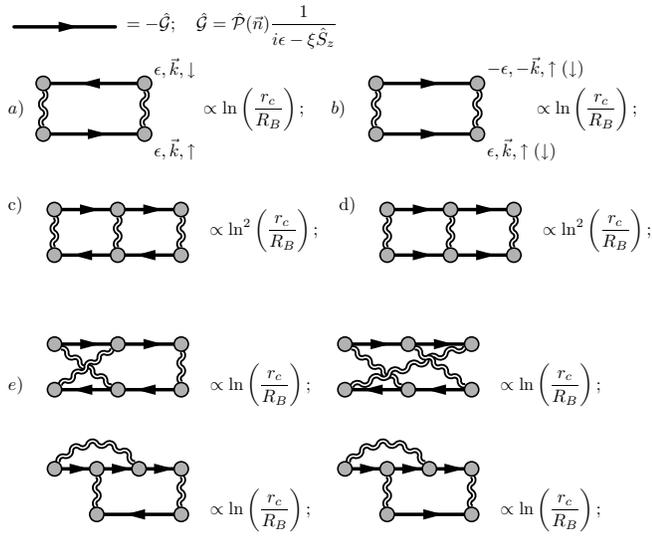}
\caption{Leading (a-d) and sub-leading (e)
logarithmic divergences at $r_c >R_B$.
Series (a,c) corresponds to the excitonic instability\cite{KeldyshKopaev}
in the electron-hole channel studied in the present series.
The interaction in the Cooper channel (b,d) is repulsive
and, therefore, renormalises to zero.
}
\label{fig10}
\end{figure}

Usual calculation of the second order correction to the interaction
vertex shown on Fig.~\ref{fig10} a,b) reveals
two logarithmically divergent contributions,
which can be readily identified as electron-hole (a) and Cooper (b) channels. 
It is worthwhile to emphasise that those contributions
are associated with the presence of the Fermi surface, and
have the structure very different from the high-energy logarithmic
terms of the previous subsection.

To collect the leading logarithmic divergences, we look
at the third order diagram, and find that only those corresponding
to the ladder series, see Fig.~\ref{fig10} b,d) are proportional
to the second power of the logarithm. The sub-leading terms, see
Fig.~\ref{fig10} e), can be combined as a perturbative (in $1/N$, or
the interaction strength) renormalization of the coefficients
in the second order diagram and, therefore, can be neglected.

Finally, it is easy to check that the signs of the diagrams
Fig.~\ref{fig10} a) and c) are the same, whereas,
the signs of the diagrams
Fig.~\ref{fig10} b) and d) are  different from each other.
The latter corresponds to the repulsive interaction in the Cooper
channel, which, thus renormalises to zero.
The former one describes the attractive interaction of  electron
and hole and leads to the excitonic instability.

Therefore, the only relevant diagrammatic series is the ladder
series in the electron-hole channel, which should
be summed up in all  orders of the perturbation theory.
The fact, that the logarithmic divergence occurs only in one channel
justifies the mean-field approximation which will be employed in the
next subsection.

\subsection{Mean field transition.}

\begin{figure}[h]
\includegraphics[width=0.455\textwidth]{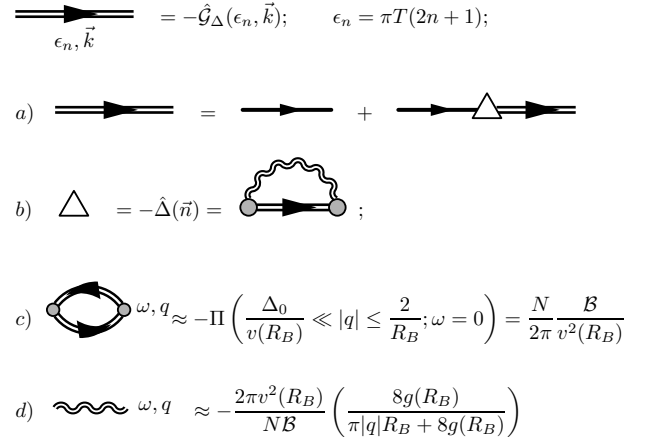}
\caption{Mean-field equations corresponding
to the summation of the most divergent series, Fig.~\ref{fig10}a), c).
}
\label{fig11}
\end{figure}

After the leading divergent series is identified, it can be summed within
the standard mean field approximation\cite{AGD}, shown
on Fig.~\ref{fig11}a-c).
Simple examination of the momentum $q$ and frequencies $\omega$
transfered through the interaction wiggly line shows
that all the momenta $q<\frac{2}{R_B}$ contribute almost equally into the formation
of the order parameter, whereas the approximation $\omega\approx 0$ is valid with
the logarithmic accuracy. The finite order parameter $\Delta$ changes
the polarisation operator only for small $q\simeq \Delta/v(R_B) \ll \frac{1}{R_B}$.
This modification may produce effect only of the order of $\Delta^2/{\cal B}^2$ and
it will be neglected, see Fig.~\ref{fig11} c-d).\cite{P}

The resulting mean field equations are obtained
from Fig.~\ref{fig11} a-b) with the interaction propagator Fig.~\ref{fig11} c)
using the approximation $|\vec{p}_1-\vec{p}| \approx 2/R_B \sin \frac{1}{2}\widehat{\vec{n};\vec{n_1}}$,
valid once again with the logarithmic accuracy, and the integration rule \rref{integration}. It results in
\be
\begin{split}
\hat{\Delta}(\vec{n})
&=T\sum_{\epsilon_n=\pi T(2n+1)}
\int\frac{dn_1}{2\pi}V\left(\widehat{\vec{n}:\vec{n_1}}\right)
\\
&
\times\hat{\cal P}(\vec{n_1})
\int_{-{\cal B}}^{\cal B}\frac{{\cal B}d\xi}{2\pi v^2(R_B)}
\frac{i\epsilon_n-\xi\hat{S}_z+\Delta(\vec{n}_1)}
{\epsilon_n^2+\xi^2+\hat{\cal P}(\vec{n_1})\Delta(\vec{n}_1)^2}
\end{split}
\label{MF}
\ee
where
\be
\begin{split}
V(\alpha)=\frac{2\pi v^2(R_B)}{N {\cal B}}
\left(\frac{8g}{2 \pi\sin\frac{|\alpha|}{2} +8g}\right);
\\
V_n\equiv \int_{-\pi}^\pi\frac{d\alpha}{2\pi}V(\alpha)\cos n\alpha.
\end{split}
\label{VMF}
\ee

Substituting $\hat{\Delta}(\vec{n})$ of the form
\be
\hat{\Delta}(\vec{n})=
\hat{S}_x\hat{\Sigma}_z\left\{\Delta_0(T)
\hat{\cal P}(\vec{n})
+ \Delta_1(T)\left[\openone-\hat{\cal P}(\vec{n})\right]\right\}
\label{solMF}
\ee
into \req{MF}, we find
\[
 \Delta_1(T)=\frac{V_1-V_0}{V_1+V_0}\Delta_0(T),
\]
and the self-consistency relation involving 
$\Delta_0(T)$ only.
At $T=0$ the gap is given by
\be
1=\frac{\left(V_0+V_1\right){\cal B}}{4\pi v^2(R_B)}\ln\frac{2{\cal B}}{\Delta_0},
\label{MFDelta}
\ee
where the constants $V_{0,1}$ are given by \req{VMF}.

Substituting \req{VMF} into \req{MFDelta}, we obtain after the integration
\be
\Delta_0(T=0)=
2{\cal B}\exp\left(-\frac{2N}{f_{\Delta}
\left(\frac{\pi}{4g(R_B)}\right)
}\right),
\label{Delta-zeroT}
\ee
where $f_\Delta(x) < 1$ is a dimensionless function of the 
scale dependent interaction strength \rref{gdef}, and $R_B$
is defined by \req{R_B}.

The explicit expression for this function is
\be
\begin{split}
f_\Delta (x)=
\left\{
\begin{matrix}
\displaystyle{\frac{4}{\pi}
\left[
\frac{\sqrt{x^2-1}}{x^2}\,{\rm arccosh}\,x-\frac{1}{x}+\frac{\pi}{2x^2}
\right]}, & x\geq 1\\\\
\displaystyle{\frac{4}{\pi}
\left[-
\frac{\sqrt{1-x^2}}{x^2}\,{\rm arccos}\,x-\frac{1}{x}+\frac{\pi}{2x^2}
\right]}, & x\leq 1
\end{matrix}
\right.
\end{split}
\label{fDelta}
\ee
and its asymptotic behaviour is given by
\be
\begin{split}
f_\Delta (x)\approx
\left\{
\begin{matrix}
\displaystyle{1-\frac{4x}{3\pi} + \frac{x^2}{4}
+{\cal O}\left(x^3\right)}
, & x\ll 1\\ \\
\displaystyle{\frac{4}{\pi x} \ln\frac{2x}{e}+
\frac{2}{x^2}
+{\cal O}\left(\frac{1}{x^3}\right)
}, & x\gg 1
\end{matrix}
\right..
\end{split}
\tag{\ref{fDelta}$^\prime$}
\label{fDeltaprime}
\ee

To write down the explicit expression for the zero-temperature gap,
we use the scale of the magnetic field introduced in \req{B0}.
If the magnetic field is weak, ${\cal B} \ll {\cal B}_0$, then  $R_B \gg {\cal R}$, and
the effective interaction is also weak, $g(R_B) \ll 1$. Using 
asymptotics \reqs{Solg} and \rref{fDeltaprime} in  \req{Delta-zeroT}, we find
\begin{subequations}
\label{DeltaB}
\be
\Delta_0(T=0)
\approx 2 {\cal B}
\exp\left(
-\frac{\pi
\left[\ln\frac{{\cal B}_0}{\cal B}+\frac{N\pi}{2}\right]}
{4\ln\left[\ln\frac{{\cal B}_0}{\cal B}+\frac{N\pi}{2}\right]}
\right).
\label{DeltaweakB}
\ee
For the strong magnetic field, ${\cal B} \gg {\cal B}_0$,
the result reads
\be
\Delta_0(T=0)
\approx 2 {\cal B}
\exp\left\{
-2N
\left[1+
\frac{1}{3}\left(
\frac{{\cal B}_0}{\cal B}
\right)^{\frac{8}{\pi^2N}}
\right]
\right\},
\label{DeltastrongB}
\ee
i.e. the dependence slowly approaches the linear function.
\end{subequations}

To complete this subsection, we consider the effect of finite temperature.
As for the usual BCS mean field, 
\req{MF} gives the temperature dependence of the
width of the mean-field gap $\Delta_0(T)$  which
contains the scale $\Delta_0(0)$ only:
\be
\ln \frac{T_{MF}}{T}+
\sum_{n=0}^\infty
\left[
\frac{1}{\sqrt{\left(n+\frac{1}{2}\right)^2+
\frac{\Delta_0(T)^2}{4\pi^2 T^2}
}}-\frac{1}{n+\frac{1}{2}}
\right]=0,
\label{tcDelta}
\ee
where $T_c$ is related to  $\Delta_0(0)$ by the
usual weak coupling BCS relation
\be
T_{MF}=\pi e^{-{\mathbb C}}\Delta_0(0) \approx 1.76\Delta_0(0).
\label{TMF}
\ee
and ${\mathbb C}$ is the Euler constant.

Equations \rref{TMF} and \rref{tcDelta} enable us to estimate the upper bound for the mean field
crossover temperature. For $N=4$ we find
\be
T_{MF}\lesssim 10^{-3}{\cal B},
\label{estimate}
\ee
where ${\cal B}$ is the Zeeman splitting. As the electron $g$-factor in graphene equals to $2$,
we estimate for the parallel magnetic field $B\simeq 40T$, $T_{MF}\simeq 60mK$, which does not seem to be non-realistic.
The other experimental realization could be putting the appropriate insulating ferromagnet on the
top of the graphene film, so that the Zeeman splitting is caused by the corresponding exchange fields.
The effective Zeeman splitting in this case may reach thousands of Kelvin-s\cite{Zaliznyak}.

In the vicinity of the mean-field crossover $T_c-T \ll T_c$, we obtain from \req{tcDelta}
\be
\Delta_0(T)^2=\frac{8\pi^2}{7\zeta(3)}T_c(T-T_c)
\approx 9.38\, T_{MF}(T-T_{MF}),
\label{smallDelta}
\ee
and $\zeta(x)$ is the Riemann $\zeta$-function.

\subsection{Microscopic calculation of the
coefficients in the Free energy and effective action.}
\label{microscopic-coefficients}

Using the symmetry arguments of Sec.~\ref{sec:3} or by the explicit calculation,
one finds that the solution of the form \rref{solMF} is not unique, and, in fact
any order parameter of the form
\be
\hat{\Delta}(\vec{n})=
\hat{\sigma}_z^{AB} \otimes \hat{Q} \left\{\Delta_0(T)
\hat{\cal P}(\vec{n})
+ \Delta_1(T)\left[\openone-\hat{\cal P}(\vec{n})\right]\right\}
\label{solMFQ}
\ee
solves \req{MF}, i.e. the fluctuation effects, studied in Sec.~\ref{sec:4}, 
are important. Here $4\times 4$ matrix $\hat{Q}$ is given by \reqs{Q}.

The free energy and the effective action describes the contribution
of  configurations of the order parameter slowly varying in time and space.
The time and space gradient terms describes the cost of creating an inhomogeneous configuration,
whereas the anisotropy leads  the certain modes to become massive.
To calculate the latter ones it is sufficient to consider the homogeneous and time
independent configurations of the order parameter $\hat{Q}$, see \reqs{Q},
whereas to find the former, one has to expand in small gradients 
of $\hat {Q}$.\footnote{In principle one can perform the gauge transformation 
${\hat Q} \to \hat{U}{\hat Q}\hat{U^\dagger} = \hat{S}_x$,
and expand the fermionic determinant in terms of the non-abelian
scalar, $\hat{U}^\dagger\partial_\tau \hat{U}$, 
and vector$, \hat{U}^\dagger\partial_\mu \hat{U}$, potentials. We have chosen not to do it here
to avoid careful considerations of the terms arising from the non-gauge invariant integration cut-off in
\req{integration}.}

\subsubsection{Stiffness.}

To find the gradient terms we expand the order
parameter as
\be
\begin{split}
Q(\r)=\hat{Q}\left[1-\frac{1}{2}\hat{\delta Q}^2(\r)\right]+\hat{\delta Q}(\r);\\
\hat{\delta Q}(\r)=\int\frac{d^2k}{(2\pi )^2}
e^{i\vec{k}\vec{r}}
\hat{\delta Q}_{\k}.
\end{split}
\label{Qexp}
\ee

 In order to preserve the constraints \rref{qconstraints} up to the second order perturbation
theory in $\delta \hat{Q}$ we require
\be
\begin{split}
&\hat{Q}=\hat{Q}^\dagger;\
\hat{Q}\delta\hat{Q}+\delta\hat{Q}\hat{Q}=0;
\\
&\left(\openone^{KK'}\otimes \hat{\tau}_{z}^s\right)
\delta\hat{Q}
\left(\openone^{KK'}\otimes \hat{\tau}_{z}^s\right)
=-\delta\hat{Q}.
\end{split}
\label{deltaqconstraints}
\ee

\begin{figure}[h]
\includegraphics[width=0.4\textwidth]{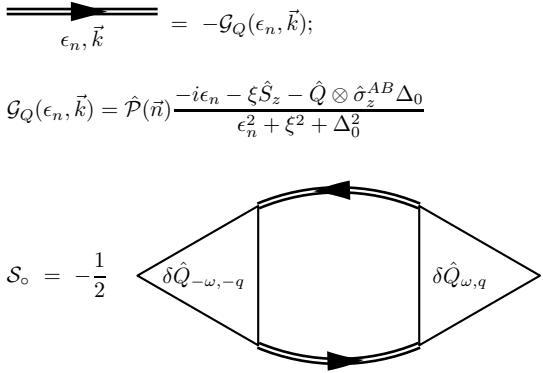}
\vspace*{1mm}
\caption{Microscopic calculation of the
stiffness and the collective modes velocity $v_*$
in the effective action. Notation is defined in \reqs{calG} and \rref{Projection}.
}
\label{fig12}
\end{figure}

Calculating the diagram on Fig.~\ref{fig12},
we obtain for the isotropic part of the Free energy
\be
\begin{split}
&{\mathbb F}_{\circ}=\frac{1}{2}
\int\frac{d^2q}{(2\pi)^2}
T\sum_{\epsilon_m}\int\frac{d^2k}{(2\pi)^2}
\\
&\times
\Big[{\rm Tr}\,\hat{\cal G}_Q(\epsilon_m,\vec{k}+\vec{q})
\hat{\delta Q}_{q}\hat{\cal G}_Q(\epsilon_m,\vec{k})\hat{\delta Q}_{,-q}
\\
&\quad -
{\rm Tr}\,\hat{\cal G}_Q(\epsilon_m,\vec{k})
\hat{\delta Q}_{q}\hat{\cal G}_Q(\epsilon_m,\vec{k})\hat{\delta Q}_{-q}
\Big],
\end{split}
\label{micro-action1}
\ee
where $\epsilon_m$ are the fermionic Matsubara frequencies.
The  term in the last line is local and it comes from the interaction part of
the Hamiltonian. Its explicit calculation is not
necessary because $\delta Q_{\omega=0,k=0}$ corresponds to the
motion along the degenerate manifold which can not produce any
contribution to the Free energy --
this requirement fixes the term unambiguously.

Expanding the Green functions in powers of small  momentum $q$
we find
\[
\begin{split}
&{\mathbb F}_{\circ}=-\frac{1}{4}
\int\frac{d^2q}{(2\pi)^2}
T\sum_{\epsilon_m}\int\frac{d^2k}{(2\pi)^2}
\\
&\times
\Big[q^2\sum_{\mu=x,y}
{\rm Tr}\,\partial_{k_\mu}\hat{\cal G}_Q(\epsilon_m,{\vec{k}})
\hat{\delta Q}_{q}\partial_{k_\mu}
\hat{\cal G}_Q(\epsilon_m,\vec{k})\hat{\delta Q}_{-q}
\Big],
\end{split}
\]
Then, the simple power counting shows that the integral is contributed by
$\epsilon,vk \ll {\cal B}$ and therefore the integration
rule \rref{integration} may be used. Finally, using the explicit expression
for the Green function from Fig.~\ref{fig12} and  \req{deltaqconstraints},
we obtain after simple algebra
\be
\begin{split}
&{\mathbb F}_{\circ}=\frac{\rho_K(T)}{8}
{\rm Tr}\,\int d^2r
 \left(\nnabla \hat{Q}\right)^2;
\\
& \rho_K(T)=\frac{\cal B}{4\pi}f_\rho\left(\frac{\Delta_0}{2\pi T}\right),
\end{split}
\label{microaction2}
\ee 
where
\be
\begin{split}
f(x)&=\sum_{n=0}^\infty\frac{x^2}{\left((n+1/2)^2+x^2\right)^{3/2}}
\\
&\approx 
\left\{
\begin{matrix}
1- {\pi}{x}e^{-2\pi x} + {\cal O}(e^{-4\pi x}), & x\gg 1\\
7\zeta(3)x^2-\frac{93\zeta(5)}{2}x^4 + {\cal O}(x^6), & x \ll 1,
\end{matrix}
\right.
\end{split}
\label{frho}
\ee
and $\zeta(x)$ is the Riemann $\zeta$-function.

As we have already explained, keeping the imaginary time derivative terms is valid only for
$\Delta(T) \gg T$. At such low temperatures we find
\be
\rho_K=\frac{\cal B}{4\pi}
\ee 
independently of any logarithmic renormalization from higher energies.

In the opposite case, $\Delta(T)\ll T$ only constant in time fluctuations
are important and we obtain ${\cal S}_{\circ}={\mathbb F}_\circ/T$, where ${\mathbb F}_\circ$ is given
by \req{4.2a} and the stiffnesses on the mean-field correlation length, $\xi_{MF}$ see  \req{xiMF},
are given by 
\be
\rho_K(\xi_{MF})=\rho_s(\xi_{MF})=\frac{\cal B}{2\pi}\left(\frac{T_{MF}-T}{T_{MF}}\right).
\label{microMF}
\ee
where we used \req{smallDelta}.

\subsubsection{Leading anisotropies}

\begin{figure}[h]
\hspace*{-6mm}\includegraphics[width=0.49\textwidth]{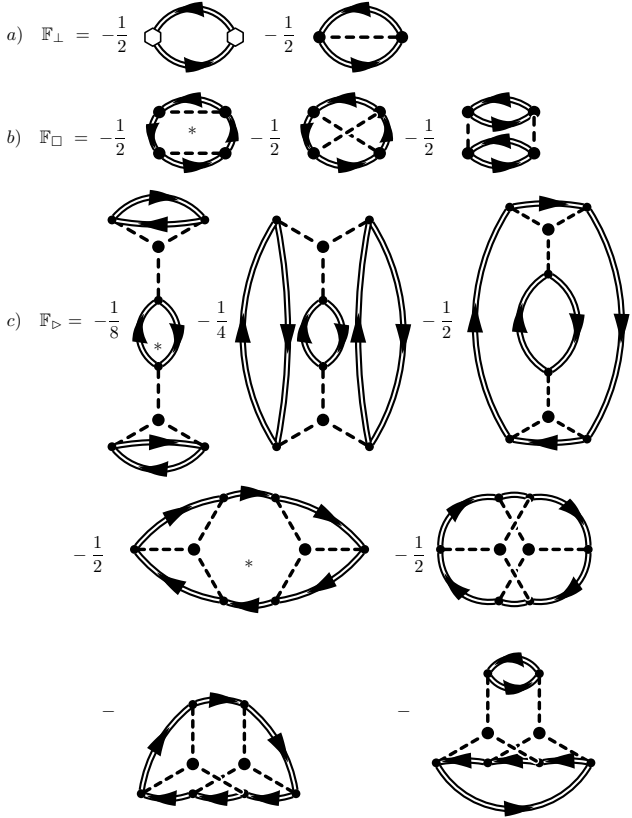}
\caption{Microscopic calculation of the
anisotropic terms in the free energy. Diagrams
containing largest power of $\ln({\cal B}/\Delta_0)$
are marked by star.
}
\label{fig13}
\end{figure}

The quadratic anisotropies \rref{4.2b} arises both due to the warping
of the single electron spectrum \rref{warping} and the short-range
interactions \rref{sr}.
The diagrammatic calculation of these anisotropies
is straightforward and it is shown on Fig.~\ref{fig13} a).

\begin{subequations}
\label{Fperp12}
We find ${\mathbb F}_\perp={\mathbb F}_\perp^{(1)}+{\mathbb F}_\perp^{(2)}$ where the
first contribution is due to the warping
\be
{\mathbb F}_\perp^{(1)} =-{\rm Tr}\,
\hat{Q}\hat{\Lambda}_z\hat{Q}\hat{\Lambda}_z
\frac{\lambda_w^2(R_B) {\cal B}^3}{16\pi v^2(R_B)}f_\rho\left(\frac{\Delta_0}{2\pi T}\right),
\label{Fperp1}
\ee
where $f_\rho$ is given by \req{frho}.

The second diagram of Fig.~\ref{fig13} b) describes the effect of the short range interaction \rref{sr}.
We  take into account only the most relevant terms as specified in \req{srRGsolution:b}.
Term proportional to $J_-^\Sigma$ do not contain the inter-valley $\Lambda$-matrices and, thus,
do not cause the anisotropy. We find
\be
\begin{split}
&{\mathbb F}_\perp^{(2)}\! =
-\left[F^z_-(R_B){\rm Tr}
\left(\hat{Q}\hat{\Lambda}_z\right)^2\!\!
+ F_-^\perp(R_B)\!\sum_{\mu=x,y}\!\!\!{\rm Tr}
\left(\hat{Q}\hat{\Lambda}_\mu\right)^2
\right]
\\
&\ \ \times R_B v(R_B)\left[
T\sum_n\int\frac{{\cal B}d\xi}{2\pi v^2(R_B)}\frac{\Delta_0(T)}{\epsilon_n^2+\xi^2+\Delta_0(T)^2}
\right]^2
.
\end{split}
\label{Fperp2}
\ee
The logarithmic integral in \req{Fperp2} is eliminated using \req{MFDelta}, and
we obtain \req{4.2b} at the scale $r_0=\xi_{MF}$, see \req{xiMF}, with the couplings
\end{subequations}
\begin{subequations}
\label{microscopic-eta}
\begin{align}
&
\eta_z=-{\cal B}
\Bigg\{\frac{\lambda_w^2(R_B) {\cal B}^2}{2\pi \Delta_0(0)^2}
\left[\frac{\Delta_0^2(0)}{\Delta_0^2(T)}
f_\rho\left(\frac{\Delta_0(T)}{2\pi T}\right)\right]
\label{microscopic-eta-a}\\
&\quad +2 F_-^z (R_B)\left(\frac{2N}{\pi f_\Delta\left(\frac{\pi}{4g(R_B)}\right)}\right)^2\Bigg\};
\nonumber
\\
&\eta_\perp=-2 {\cal B} F_-^\perp (R_B)\left(\frac{2N}{\pi f_\Delta\left(\frac{\pi}{4g(R_B)}\right)}\right)^2
\label{microscopic-eta-b}
\end{align}
where $f_\Delta (x)$ is defined in \req{fDelta}.
\end{subequations}

Equations \rref{microscopic-eta} give complete expressions for the
anisotropies in the free energy \rref{4.2b} in terms of the microscopic coupling constants
defined on the scale $R_B$. Equations \rref{Solwarping},~\rref{srRGsolution},
\rref{FestimateRG},
~\rref{Solg},  
\rref{Rbsol1},  \rref{Rbsol2}, \rref{fDelta},
and \rref{DeltaB}
enable us to find the magnetic field dependence of the anisotropy constants.
In the case of the strong magnetic field ${\cal B} \gtrsim {\cal B}_0$, see \req{B0},
it yields
\begin{subequations}
\label{asymptotics}
\begin{align}
& \eta_z \approx -{\cal B}
\Bigg\{\lambda_w^2({\cal R})
\left(\frac{68\, {\cal B}}{{\cal B}_0}\right)^{3.9}
\left[\frac{\Delta_0^2(0)}{\Delta_0^2(T)}
f_\rho\left(\frac{\Delta_0(T)}{2\pi T}\right)\right]
\nonumber\\
&
\qquad\qquad + 12.9 \,F_-^z ({\cal R})\Big\};
\nonumber\\
&\eta_\perp \approx - 12.9 {\cal B}F_-^\perp ({\cal R}).
\label{etaMagneticfieldstrong}
\end{align}
For the weak magnetic field, ${\cal B} \lesssim {\cal B}_0$, we find
\begin{align}
 \eta_z& \approx -{\cal B}
\Bigg\{\lambda_w^2({\cal R}) \left(\frac{{\cal B}}{5.1 {\cal B}_0}\right)^2
\exp\left(
\frac{\pi
\left[\ln\frac{{\cal B}_0}{\cal B}+6.28\right]}
{2\ln\left[\ln\frac{{\cal B}_0}{\cal B}+6.28\right]}
\right)
\nonumber\\
&\times
\left[\frac{\Delta_0^2(0)}{\Delta_0^2(T)}f_\rho\left(\frac{\Delta_0(T)}{2\pi T}\right)\right]
\left[
0.15 \ln \left(\frac{{\cal B}_0}{\cal B}\right)+1\right]^{-13/4}
\nonumber\\
& \quad + 12.9 \,F_-^z ({\cal R}) \left(\frac{{\cal B}}{{\cal B}_0}\right)\left[
0.15 \ln \left(\frac{{\cal B}_0}{\cal B}\right)+1\right]^4\Bigg\} ;
\nonumber\\
&\eta_\perp \approx - 12.9 {\cal B}\, F_-^\perp ({\cal R})
 \left(\frac{{\cal B}}{{\cal B}_0}\right)\left[
0.15 \ln \left(\frac{{\cal B}_0}{\cal B}\right)+1\right]^4
.\label{etaMagneticfieldweak}
\end{align}
\label{etaMagneticfield}
\end{subequations}

Comparison of the anisotropies from \reqs{etaMagneticfield} with the expressions for stiffness \rref{microMF}
together with the estimates $F_-^{z,\perp} ({\cal R}) \simeq 1$ indicates that
the weak anisotropy case is possible only for ${\cal B} \ll {\cal B}_0$ and
at the strong magnetic field the anisotropy is dominating already
at the mean-field correlation length, see Sec.~\ref{sec:anisotropies-strong}.

The very peculiar situation arises if $F_-^{z} ({\cal R}) < 0$,
as we explained before, the value and the sign of this constant is determined by the details at
the distances of the order of the lattice constant.
In this case the exchange and the warping produce the contribution of the
different sign and intersection of the Heisenberg line, $\eta_z > 0, \
  |\eta_\perp|= |\eta_z|$, see Sec.~\rref{Heisenberg} becomes possible.
Using \req{etaMagneticfieldstrong}, and estimates \rref{lambdaR} and \rref{FestimateRG},
we find the estimate for such field ${\cal B}_H$
\be
{\cal B}_H\simeq 10^{-2}{\cal B}_0\left(\frac{|
F^z_-({\cal R})+|F^\perp_-({\cal R})|
|}{\lambda_w^2\left({\cal R}\right)}\right)^{0.26}
\!\!\!\!\approx
1\div 10 {\cal B}_0 \gtrsim {\cal B}_0.
\label{BHestimate}
\ee
The manifestation of this line on the phase diagram will be considered in Sec.~\ref{sec:log4}.

Calculation of the other anisotropies is self-explanatory from diagrams Fig.~\ref{fig13} b-c).
Though, formally, they are of the same order in perturbation theory, they still can be
classified in powers of $\ln({\cal B}/\Delta) \ll 1$. Taking into account
only the leading logarithmic contributions, and reexpressing the logarithmic
expression using the mean-field equation \rref{MFDelta}, we find for anisotropy coefficients
in \reqs{4.2c} -- \rref{4.2d}
\begin{subequations}
\begin{align}
&\kappa(\xi_{MF})=
\frac{\cal B}{2\pi}
\left(\frac{N \left[F_\perp^- ({\cal R})-3J_\perp^\Lambda ({\cal R})
\right]}{\pi f_\Delta\left(\frac{\pi}{4g(R_B)}\right)}\right)^2f_\rho\left(\frac{\Delta_0(T)}{2\pi T}\right);
\label{micro-kappa}
\\
&\zeta(\xi_{MF})=
\frac{2\cal B}{\pi}\left[{\cal F}_-({\cal R})\right]^2
D\left[\frac{N }
{\pi f_\Delta\left(\frac{\pi}{4g(R_B)}\right)}\right]\nonumber\\
&\qquad \times
\left(\frac{\Delta_0(T)}{\Delta_0(0)}\right)^2
f_\rho\left(\frac{\Delta_0(T)}{2\pi T}\right), 
\label{micro-zeta}
\end{align}
where $D(x)=xe^{-x}$, the functions $f_{\rho,\Delta}$ are defined in \reqs{frho} and \rref{fDelta}.
Deriving \req{micro-kappa}, we took into account that $\kappa$ is important only
if $\eta_\perp \to 0$, see Sec.~\ref{etaperpsmall}. Thus, the irrelevant constants,
see \reqs{srRGsolution:a} had to be taken into account. 
\label{micro-zetakappa}
\end{subequations}

Expressions \rref{micro-zetakappa} together with \reqs{srRGsolution:a} and \rref{umklappSolution}
shows that those anisotropies are very small, so we will not analyse their asymptotics further.

\subsection{Phase diagram in ${\cal B}$-$T$ plane.}
\label{sec:log4}

This subsection combines the symmetry analysis of Sec.~\ref{sec:4} with the
microscopic calculation of the Free energy couplings in Sec.~\ref{microscopic-coefficients}.
As the result, we will construct the phase diagram in the plane determined
by the Zeeman splitting ${\cal B}$ and by the system temperature, $T$.
As we have discussed in Sec.~\ref{sec:4}, the interesting phase transitions
are determined by the thermal fluctuations and it is convenient to introduce
dimensionless Ginzburg-Levanyuk parameter
\be
Gi\equiv \frac{4T_{MF}}{\cal B} \ll 1,
\label{Gi}
\ee
characterising strength of such fluctuations. Here the 
mean-field transition temperature is given by \reqs{TMF} and \rref{Delta-zeroT} -- \rref{DeltaB}.

Apparently, not all the regions of the phase diagram of Fig.~\ref{fig62} can
be explored by varying ${\cal B},T$ and we will
restrict ourselves with two most realistic, as we believe, cases.
Namely, we assume that the absolute values of the interaction
constants $|F_-^{z,\perp}(a)|\simeq 1$. For the sake of concreteness,
we assume $F_-^{\perp}(a) > 0$. [Case of $F_-^{\perp}(a) < 0$ is obtained
by the replacement $N_1\leftrightarrow N_4$.]

\begin{figure}[ht]
\includegraphics[width=0.42\textwidth]{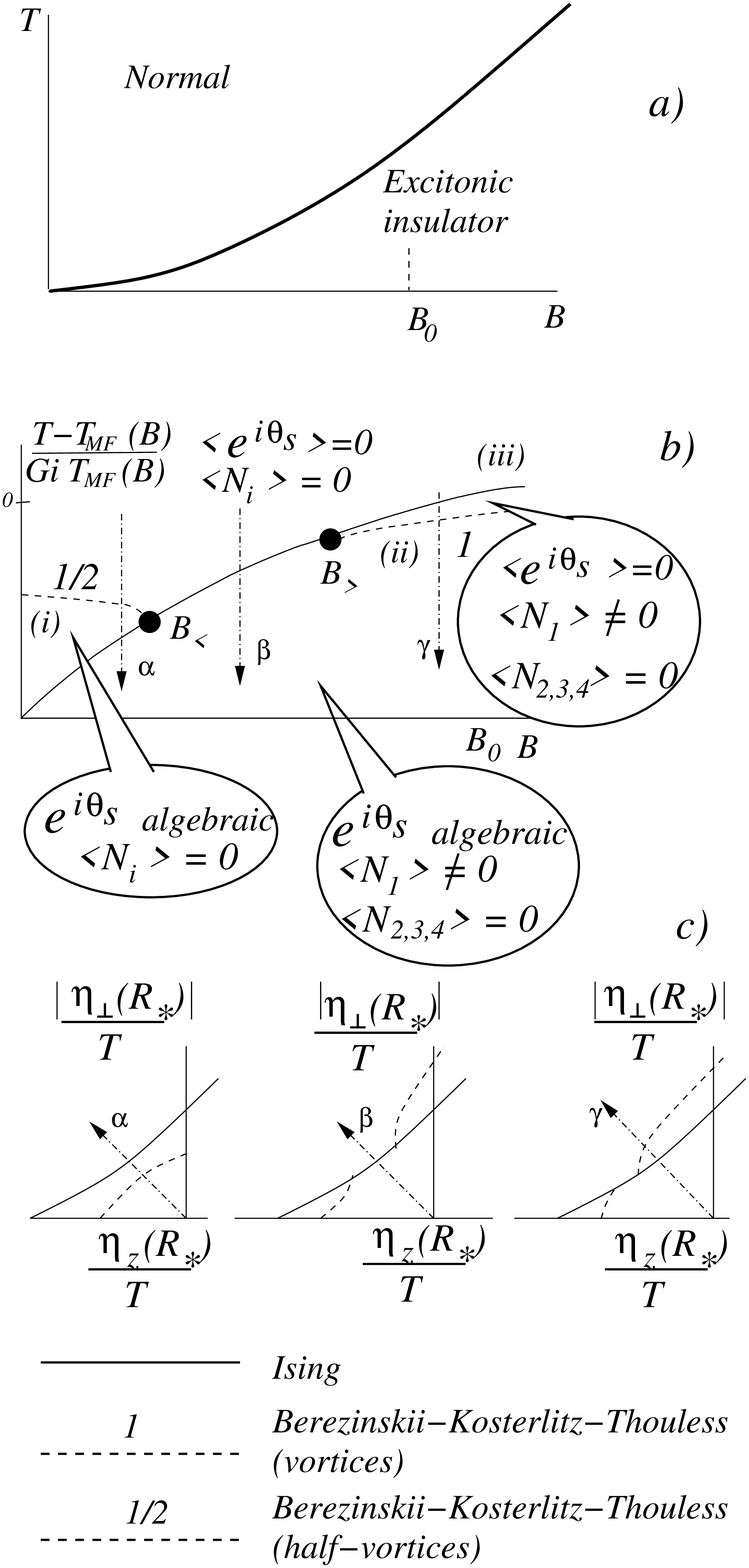}
\\
\vspace*{1cm}
\caption{Phase diagram of the graphene in the parallel magnetic field
for the short range interaction constant $F^z_->0$. a) 
Mean field structure of the phase diagram; b) The ``fine'' structure of the phase diagram
in the close vicinity of the mean-field transition temperature, $T_{MF}({\cal B})$;
c) Relation of the phase diagram (b) to the more phenomenological phase diagram of Fig.~\ref{fig62}.}
\label{fig141}
\end{figure}

Let us consider first the case of $F_-^{z} <0$.
Then, according to \reqs{microscopic-eta}, one finds $\eta_{z,\perp} < 0$.
Mean-field diagram obtained from  Fig.~\ref{fig4} is trivial
and includes the continuous transition from the disordered
normal state, to the spin-flux state, Fig.~\ref{fig3} (d), of the excitonic
insulator, see Fig.~\ref{fig141}.

The fine structure of the phase diagram, Fig.~\ref{fig141} (b), 
is obtained from the general Fig.~\ref{fig62}, by using
the microscopic expression for the Free energy couplings
derived in Sec.~\ref{microscopic-coefficients}.
The topological structure of the phase diagram
is most easily obtained by the mapping of the paths
in ${\cal B},T$ plane to the path in $(\eta_z,\eta_\perp)$ plane as shown
in Fig.~\ref{fig141} (c).

The positions of the transitions lines
on the phase diagrams are obtained by combining the
phenomenological results of Sec.~\ref{sec:4} and 
the microscopic analysis of Sec.~\ref{microscopic-coefficients}.

For instance, using \reqs{Kjump-both} and \rref{microMF},
we obtain in the limit of the small vortex fugacity
\begin{subequations}
\be
\frac{T_{MF}-T}{T_{MF}Gi}\approx 
\left\{
\begin{matrix}
4; & {\rm line\ (i)};
\\
1; & {\rm line\ (ii)}.
\end{matrix}
\right.
\label{KTposition}
\ee
Analogously, using \reqs{notweak}, \rref{rhoK} and \rref{etaMagneticfieldweak}
we obtain a position of the Ising line.
With the logarithmic accuracy, we find
for line (iii) of Fig.~\ref{fig141} (b):
\be
\frac{T_{MF}-T}{T_{MF}Gi}\approx 
\left\{
\begin{matrix}
\frac{1}{4}\ln\frac{{\cal B}_0 Gi}{\cal B}; &  
{\cal B}_0\exp\left(-1/Gi\right) \lesssim {\cal B} \ll {\cal B}_0 Gi
\\
\\
\to 0; & {\cal B} \gtrsim {\cal B}_0 Gi.
\end{matrix}
\right.
\ee
\end{subequations}

Case of $F_-^z<0$ is more sophisticated. As we noticed in Sec.~\ref{microscopic-coefficients},
coefficient $\eta_z$ changes its sign as the function of the magnetic field and
at some point crosses the Heisenberg line at field ${\cal B}_H$. At the mean-field level,
it corresponds to the continuous transition between two-kinds of excitonic insulator:
spin flux state, see Fig.~\ref{fig3} (d), and the link centered spin density wave, see
Fig.~\ref{fig3} (b,c).

\begin{figure}[ht]
\includegraphics[width=0.42\textwidth]{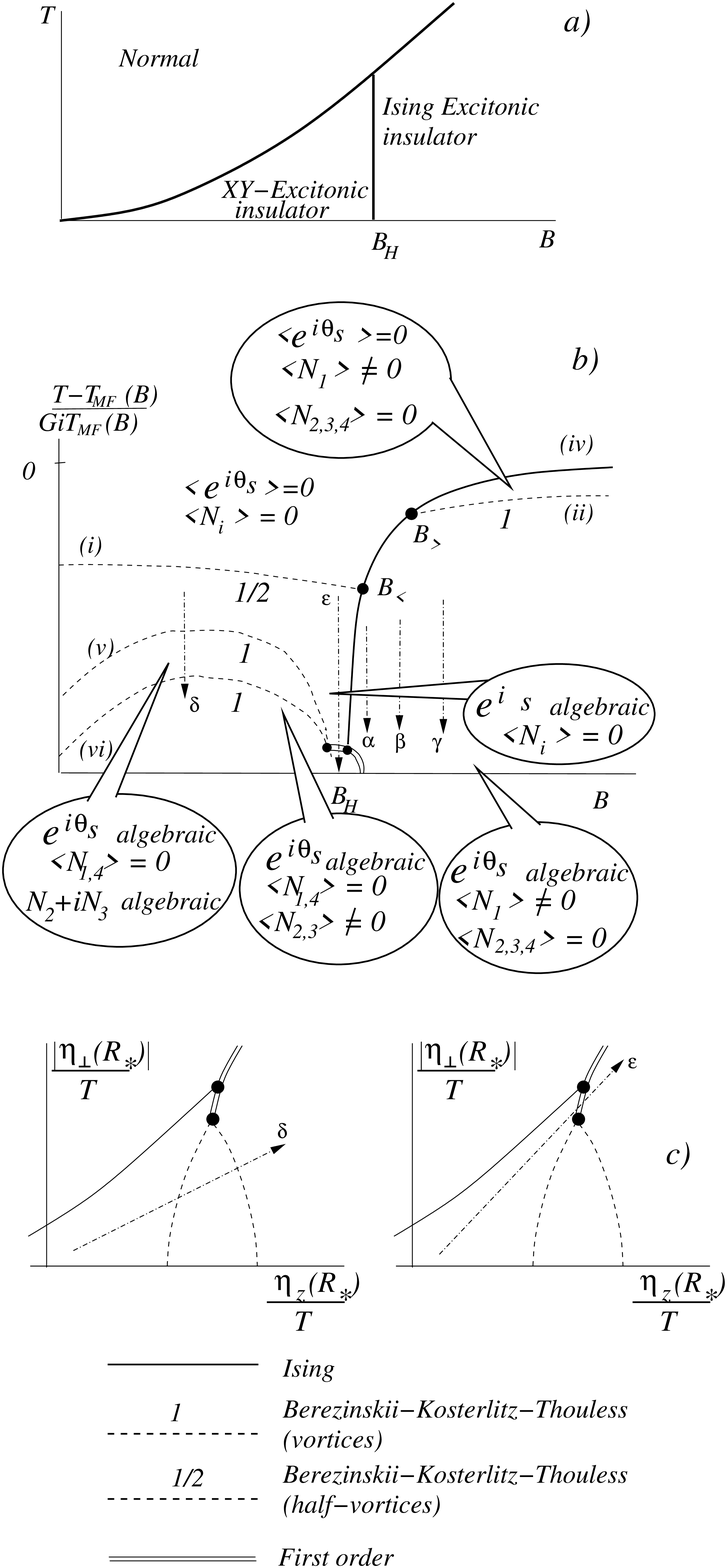}

\vspace*{0.3cm}
\caption{Phase diagram of the graphene in the parallel magnetic field
for the short range interaction constant $F^z_-<0$. a) 
Mean field structure of the phase diagram; b) The ``fine'' structure of the phase diagram
in the close vicinity of the mean-field transition temperature, $T_{MF}({\cal B})$;
c) Relation of the phase diagram (b) to the more general phenomenological phase diagram of Fig.~\ref{fig62}.
Cross-section denoted by $\alpha,\beta,\gamma$ are shown on Fig.~\ref{fig141} c).}
\label{fig142}
\end{figure}

Similarly to the previous case, 
the fine structure of the phase diagram, Fig.~\ref{fig142} (b), 
is obtained from the general Fig.~\ref{fig62}, by the mapping of the paths
in ${\cal B},T$ plane to the path in $(\eta_z,\eta_\perp)$ plane as shown
in Fig.~\ref{fig142} (c).

The Berezinskii-Kosterlitz-Thouless transition lines (i), (ii) on Fig.~\ref{fig142} (b)
are still determined by the expressions \rref{KTposition}.
Ising line (iv)   Berezinskii-Kosterlitz-Thouless
line and the are found from \reqs{positions} and \rref{microMF}.
Expanding 
\[
\Delta\eta \approx \left(\frac{B-B_H}{B_H}\right) \eta_\perp
\]
we obtain with the logarithmic accuracy
for $ {\cal B}_H\exp(-Gi) \ll |{\cal B}-{\cal B}_H|\ll {\cal B}_H$:
\begin{subequations}
\be
\frac{T_{MF}-T}{T_{MF}Gi}\approx 
\left\{
\begin{matrix}
\displaystyle{\frac{1}{8}
\ln\frac{{\cal B}_H}{{\cal B}-{\cal B}_H }
;} &  
{\rm line\ (iv)};
\\
\\
\displaystyle{\frac{1}{8}\ln\frac{{\cal B}_H}{{\cal B}_H-{\cal B}};} &
{\rm line\ (v)}.
\end{matrix}
\right.
\ee
For larger ${\cal B}$ the Ising line (iv) approaches the mean-field temperature.
The Berezinskii-Kosterlitz-Thouless transitions lines (v), for ${\cal B} \ll {\cal B}_H$ can be found using
 \reqs{notweak}, \rref{rhoK}, and \rref{etaMagneticfieldweak}.
For the fields ${\cal B}_0\exp\left(-1/Gi\right) \lesssim {\cal B} \ll {\cal B}_0 Gi$
this yields
\be
\frac{T_{MF}-T}{T_{MF}Gi}\approx 
\frac{1}{4}\ln\frac{{\cal B}_0 Gi}{\cal B}; \quad 
{\rm line (v)};
\ee
The Berezinskii-Kosterlitz-Thouless line (vi) turns
out to lie outside the fluctuation region due to the large numerical factor 
in \req{alphaperp} and smallness of $\zeta$ in \req{micro-zeta}. We will not write-down its asymptotic behaviour. 
\end{subequations}

This completes our analysis of the structure of the phase diagram
of graphene in the parallel magnetic field, characterised by the Zeeman splitting
${\cal B}$.

\section{Summary and conclusions}

   In this paper we have discussed two problems concerning clean graphene: (i) possible  effects of in-plane magnetic field in facilitating a formation of excitonic condensate and (ii) a  role of the long range Coulomb interaction and its influence on other interactions in the system. The second topic is more general than the first and has a broader significance, though, as far the paper goes, it was discussed in the second part. 

 In zero magnetic field graphene is a gapless semiconductor with two
 Fermi points in the Brillouin zone (valleys). In-plane magnetic field
 pushes up- and down-spin bands in opposite directions transforming
 the system into a metal with extended Fermi surfaces for  electrons
 and holes of opposite spin.  There are two such Fermi surfaces
 corresponding to two possible valley indices. Electrons and holes
 attract through the Coulomb interaction which creates a possibility
 of exciton condensation along the lines first described by Keldysh
 and Kopaev \cite{KeldyshKopaev}. The maximal possible symmetry of the
 order parameter is U(2), lattice effects bring it down to U(1). The
 system in its low temperature phase is an insulator with a gapless
 collective mode corresponding to fluctuations of spin density in the
 directions transverse to the applied magnetic field. The above is a
 brief summary of the discussion of Sections II-IV. Section IV also
 contains a detailed phase diagram. The discussion in these sections
 dealt with the Landau-Ginzburg 
free energy functional written purely on symmetry grounds where  various energy scales of the system enter as parameters. 

 Section V contains a microscopic analysis tailored especially for
 graphene. The ultimate goal was to obtain estimates for the critical
 temperature and various parameters of the phase diagram, but a
 byproduct of the analysis is a study of   how the strongest
 interaction in graphene - the Coulomb interaction affects the
 spectrum  and renormalizes other interactions (such as the short
 range exchange). As is well known, the Coulomb interaction in
 graphene, measured by its dimensionless value $g = \pi e^2/2v$ is
 quite strong at energies of the order of the bandwidth. We have
 found, however, that the effective coupling steadily diminishes at
 low energies and asymptotically vanishes at $E=0$. The scale
 dependence of $g$ is rather slow and is given by
 Eq.(\ref{Solg}). This formula includes an important scale ${\cal R}$,
 which we estimate for graphene as being of order of $10^2-10^3$
 lattice constants. This scale separates the region of relatively
 strong interaction where $g(r)$ decreases as a power law, from the
 region of weak coupling where $g(r) \sim 
[\ln r]^{-1}$. It also sets the scale ${\cal B}_0$ for the magnetic field (\ref{B0})[our estimate is ${\cal B}_0 \approx 10-100$K].  

The renormalization process is drastically altered at energies of order of the applied magnetic field ${\cal B}$. The field sets the ultraviolet cut-off for the physics of excitonic insulator. However, the upper cut-off for its collective excitations is much lower and is set by the value of the mean field gap $\Delta_0$. One may anticipate that the latter energy scale is exponentially small in comparison with the cut-off ${\cal B}$. This is indeed the case, but fortunately the inverse coupling constant $1/g(r \sim {\cal B}^{-1})$, which stays in the exponent, depends rather weakly on the magnetic field [see (\ref{DeltaweakB},\ref{DeltastrongB})] so that the magnitude of $\Delta_0$ is not that small. Our estimate is that in fields of the order of or stronger than 10T the mean field temperature is $T_{MF} \leq 10^{-3}{\cal B}$. This makes it possible to observe the excitonic effects described in this paper in the temperature range of tens of mK (see more discussion in Section V C). 

 Though the long range Coulomb interaction is certainly the main
 player, its renormalization drags with itself weaker interactions,
 such as the short range exchange, strengthening them at low
 energies. Such interactions break the U(2) symmetry present at low
 energies when only the long range Coulomb interaction is taken into
 account. The analysis of Section V C.2 demonstrates that the U(2)
 symmetry survives in the excitonic insulator only at ${\cal B} <<
 {\cal B}_0$ when the estimated transition temperatures are probably
 too low for the effect to be observed. In the realistic region ${\cal
   B} \geq {\cal B}_0$ the anisotropy is strong. The expected phase
 diagrams in ${\cal B}-T$ plane are given on
 Figs.~\ref{fig141},\ref{fig142} [they differ by a sign of a certain
 interaction parameter which on the current stage remains
 unknown]. Strong fields of order of 10T or more, 
which are required to make the excitonic insulator observable at
 temperatures above tens of mK, 
will probably put one in the regime marked by $\beta$ or $\gamma$ on
 the phase diagram Figs.~\ref{fig141}, \ref{fig142}. The spin
 configuration corresponding to this regime is depicted on 
Fig. \ref{fig3} d) and corresponds to the spin-flux phase.

\acknowledgments
AMT was  supported  by 
the DOE under contract number DE-AC02 -98 CH 10886.  We acknowledge
inspirational conversations with I. Zaliznyak, 
and interesting discussions with L. Levitov, D. Khveshchenko, and M. Foster.

\appendix

\section{Derivation of \req{dual}.}
\label{appendixA}

Let us re-write \req{4.2aprime} in a form
\be
{\mathbb F}_\circ
= 
\frac{\rho_K}{4}
{\mathrm Tr}\hat{j}_\mu^2+
\frac{\rho_s-\rho_K}{8}
\left[{\mathrm Tr}\hat{j}_\mu\right]^2,
\label{apa1}
\ee
where
\be
\hat{j}_\mu\equiv -i \hat{V}^\dagger\partial_\mu \hat{V}. 
\label{apa2}
\ee
The topological defects \rref{1v} and \rref{halfv}
are determined by the condition
\be
\frac{1}{2\pi}\oint dx_\mu {\mathrm Tr}\hat{j}_\mu=\pm 1;  \pm \frac{1}{2},
\label{apa3}
\ee
i.e. 
\be
\theta= \frac{-i}{2} {\mathrm Tr} \ln \hat{V}
\label{apa32} 
\ee
must be multi-valued function of the coordinate.

\begin{figure}
\includegraphics[width=0.3\textwidth]{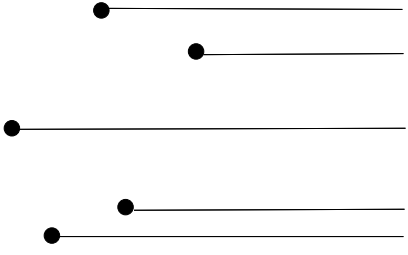}
\caption{Cuts on the $x-y$ plane attached to each topological defect
in the five-vortex configuration.}
\label{figapA}
\end{figure}

In order to avoid the consideration of the multi-valued field
we introduce cuts parallel to, say, $x$ axis 
connecting each vortex or half-vortex with the boundary of the system
[any physical quantity, obviously, does not depend on the choice of
the cut],
see Fig.~\ref{figapA},
and consider all the matrices to be single valued function but
the phases experiencing the discontinuity on the cuts.

The current \rref{apa2} should be continuous thus we modify
the definition as
\be
\begin{split}
&\hat{j}_x = -i \hat{V}^\dagger\partial_x\hat{V};
\\
&\hat{j}_y = -i \hat{V}^\dagger\partial_y\hat{V}
+\pi
\sum_{j=1}^{{\cal N}^{(1)}}l_j^{(1)}\delta(y-y_j^{(1)}){\mathrm sgn}(x-x_j^{(1)})
\\
&
+\pi
\sum_{j=1}^{{\cal N}^{(1/2)}}l_j^{(1/2)}
\frac{1+\mm_j\cdot{\boldsymbol \sigma}}{2}
\delta(y-y_j^{(1/2)}){\mathrm sgn}(x-x_j^{(1/2)})
; 
\end{split}
\label{apa4}
\ee
where ${\cal N}^{(1)}$ and ${\cal N}^{(1/2)}$ are the number
of the vortices and the vertices respectively, $x_j,y_j$ are
their coordinates, $l_j=\pm 1$ is the corresponding vorticity
and $\mm_j$ is the unit vector characterising the spin of the
$j$th half-vortex.

Writing the summation over the vortex and half-vortex coordinates
explicitly, we obtain 
\begin{widetext}
\be
\begin{split}
{\cal Z}&\propto \sum_{{\cal N}^{(1)}=0}^\infty 
\frac{
\mu_{1}^{{\cal N}^{(1)}}
}
{{\cal N}^{(1)}!}
\prod_{j=1}^{{\cal N}^{(1)}}
\sum_{l_j^{(1)}=\pm 1}
\int \frac{dx_j^{(1)}dy_j^{(1)}}{r_0^2}
 \sum_{{\cal N}^{(1/2)}=0}^\infty 
\frac{
\mu_{1/2}^{{\cal N}^{(1/2)}}
}
{{\cal N}^{(1/2)}!}\prod_{j=1}^{{\cal N}^{(1/2)}}
\sum_{l_j^{(1/2)}=\pm 1}
\int \frac{dx_j^{(1/2)}dy_j^{(1/2)} d\mm_j}{4\pi r_0^2}
 \int {\cal D}{\hat V} 
\\
&\times
\exp
\left\{-\int dxdy
\left[
\frac{\rho_K}{4T}
{\mathrm Tr}\hat{j}_\mu^2+
\frac{\rho_s-\rho_K}{8T}
\left({\mathrm Tr}\hat{j}_\mu\right)^2
\right]
\right\},
\end{split}
\label{apa5}
\ee
where the matrix current $\hat{j}_\mu$ is given by \req{apa4}.

After introducing the dual $2\times 2$ matrix field
$\hat{h}=\hat{h}^\dagger$, \req{apa5} acquires the form
\be
\begin{split}
{\cal Z}&\propto \sum_{{\cal N}^{(1)}=0}^\infty 
\frac{
\mu_{1}^{{\cal N}^{(1)}}
}
{{\cal N}^{(1)}!}
\prod_{j=1}^{{\cal N}^{(1)}}
\sum_{l_j^{(1)}=\pm 1}
\int \frac{dx_j^{(1)}dy_j^{(1)}}{r_0^2}
 \sum_{{\cal N}^{(1/2)}=0}^\infty 
\frac{
\mu_{1/2}^{{\cal N}^{(1/2)}}
}
{{\cal N}^{(1/2)}!}\prod_{j=1}^{{\cal N}^{(1/2)}}
\sum_{l_j^{(1/2)}=\pm 1}
\int \frac{dx_j^{(1/2)}dy_j^{(1/2)} d\mm_j}{4\pi r_0^2}
 \int {\cal D}{\hat V}{\cal D}{\hat h} 
\\
&\times
\exp\left\{
-\int dxdy
\left[
 \frac{\rho_K}{4T}{\mathrm Tr}\hat{j}_x^2+
 \frac{T}{4\rho_K} {\mathrm Tr}\left(\partial_x\hat{h}\right)^2+
\frac{\rho_s-\rho_K}{8T}
 \left({\mathrm Tr}\hat{j}_x  \right)^2+
\left(\frac{T}{8\rho_s}-\frac{T}{8\rho_K}\right)
 \left({\mathrm Tr}\partial_x\hat{h} 
 \right)^2
+i {\mathrm Tr}\hat{j}_y\partial_x\hat{h} 
\right]
\right\},
\end{split}
\label{apa6}
\ee
Substituting \req{apa4} into \req{apa6}, integrating the
terms with $\delta$ functions by parts and summing over $l_j$, we find
\be
\begin{split}
{\cal Z}\propto
 \int {\cal D}{\hat V}{\cal D}{\hat h} 
& 
\sum_{{\cal N}^{(1)}=0}^\infty 
\frac{
\mu_{1}^{{\cal N}^{(1)}}
}
{{\cal N}^{(1)}!}\prod_{j=1}^{{\cal N}^{(1)}}
\int \frac{dx_j^{(1)}dy_j^{(1)}}{r_0^2}
2\cos \pi{\mathrm Tr}\hat{h}(r_j^{(1)})
\\
&\times
 \sum_{{\cal N}^{(1/2)}=0}^\infty 
\frac{
\mu_{1/2}^{{\cal N}^{(1/2)}}}
{{\cal N}^{(1/2)}!}\prod_{j=1}^{{\cal N}^{(1/2)}}
\int \frac{dx_j^{(1/2)}dy_j^{(1/2)d\mm_j} }{ 4\pi r_0^2}
2 \cos \pi \left[{\mathrm Tr}\hat{h}(r_j^{(1/2)})
\frac{1+\mm_j\cdot\vec{\sigma}}{2}
\right]
\\
&\times
\exp\Bigg\{
-\int dxdy
\Big[
 \frac{\rho_K}{4T}{\mathrm Tr}\partial_x\hat{V}^\dagger\partial_x\hat{V}+
 \frac{T}{4\rho_K} {\mathrm Tr}\left(\partial_x\hat{h}\right)^2+
\frac{\rho_s-\rho_K}{8T}
 \left( -i{\mathrm Tr}
\hat{V}^\dagger\partial_x\hat{V} \right)^2
\\
&
\qquad\qquad\qquad
+
\left(\frac{T}{8\rho_s}-\frac{T}{8\rho_K}\right)
 \left({\mathrm Tr}\partial_x\hat{h} 
 \right)^2
+ {\mathrm Tr}\hat{V}^\dagger\partial_y\hat{V}\partial_x\hat{h} 
\Big]
\Bigg\}.
\end{split}
\label{apa7}
\ee
\end{widetext}
After the summation over ${\cal N}^{(1/2)}$, ${\cal N}^{(1)}$,
and integration over $\mm_j$, we obtain \req{dual}.

\section{Analysis of the Ising phase transition.}
\label{AppendixB}

In the vicinity of $T=\pi\rho_K$, where the mutually dual cosines have
the same scaling dimension 1, partition \rref{N1N2phi} can be
mapped to the quantum many-body problem at zero temperature and then
refermionized. The reader can consult Ref.~\onlinecite{book} where the necessary
information about 2D Ising model is provided.
Choosing $y$ coordinate for imaginary time, we 
re-write the classical \req{N1N2phi}
\be
\begin{split}
 &{\cal Z}\propto 
 {\rm Tr}\exp\left(-L_y\int d x \hat{\cal H}\right)
 \\
 &\hat{\cal H}
  = 
 \frac{\rho_K}{2T}\left(\partial_{x}  \hat{\phi}\right)^2 +
 \frac{T}{2\rho_K}\left(\partial_x  \hat{\theta}\right)^2
 \\
&\quad +
 \frac{\eta_\perp}{R_*^2 T}\cos 2\hat{\phi}+
% %\frac{\mu_{14}}{R_*^2}\cos \left(2\pi T\theta/\rho_K\right),
 \frac{\mu_{14}}{R_*^2}\cos \left(2\pi\hat{\theta}\right),
 \\
& \left[\partial_x\hat{\phi}(x);\hat{\theta}(x')\right]=i\delta(x-x')
\end{split}
\label{N1N2phi1}
\ee 

Then using the fermionization rules
\be
\begin{split}
&\hat{R}(x)\propto \exp\left[i\hat{\phi}(x)+i\pi\hat{\theta}\right];
\\
&\hat{L}(x)\propto \exp\left[-i\hat{\phi}(x)+i\pi\hat{\theta}\right];
\end{split}
\ee
we write down the corresponding 1D quantum fermionic Hamiltonian
density as
\begin{eqnarray}
&& 
\hat{\cal H} = i(L^+\partial_xL - R^+\partial_x R) + (\pi T/\rho_K -1)R^+RL^+L \nonumber\\
&& + \eta_{\perp}(R^+L + L^+R) + \mu_{34}(R^+L^+ + LR)\label{ferm}
\end{eqnarray}
It is convenient to decompose the Dirac spinor into the real (Majorana) components:
\be
R = r_1 + ir_2, \quad L = l_1 + il_2
\ee
where the corresponding operators are real ($r_a = r^+_a, l_a = l_a^+)$ and satisfy the following commutation relations:
\begin{eqnarray}
&& \{r_a(x_1),r_b(x_2)\} = \delta_{ab}\delta(x_{12}), \\
&&\{l_a(x_1),l_b(x_2)\} = \delta_{ab}\delta(x_{12}),  \{r_a(x_1),\l_b(x_2)\} =0\nonumber
\end{eqnarray}
Then Eq.~(\ref{ferm}) becomes
\begin{eqnarray}
&& {\cal H} = \frac{i}{2}(l_a\partial_x l_a - r_a\partial_x r_a) +  \tau(\l_1r_1)(l_2r_2)\nonumber\\
&& i(\eta_{\perp}/T + \mu_{34})r_1l_1 + i(\eta_{\perp}/T - \mu_{34})r_2l_2 \label{2Ising}
\end{eqnarray}
where $\tau = [T/(\pi\rho_K) -1]$. Hamiltonian (\ref{2Ising}) describes
two quantum Ising models coupled by the energy density operators. The
original order parameter operator can be expressed in terms of order
and disorder parameters of the Ising models $\sigma$ and $\mu$,
see Ref.~\onlinecite{book} for the corresponding definitions:
\be
e^{i\phi} = \sigma_1\sigma_2 + i\mu_1\mu_2 \label{order}
\ee

The sign of Majorana mass in the Ising model plays an important role
determining what operator ($\sigma$ or $\mu$) acquires a vacuum
expectation value. Then from (\ref{order}) it is clear that this
operator acquires a finite expectation value when the masses of the
two species of Majorana fermions have the same sign so that either
$\sigma_a$ or $\mu_a$ fields condense simultaneously. The high
temperature phase is characterised by masses of different sign.

At small $|\tau| << 1$ we can use perturbation theory 
to write the equations for the masses (let us choose $\eta_\perp>0$):
\begin{eqnarray}
&& m_1 = (\eta_{\perp}/T + \mu_{34}) > 0;\nonumber\\
&& m_2 = (\eta_{\perp}/T - \mu_{34}) + \frac{\tau}{2\pi}m_1\ln(1/R^*|m_1|);
\end{eqnarray} 
It follows that $m_2$ changes sign at
\be
T_c/\pi\rho_K -1 = 
\frac{\eta_{\perp}/T - \mu_{34}}{\eta_{\perp}/T + \mu_{34}}
\ln\left(\frac{1}{R^*(\eta_{\perp} + \mu_{34})}\right)
\label{tautau}
\ee
where the second order phase transition from disordered (high
temperature) 
to the ordered (low temperature) state takes place. Equation
\rref{tautau} agrees with  \req{criticalline1}.

\end{document}